%% file: toloba_accepted.tex
\documentclass[preprint2]{emulateapj}
\usepackage{multirow}

\newcommand{\Reff}{$R_e$}

\newcommand{\kms}{km~s$^{-1}$}

\newcommand{\Vsys}{$V_{\rm sys}$}
\newcommand{\VGC}{$V_{\rm GC}$}
\newcommand{\Vmax}{$V_{\rm rot}$}
\newcommand{\PAmax}{${\rm PA}_{\rm max}$}
\newcommand{\sigGC}{$\sigma_{\rm GC}$}
\newcommand{\NGC}{$N_{\rm GC}$}

\shorttitle{NGVS XVI. Angular Momentum of dEs from GC Satellites}
\shortauthors{Toloba et al.}

\begin{document}

\title{The Next Generation Virgo Cluster Survey XVI. The Angular Momentum of Dwarf Early-Type Galaxies from Globular Cluster Satellites}


\author{Elisa~Toloba\altaffilmark{1,2}\footnote{Fulbright Postdoctoral Fellow}}\email{toloba@ucolick.org}

\author{Biao~Li\altaffilmark{3,4}}
\author{Puragra~Guhathakurta\altaffilmark{1}}
\author{Eric~W.~Peng\altaffilmark{3,4}}
\author{Laura~Ferrarese\altaffilmark{5}}
\author{Patrick~C$\hat{{\rm o}}$t\'e\altaffilmark{5}}
\author{Eric~Emsellem\altaffilmark{6,7}}
\author{Stephen~Gwyn\altaffilmark{5}}
\author{Hongxin~Zhang\altaffilmark{3,4}}
\author{Alessandro~Boselli\altaffilmark{8}}
\author{Jean-Charles~Cuillandre\altaffilmark{9}}
\author{Andres~Jordan\altaffilmark{10}}
\author{Chengze~Liu\altaffilmark{11}}
\affil{$^1$UCO/Lick Observatory, University of California, Santa Cruz, 1156 High Street, Santa Cruz, CA 95064, USA}
\affil{$^2$Texas Tech University, Physics Department, Box 41051, Lubbock, TX 79409-1051, USA}
\affil{$^3$Department of Astronomy, Peking University, Beijing 100871, China}
\affil{$^4$Kavli Institute for Astronomy \& Astrophysics, Peking University, Beijing 100871, China}
\affil{$^5$National Research Council of Canada, Herzberg Astronomy and Astrophysics, 5071 West Saanich Road, Victoria, BC V9E 2E7, Canada}
\affil{$^{6}$European Southern Observatory, Karl-Schwarzschild-Str. 2, 85748, Garching, Germany}
\affil{$^{7}$Universit\'e Lyon 1, Observatoire de Lyon, Centre de Recherche Astrophysique de Lyon and Ecole Normale Sup\'erieure de Lyon, 9 Avenue Charles Andr\'e, F-69230, Saint-Genis Laval, France}
\affil{$^{8}$ Laboratoire d'Astrophysique de Marseille-LAM, Universit\'e d'Aix-Marseille \& CNRS, UMR 7326, 38 rue F. Joliot-Curie, 13388 Marseille Cedex 13, France}
\affil{$^9$CEA/IRFU/SAP, Laboratoire AIM Paris-Saclay, CNRS/INSU, Universit\'e Paris Diderot, Observatoire de Paris, PSL Research University, F91191 Gif-sur-Yvette Cedex, France}
\affil{$^{10}$Instituto de Astrof\'isica, Pontificia Universidad Cat\'olica de Chile, Vicuna Mackenna 4860, 7820436 Macul, Santiago, Chile}
\affil{$^{11}$Department of Physics and Astronomy, Shanghai Jiao Tong University, Shanghai 200240, China}

\begin{abstract}

We analyze the kinematics of six Virgo cluster dwarf early-type galaxies (dEs) from their globular cluster (GC) systems. We present new Keck/DEIMOS spectroscopy for three of them and reanalyze the data found in the literature for the remaining three. We use two independent methods to estimate the rotation amplitude (\Vmax) and velocity dispersion (\sigGC) of the GC systems and evaluate their statistical significance by simulating non-rotating GC systems with the same number of GC satellites and velocity uncertainties. Our measured kinematics agree with the published values for the three galaxies from the literature and, in all cases, some rotation is measured. However, our simulations show that the null hypothesis of being non-rotating GC systems cannot be ruled out. 
In the case of VCC~1861, the measured \Vmax\ and the simulations indicate that it is not rotating. In the case of VCC~1528, the null hypothesis can be marginally ruled out, thus, it {\it might} be rotating although further confirmation is needed. 
In our analysis, we find that, in general, the measured \Vmax\ tends to be overestimated and the measured \sigGC\ tends to be underestimated by amounts that depend on the intrinsic \Vmax/\sigGC, the number of observed GCs (\NGC), and the velocity uncertainties. The bias is negligible when \NGC~$\gtrsim 20$. In those cases where a large \NGC\ is not available, it is imperative to obtain data with small velocity uncertainties. For instance, errors of $\leq 2$~\kms\ lead to \Vmax~$< 10$~\kms\ for a system that is intrinsically not rotating.

\end{abstract}


\section{Introduction}

Dwarf early-type galaxies (dEs) are historically defined as having low luminosity ($M_B \gtrsim -18$), smooth and faint surface brightness distribution ($\mu_B \gtrsim 22$~mag~arcsec$^-2$), small amounts of interstellar gas and dust, and red colors that indicate the presence of an old stellar population. Dwarf early-type galaxies are also the most common galaxy class in high density environments \citep[e.g.][]{Sand85,Bing88} and they are rarely found in isolation \citep{Gavazzi10,Geha12}. 

Morphologically, dEs are apparently very simple systems. However, the dE classification, as any other morphological classification, is mainly based on the visual inspection of images. This makes the morphological classification strongly dependent on the data quality, depth, and resolution. Therefore, it is not surprising that, as better data becomes available, many galaxies formerly classified as dEs show a substantial amount of substructure  in form of spiral arms, disks, and/or irregular features \citep{Jerjen00,Barazza02,Geha03,Graham03,DR03,Lisk06a,Lisk06b,Lisk07,Ferrarese06,Janz12,Janz14}.

Kinematics can be more revealing than morphology, and indeed dEs with similar structural properties can have very different rotation speeds \citep{Ped02,SimPrugVI,Geha02,Geha03,VZ04,Chil09,etj09,etj11,etj14a,etj14b,etj14c,Rys13,Rys14}. Regarding their stellar populations, they span a wide range of subsolar metallicities, from [M/H]~$\sim -0.1$ to $-1.5$, and old ages, from 1 to 14~Gyr \citep{Mich08,Paudel10,Koleva11,etj14b}.

Classically, dEs have been considered primordial objects that formed early in dense environments and evolved passively through out the cosmic time \citep[e.g.][]{WR78,WF91}. However, this galaxy class is rather heterogeneous. Thus, some of the dEs --- for instance those that show subtle spiral structures ---are likely to come from a progenitor population that was transformed by the cluster environment where they reside \citep[e.g.][]{Kormendy85,Kormendy12,Boselli08a,Boselli14,Moore98,Mast05,etj09,etj12,etj14c}. Which galaxies are the progenitors of dEs and which mechanism dominates their transformation is still under debate.

Kinematics is a powerful tool to investigate the physical processes involved in the formation and evolution of galaxies. One extreme example is the kinematically decoupled core found in some massive early-type galaxies \citep[e.g.][]{Kraj11} and in a few low mass early-type galaxies \citep[e.g.][G\'erou et al., submitted]{Thomas06,Chil07,etj14a}. These could be remnants of galaxy mergers or gas accretion, and are thus good indicators of the formation history of the galaxy.

Halos of galaxies are particularly interesting because they contain precious clues of their galaxy host assembly history, like dynamical signatures of mergers, and can trace their dark matter content \citep[e.g.][]{Hoffman10,Hopkins10,Oser10,Pillepich14}. Unfortunately, due to the rapid decrease of the galaxy surface brightness with radius, stellar kinematics can typically probe the galaxy dynamics and substructure only up to $1-2$ half-light radius (\Reff). At a larger radii, it is necessary to employ bright dynamical tracers. Planetary nebulae and globular clusters (GCs) have been used to analyze the dynamical field and dark matter halo of massive early-type galaxies \citep[e.g.][]{Cote01,Cote03,napolitano09,coccato09,schuberth10,Strader11,Foster11,morganti13,Pota13}.

While most work to-date has focused on the dynamical signatures of halos of massive early-type galaxies, only seven low mass early-type galaxies have been explored out to large radii: the M31 satellites M32, NGC~205, NGC~187, and NGC~145, studied in great detail using several hundreds of resolved stars thanks to their proximity \citep{Geha06,Geha10,Howley13}, and three bright dE Virgo cluster members, VCC~1087, VCC~1261, and VCC~1528. Due to the distance to the Virgo cluster \citep[$\sim$16.5~Mpc;][]{Mei07} these last three halos have been studied through the analysis of GCs \citep{Beasley06,Beasley09}. In all seven cases, it was found that some rotation was present in the outer halos of these galaxies. Li et al. (2016, in preparation) uses the GCs of Virgo cluster dEs to infer the properties of the average dark matter halo that hosts these low mass galaxies.

In this work, we explore the dynamical properties of the halos of the Virgo cluster dEs VCC~1539, VCC~1545, and VCC~1861 using their GCs as tracers. These three dEs are fainter than the previous dEs targeted in the Virgo cluster by \citet{Beasley06,Beasley09}. 
In addition, we reanalyze the GC systems of VCC~1087, VCC~1261, and VCC~1528 using a method to obtain simultaneously the rotation and velocity dispersion of the GC system, described by \citet{Strader11} and further exploited for massive early-type galaxies by \citet{Pota13}.

To provide a solid statistical basis to this work, we explore the effects on the measured kinematics when we use finite samples of objects, which is the case of the analysis of GCs. We simulate rotating and non-rotating GC systems of typical dEs and analyze the statistical significance of the measured kinematics as a function of the rotation speed and velocity dispersion of the GC system, the number of GCs observed, and the observed velocity uncertainties.

The GCs targeted are selected from the Next Generation Virgo Cluster Survey \citep[NGVS;][]{Ferrarese12} and from the ACS Virgo Cluster Survey \citep[ACSVCS;][]{Peng06,Jordan09}. The NGVS is $\sim 900-$hour program carried out at the Canada-France-Hawaii Telescope (CFHT) between the years 2008 and 2013. The survey used MegaCam to make a panoramic map in the $u^*giz$ filters of the region contained within the cluster virial radius covering a total area of 104~deg$^2$. A subset of this area has been followed-up in the infrared \citep{Munoz14}. The NGVS has a point-source completeness limit of $g\sim25.9$~mag and a corresponding surface brightness limit of $\mu_g\sim 29$~mag~arcsec$^{-2}$. The exquisite data quality of the survey relies on an $i$ band median seeing of FWHM~$\sim 0.54''$.

One of the most important data products the NGVS will provide is an updated catalog of Virgo cluster galaxy members, including new low surface brightness objects and a full census of point-like sources in the direction of Virgo, discriminating between Milky Way, Virgo, and background objects based on color and sometimes spectroscopic information. Some papers already available in the NGVS series discuss the globular cluster population in Virgo \citep{Durrell14}, the dynamical properties of star clusters, ultra-compact dwarf galaxies (UCDs) and galaxies in core of the cluster \citep[][]{Zhu14,Zhang15}, the possible environmental origin of UCDs in Virgo's three main subclusters (Liu et al., in preparation), the dynamical and stellar population properties of early-type galaxies with similar masses to the ones presented here but with higher central surface brightnesses \citep{Guerou15}, the tidal interactions between galaxies in the cluster environment \citep{Arrigoni12,Paudel13}, and a catalog of photometric redshift estimations for background sources \citet{Raichoor14}.

This paper is organized as follows. The selection of GC candidates, observations, data reduction, velocity measurements, and membership analysis are described in Section \ref{data}. The statistical methods used to estimate the kinematics of GC systems are described in Section \ref{measurements}. The analysis of the effects of having discrete samples of GCs for rotating and non-rotating GC systems is presented in Section \ref{finite_sampling_effects}. The kinematics of the three new dEs and the reanalysis of the kinematics of the three dEs found in the literature are presented in Section \ref{our_dEs}. We discuss our findings in Section \ref{discussion}, and summarize our conclusions in Section \ref{concl}.

\section{Data}\label{data}

This paper is focused  on the analysis of the rotation speed of the globular cluster systems of VCC~1087, VCC~1261, VCC~1528, VCC~1539, VCC~1545, and VCC~1861. The three former dEs are previously analyzed by \citet{Beasley06,Beasley09}. The three latter dEs are new observations presented here for the first time. This Section focuses on the description of the new data. Table \ref{dEs_properties} shows the properties of the six dEs.

VCC~1539, VCC~1545, and VCC~1861 are a subsample of the larger Keck/DEIMOS spectroscopic survey of 21 dEs and their associated globular clusters described in Guhatakurta et al. (2016, in preparation). In this paper we focus on the three galaxies with the largest number of observed GC satellites ($\geq 9$). The remaining 18 dEs in our sample have $\leq 6$ observed GC satellites, thus they are not suitable for this kind of analysis. However, Li et al. (2016, in prep.) combines the 82 GC satellites of all 21 dEs together to analyze the dark matter profile of an average dE.

Below we summarize the criteria used to select the targets, the observation strategy, and the reduction procedure  for VCC~1539, VCC~1545, and VCC~1861 (full details will be provided in Guhatakurta et al., in prep.).
The details of the target selection, observations, and data reduction for VCC~1087, VCC~1261, and VCC~1528 are described in \citet{Beasley06,Beasley09}.

\begin{table*}
\begin{center}
\caption{Properties of the dwarf early-type galaxies and their nuclei \label{dEs_properties}}
\begin{tabular}{|c|c|c|c|c|c|c|c|c|c|c|c|c|c|}
\hline \hline
   Galaxy      &     m$_g$       & $n$   & \Reff  &  $\epsilon$ &  $R_{80}/R_{20}$ & $D$     &  $D_{\rm M87}$ & \Vsys & $V_{\rm *,rot}$ &$R_{\rm nuc}$ &  m$_{g,\rm nuc}$  & $R_{0}$ & $\sigma_0$  \\
                   &      (mag)       &           & ($''$)  &                     &                         &(Mpc)  & (deg)  &  \kms  &\kms      &  ($''$)  &  (mag)      &  ($''$)  &\kms            \\
     (1)          &        (2)          &   (3)    &   (4)    & (5)               &   (6)                  &   (7)   &   (8)     &   (9)        &   (10)     & (11)       & (12)     & (13)     & (14)         \\     
\hline
VCC~1087   &  13.8            &  1.77 & 19.26  & 0.27 & 4.6 &16.7 & 0.88 &  658.6$\pm$0.7$^1$  & 4.6$\pm$2.7$^1$ & N/A$^a$  &  20.1 & 1.2 & 32.5$\pm$2.6$^1$  \\
VCC~1261   &  13.2            &  2.03 &  20.09 & 0.27 & 5.7 &18.1 & 1.62 & 1825.3$\pm$0.7$^1$ & 1.8$\pm$3.8$^1$ & 0.23         & 19.1 &  1.2 & 41.6$\pm$1.4$^1$\\
VCC~1528   &  14.3            &  2.17 &  10.11 & 0.19 & 6.0 & 16.3 & 1.20& 1615.4$\pm$0.7$^1$ & 0.8$\pm$1.5$^1$ &  ---          & ---  &  1.2 & 47.8$\pm$3.5$^1$ \\
VCC~1539   &  15.5            &  1.35 &  17.24 & 0.09 & 4.3 &16.9 & 0.88 & 1526.0$\pm$6.1$^2$ & ---       &  0.52         & 20.4 &  0.6 & 30.9$\pm$11.1$^2$ \\
VCC~1545   &  14.5            &  2.67 &  12.53 & 0.14 & 7.7 &16.8 & 0.89 & 2073.1$\pm$1.0$^2$ & $\sim$ 25$^3$       &  0.05$^b$   & 25.0 &  0.6 & 22.5$\pm$2.6$^2$ \\
VCC~1861   &  13.8            &  1.55 &  20.29 & 0.03 & 5.6 &16.1 & 2.76 & 627.2$\pm$1.2$^2$   &  5.3$\pm$2.5$^1$    &  N/A$^a$   & 20.2 &  0.6 & 20.8$\pm$2.8$^2$ \\
\hline
\end{tabular}
\end{center}
\tablecomments{Column 1: galaxy name. Column 2: total apparent magnitude in the $g$ band. Column 3: best fit S\'ersic index. Column 4: radius that contains half of the light in the $g$ band. Column 5: mean $g$ band isophotal ellipticity within the \Reff. Column 6: concentration index measured as the ratio between the radii that contain $80\%$ and $20\%$ of the light in the $g$ band. Column 7: distance estimated from surface brightness fluctuations by \citet{Mei07}. Typical uncertainties are $\sim 0.5$~Mpc. Column 8: projected distance to M87, considered to be the center of the Virgo cluster. Column 9:  heliocentric corrected systemic velocity. Column 10: stellar rotation at the \Reff\ from the literature. Column 11: radius of the nucleus in the $g$ band. Column 12: apparent $g$ band magnitude of the nucleus. Column 13: radius of the central region used to measure the kinematics. Column 14: central velocity dispersion.\\
$^1$ Measurements from \citet{etj14b}. \\
$^2$ Measurements done in the spectra presented here. \\
$^3$ Estimation based on the velocity map by \citet{Chil09}. This map only extends from the center to $\pm \sim5.5''$ along the major axis, thus this value should be considered a lower limit.\\
$^a$ Indicates not resolved nucleus (value not available).\\
$^b$ Indicates marginal detection of the nucleus.}
\end{table*}

\subsection{Photometric Selection of Spectroscopic Candidates}\label{sample}

Our parent sample comprises 21 low luminosity and low surface brightness early-type galaxies that fall into the NGVS footprint. These 21 dEs have $g$ band magnitudes in the range $-17.2 < {\rm M}_g < -13.7$ (see Guhathakurta et al. 2016).

Nine out of the 21 dEs are observed with HST as part of the ACS Virgo Cluster Survey \citep[ACSVCS;][]{Cote04}. This survey is conducted in the $g$ and $z$ bands and, due to the high spatial resolution of HST/ACS, nearly all the GCs appear as extended sources (the spatial resolution of HST/ACS, 0.05~$''/$pix,  corresponds to 3.9~pc at the distance of the Virgo cluster). \citet{Peng06} demonstrates that the color-magnitude diagram based on $g$ and $z$ band magnitudes in combination with size information provide a clean separation between foreground stars, GCs, and background galaxies. Thus, we use the ACSVCS GC catalog of \citet{Peng06} and \citet{Jordan09} to select candidates of GC satellites of the available nine dEs. 

For the remaining 12 dEs not included in the ACSVCS we use the NGVS GC catalog to select our sources. Due to the large field of view of the Keck/DEIMOS spectrograph ($16.3' \times 5'$) in comparison with the HST/ACS camera ($3.4' \times 3.4'$) we also include NGVS GC candidates at even larger radii for the dEs included in the ACSVCS when possible. All the GC candidates are selected to have colors of $0.5 < g-i < 1.4$ which is the expected color range for low luminosity early-type galaxies \citep{Peng06}.

We restrict our selection to $g<24.5$ even though we know that the contamination below $g=23$ is high. The main source of contaminants at those faint magnitudes are background galaxies, which are easily separated from the GCs by their spectroscopic characteristics, i.e. emission lines. In addition, to fill the Keck/DEIMOS slitmasks in the most efficient way, we also include some point-like objects with bluer colors than the GC candidates ($0.25 < g-i < 0.45$). These are candidate Milky Way foreground stars belonging to the Sagittarius stream and the Virgo overdensity. The data related to these foreground targets will be presented in a future paper (Peng et al., in prep.). Table \ref{breakdown} shows a summary of all the target objects in the parent sample and a breakdown indicating whether they are spectroscopically confirmed, contaminants, or failed spectra due to instrumental artifacts. A number of objects are serendipitously intercepted by the slits. In Table \ref{breakdown}  we also indicate whether these serendipitous detections are GC or Milky Way star candidates.

\begin{figure}
\centering
\includegraphics[bb= 220 422 594 820,angle=90,width=7.5cm]{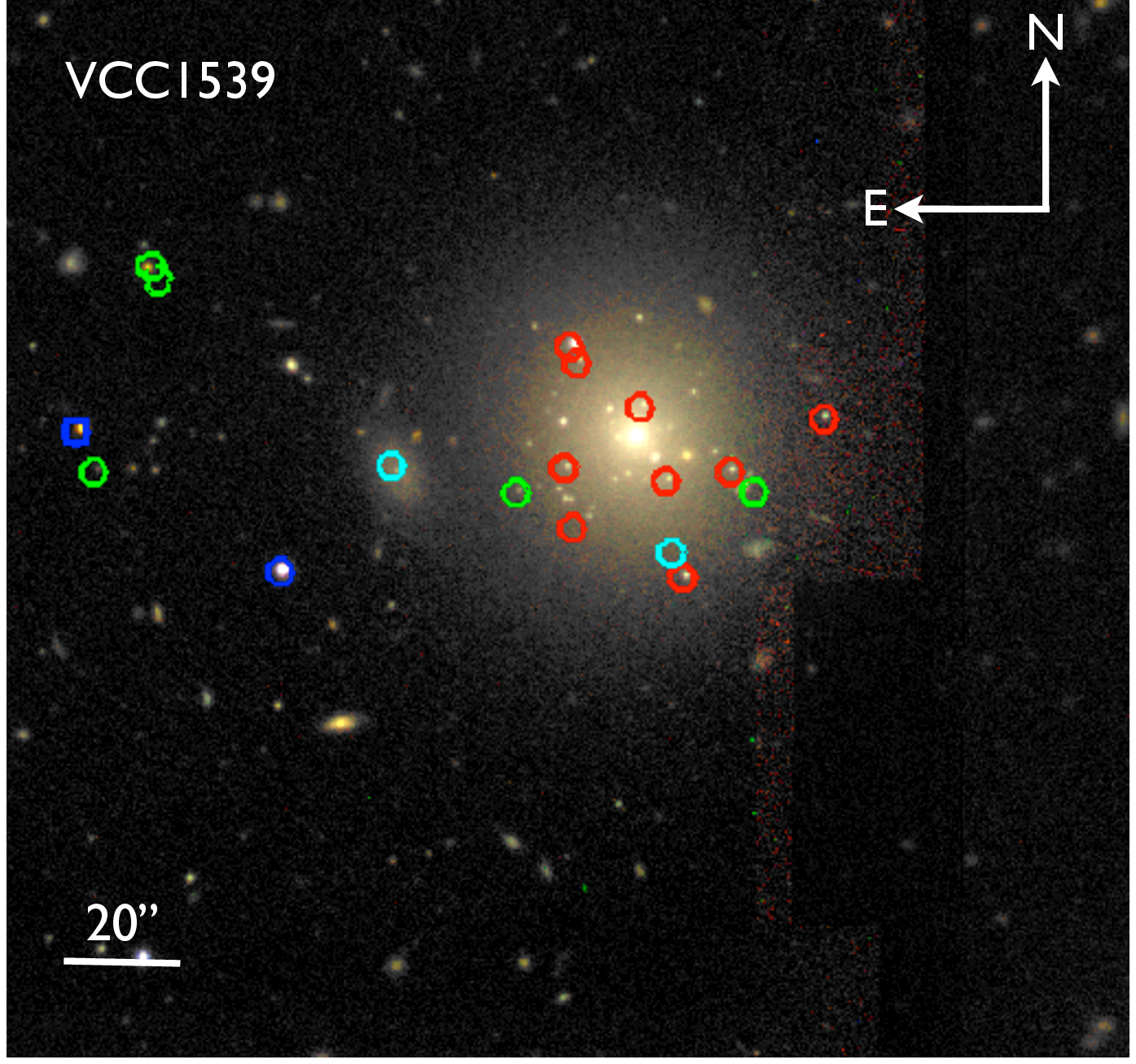}
\includegraphics[bb= 190 422 594 820,angle=90,width=7.5cm]{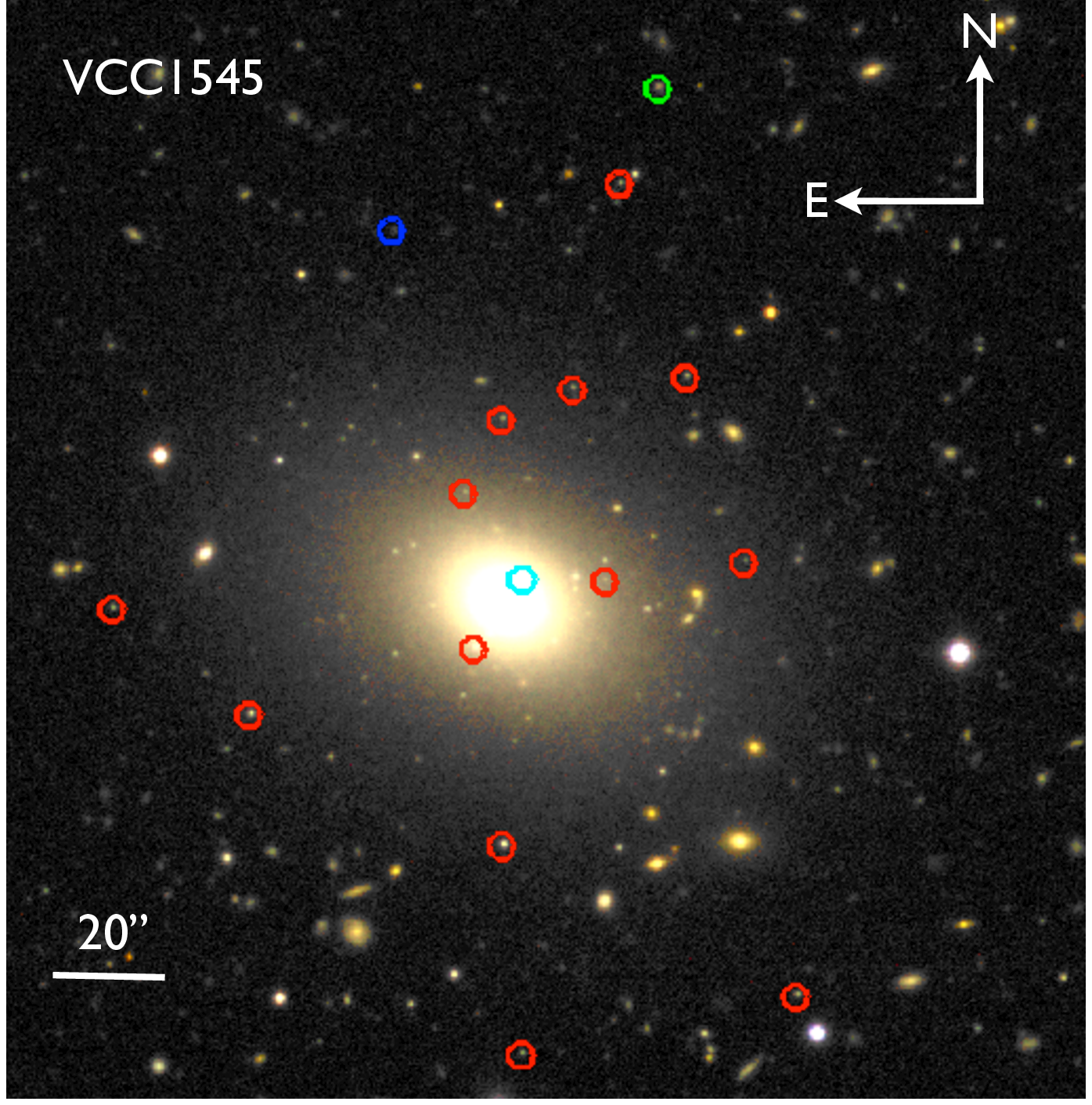}
\includegraphics[bb= 175 340 594 817,angle=90,width=7.5cm]{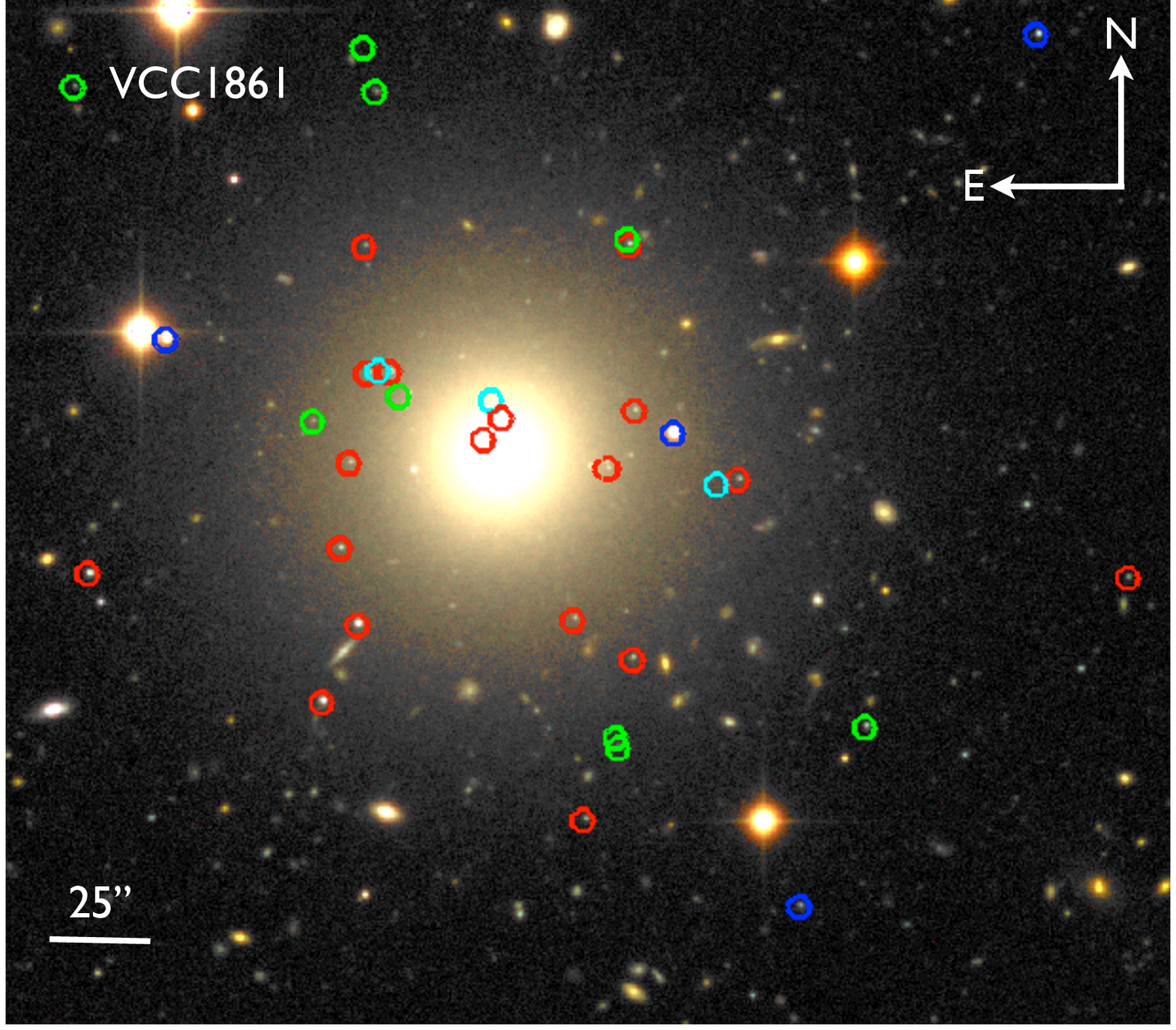}
\caption{Next Generation Virgo Cluster Survey $giz$ color images for VCC~1539, VCC~1545, and VCC~1861. These three panels show a zoom-in of the much larger Keck/DEIMOS mask, whose field of view is $16.3' \times 5'$. Red circles indicate the spectroscopically confirmed target GC satellites of the dE (see Section \ref{member_criteria} for a description of the membership criteria). Blue circles indicate foreground stars. Green circles indicate background galaxies. Cyan circles indicate those objects for which the spectrum is not retrieved due to instrumental artifacts. }\label{candidates}
\end{figure}

Figure \ref{candidates} shows the NGVS $giz$ color images for VCC~1539, VCC~1545, and VCC~1861. The color symbols indicate GC that are confirmed members of the dE (see Section \ref{member_criteria} for a description of the membership criteria), foreground stars, background galaxies, and targeted objects for which we could not measure a radial velocity because of instrumental artifacts affecting the spectrum.

\begin{table*}
\begin{center}
\caption{Summary of the Observed GC and Milky Way Star Candidates in the Parent Sample\label{breakdown}}
{\renewcommand{\arraystretch}{1.}
\resizebox{15cm}{!} {
\begin{tabular}{c|c|c|c|c|c|c}
\hline \hline
 Candidates   &    Targets      & Confirmed   & Galaxy               & Other               & Serendipitous & Failures \\
                     &                      &                    & Contaminants    & Contaminants  &                       &               \\
   (1)             &        (2)          &      (3)          &   (4)                     &   (5)                  &  (6)                 &     (7)       \\     
\hline
Virgo cluster GCs              &         317        &    142          &  40                     &   117      &  28        & 17         \\
Milky Way Stars             &         244        &    192          &  13                     &     28          &  7          &   9          \\
\hline
\end{tabular}
}}
\end{center}
\tablecomments{Column 1: target type based on photometry. Column 2: number of photometric target candidates. Column 3: spectroscopically confirmed targets that correspond to the classification in Column 1. Column 4: number of emission-line galaxy contaminants. Column 5: number of other types of contaminants. In the case of GC candidates, these are unknown objects that could be either stars redder than $g-i=0.45$ or GCs because their systemic velocities are in the range of velocity overlap between the Virgo cluster and the Milky Way objects ($-300 <$~\Vsys~$<300$~\kms). In the case of star candidates, these contaminants have stellar colors ($0.25 < g-i < 0.45$) but their systemic velocities are well beyond the range of the typical Milky Way objects (\Vsys~$\ll -300$~\kms\ or \Vsys~$\gg 300$~\kms). For more details on the nature of these objects see Guhathakurta et al. (in prep.). Column 6: serendipitous objects intercepted by our designed slits with radial velocities consistent with the object classification indicated in Column 1. Column 7: the failures are instrumental artifacts, objects for which we could not obtain a spectrum.}
\end{table*}

\begin{figure*}
\centering
\includegraphics[angle=0,width=8.5cm]{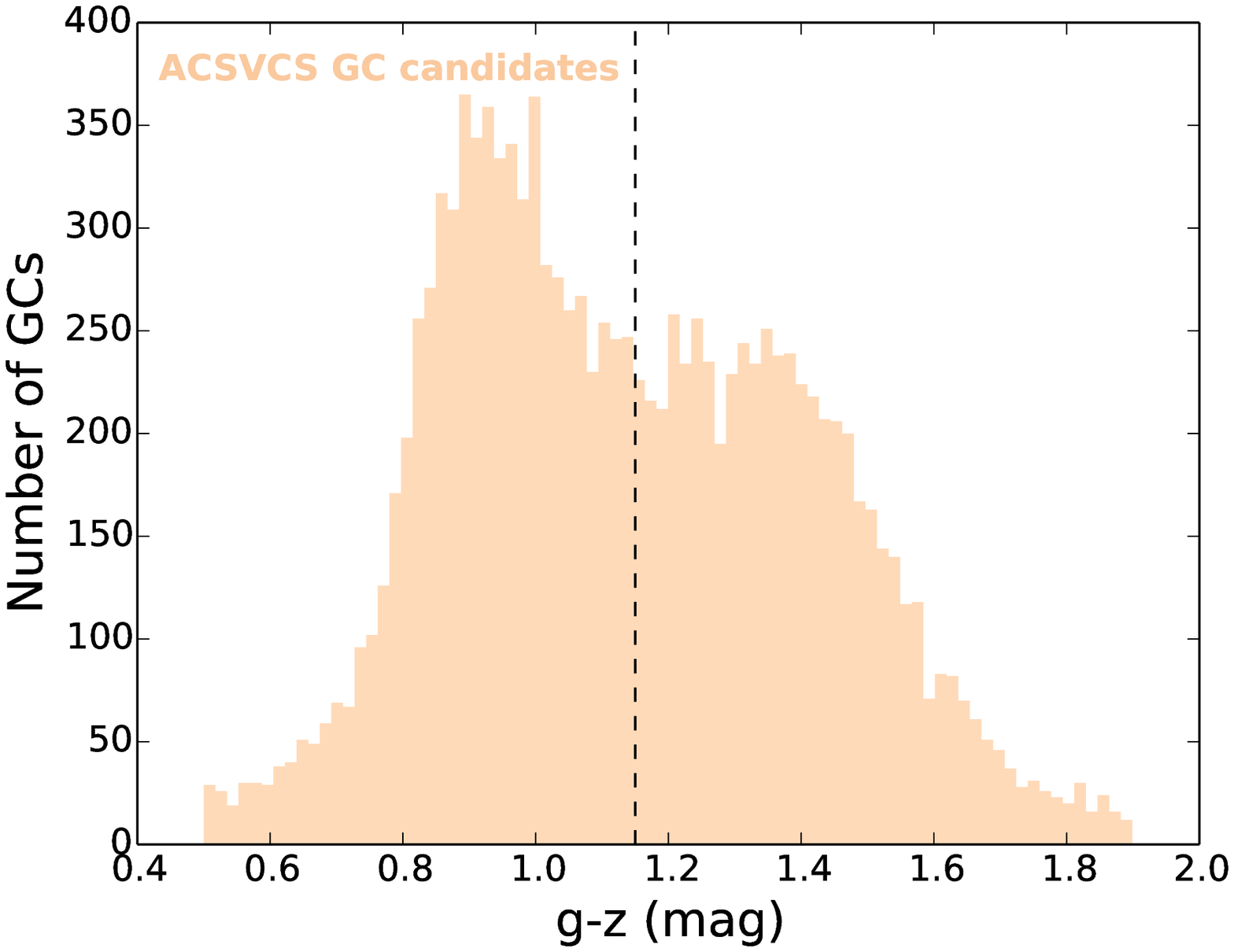}
\includegraphics[angle=0,width=8.5cm]{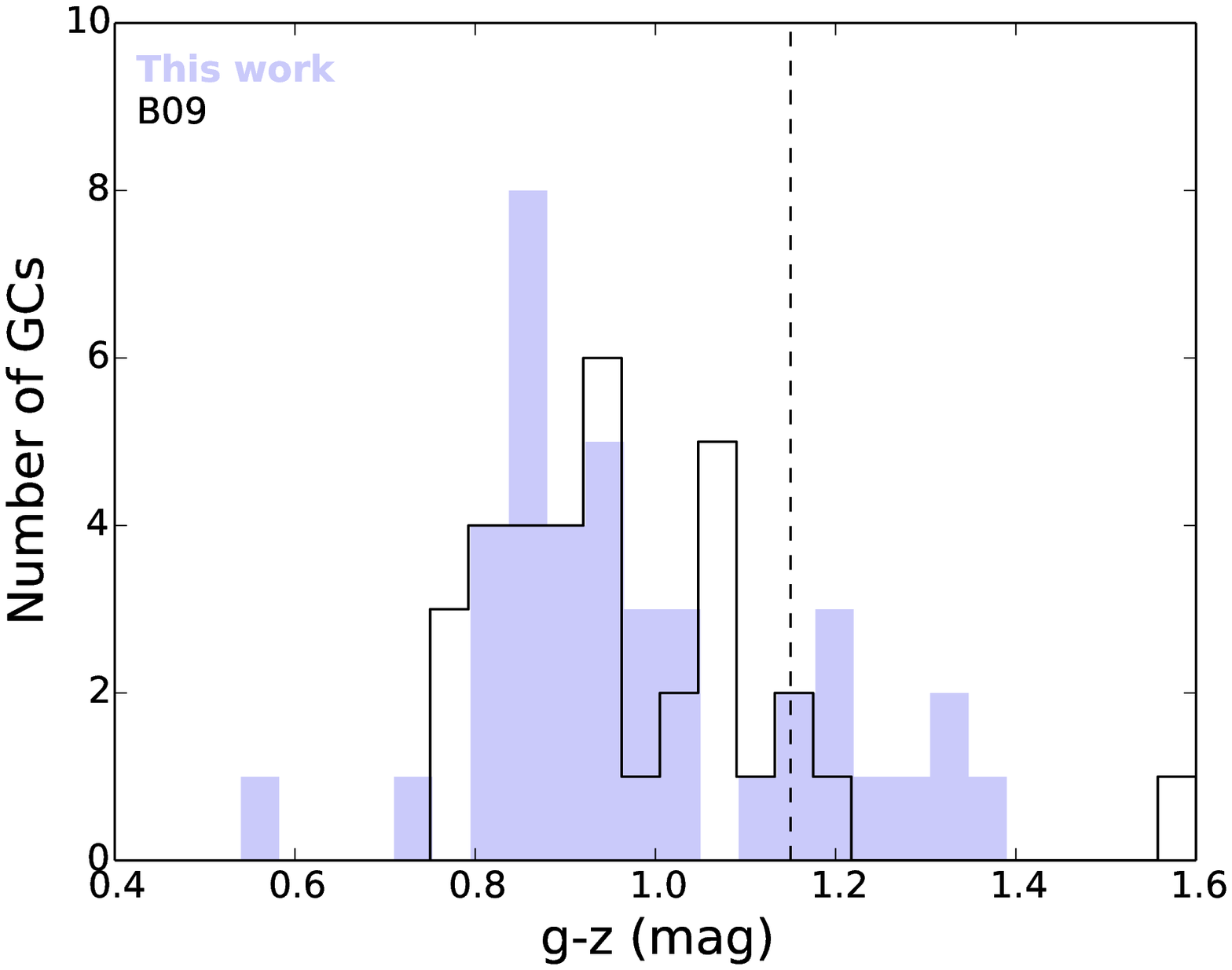}
\caption{Color distribution of GC candidates for the 100 Virgo cluster early-type galaxies of the ACSVCS \citep[these galaxies cover the luminosity range $-22<M_B<-15$;][]{Peng06,Jordan09} on the left, and color distribution of GC satellites for the six galaxies analyzed here on the right. The purple histogram shows the GC candidates of the three dEs studied here for the first time, and the black contour shows the GC candidates for the three dEs studied by \citet{Beasley06,Beasley09}. The vertical dashed line in both panels indicate the rough separation between the blue and red GC populations \citep[see e.g.;][]{Peng06}.} \label{colors}
\end{figure*}

Figure \ref{colors} shows the color distribution of the ACSVCS sample of GC candidates for 100 Virgo cluster early-type galaxies that span a luminosity range $-22<M_B<-15$. The fraction of metal-rich GCs, i.e. red GCs, those with $g-z>1.1$ increases with luminosity, thus, the GC satellites of dwarf early-type galaxies are mainly blue or metal-poor \citep{Peng06,Jordan09}. Figure \ref{colors} also shows the color distribution of the observed GC satellites for the six galaxies analyzed here. The number of red GC satellites is too small to analyze the blue and red GCs separately for these galaxies.

\subsection{Spectroscopy and Data Reduction}\label{obs}

We use the DEIMOS spectrograph \citep{DEIMOS} located at the Keck~II 10~m telescope in the Mauna Kea Observatory (Hawaii) to observe 9 slitmasks. The dEs VCC~1539 and VCC~1861 are targeted in two different slitmasks while VCC~1545 is targeted in just one slitmask.

The observations are carried out using the 600~l/mm grating centered at 7000~\AA~ with slit widths of 0.8$''$ and the GG455 filter to block shorter wavelength light. All slits are aligned with the mask position angle. This instrumental configuration provides a wavelength coverage of $4800-9500$~\AA~ with a spectral pixel scale of 0.52~\AA/pixel, and a spectral resolution of 2.8~\AA\ (FWHM) or an instrumental resolution of $\sim 50$~\kms. The exposure time for each mask ranges between 3600~s and 4800~s under seeing conditions of $0.6''-0.9''$ (FWHM). The different exposure time for each mask is an attempt to compensate for the non-uniformity of the observing conditions, such as transparency, seeing, moon light, etc. In Table \ref{Tobs} we summarize the observations.

The raw two-dimensional spectra are reduced and extracted into one-dimensional spectra using the standard {\sc spec2d} pipeline for DEIMOS designed by the DEEP Galaxy Redshift Survey team \citep{DEIMOSpipeline1,DEIMOSpipeline2} and modified by \citet{SimonGeha07} to optimize the reduction of unresolved targets and by \citet{Kirby15} to make it more suitable for bluer DEIMOS spectra, given that the standard DEIMOS pipeline was designed for the 1200~l/mm grating. The main steps in the reduction process consist of flat-field and fringe corrections, wavelength calibration, sky subtraction, and cosmic ray cleaning. Due to the large tilt of the grating to achieve blue wavelengths (central wavelength 7000~\AA~ instead of the nominal 7200~\AA), some ghosts appear in the two-dimensional spectra for some of the slits resulting in inaccurate wavelength calibration in the region $< 6500$~\AA. We use the sky lines located at 5200~\AA, 5577~\AA, and 6300~\AA\ to make small linear adjustments in the wavelength calibration, i.e. the adjustment ($\Delta\lambda$) is a first order polynomial that depends on the wavelength ($\Delta\lambda = a\lambda+b$, $a$ and $b$ are the best fit slope and intercept for the observed wavelength of the three sky lines with respect to their intrinsic wavelength). After applying this correction, the accuracy of the wavelength calibration is 0.03~\AA. However, the region of the spectra with $\lambda < 5200$~\AA\ is not used to measure radial velocities because our data has a low signal-to-noise ratio (a median of S/N~4~\AA$^{-1}$ in the calcium triplet region and even lower at wavelengths bluer than 7500~\AA\ where the detector blazes) and because the bluest wavelengths are not covered for all slits due to their placement in the slitmask. We assess the reliability of the measured radial velocities performing two tests: 1) we split the wavelength coverage in two pieces at $\sim7000$~\AA\ and measure the radial velocity independently in both pieces; 2) we compare the radial velocities obtained for repeated measurements of target objects.  In both cases, the resulting difference in the measured radial velocities divided by the square root of the quadratic sum of their errors follows a Gaussian distribution whose width is unity (see the Appendix for more details on how the radial velocities are measured and how their uncertainties are estimated).

The reduced one-dimensional spectrum is obtained by identifying the target in the reduced two-dimensional spectrum and extracting a small window centered on it. The target is identified by finding the peak of the spatial intensity profile obtained by collapsing the two-dimensional spectrum in the wavelength direction. A Gaussian function is fitted to the target and its width is used as extraction window. The one-dimensional spectrum is obtained by extracting the spectra within this window weighting by the Gaussian distribution, i.e. each pixel is weighted by the value of the Gaussian distribution in that same pixel. This follows the standard optimal extracting scheme of \citet{Horne86}. In some cases this technique fails due to instrumental effects such as bad columns. When that happens the extraction window is a boxcar centered on the target.

\begin{table}
\begin{center}
\caption{Summary of the Spectroscopic Observations\label{Tobs}}
\begin{tabular}{c|c|c|c|c}
\hline \hline
   Galaxy      &     Slitmask       & Exposure Time   & PA       & Seeing (FWHM) \\
                   &                          &      (s)                  & (deg)   &  ($''$)                \\
     (1)          &        (2)              &      (3)                  &   (4)     &   (5)                  \\     
\hline
VCC~1539   &         4               &4800               &  119.2  &   0.8    \\
VCC~1539   &         5               &3600               &   23.9   &   0.6    \\
VCC~1545   &         6               &3600               & $-20.0$   &   0.6    \\
VCC~1861   &         8               &3600               &   14.9   &   0.7    \\
VCC~1861   &         9               &4200               &  103.0  &   0.5    \\
\hline
\end{tabular}
\end{center}
\tablecomments{Column 1: galaxy name. Column 2: slitmask number. Column 3: exposure time in seconds. Column 4: position angle of the long side of the DEIMOS mask in degrees measured North-East. Column 5: average seeing during the observation of that slitmask.}
\end{table}

\begin{figure*}
\centering
\includegraphics[width=18cm]{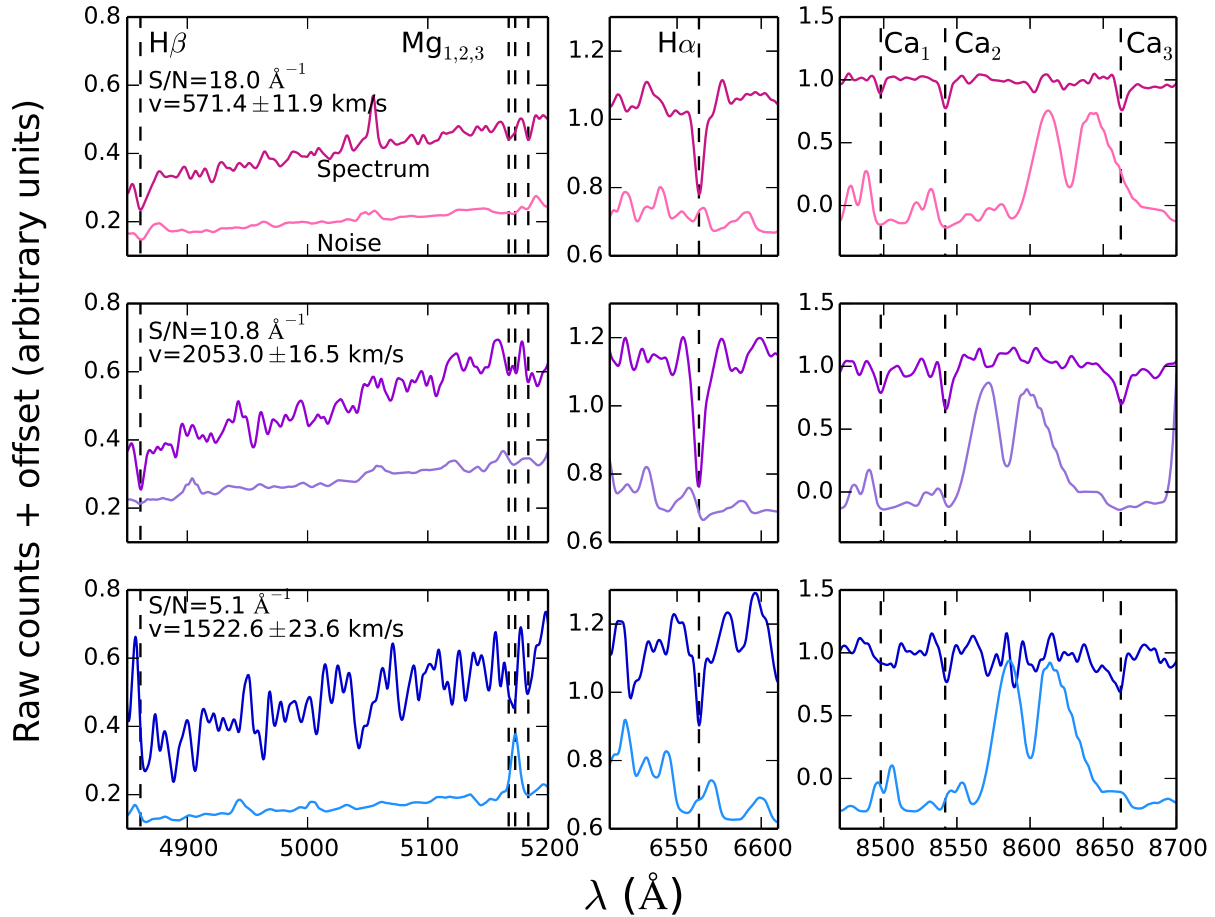}
\caption{Examples of three GC spectra with different S/N  put into the rest frame using their radial velocities. The upper spectrum in each panel shows the target spectrum and the lower spectrum is the noise. The panels are arranged in order of decreasing S/N from top to bottom. Three wavelength regions are shown for each spectrum: the region that includes the H${\beta}$ and the Mg triplet lines, the region that includes the H$\alpha$ line, and the region that includes the Ca triplet lines. These lines are indicated with vertical dashed black lines.}\label{spectra}
\end{figure*}

Figure \ref{spectra} shows three examples of GC spectra  with different S/N. The spectra are put into the rest frame using their radial velocities estimated as described in the Appendix. 

\subsection{Membership Criteria}\label{member_criteria}

\begin{figure}
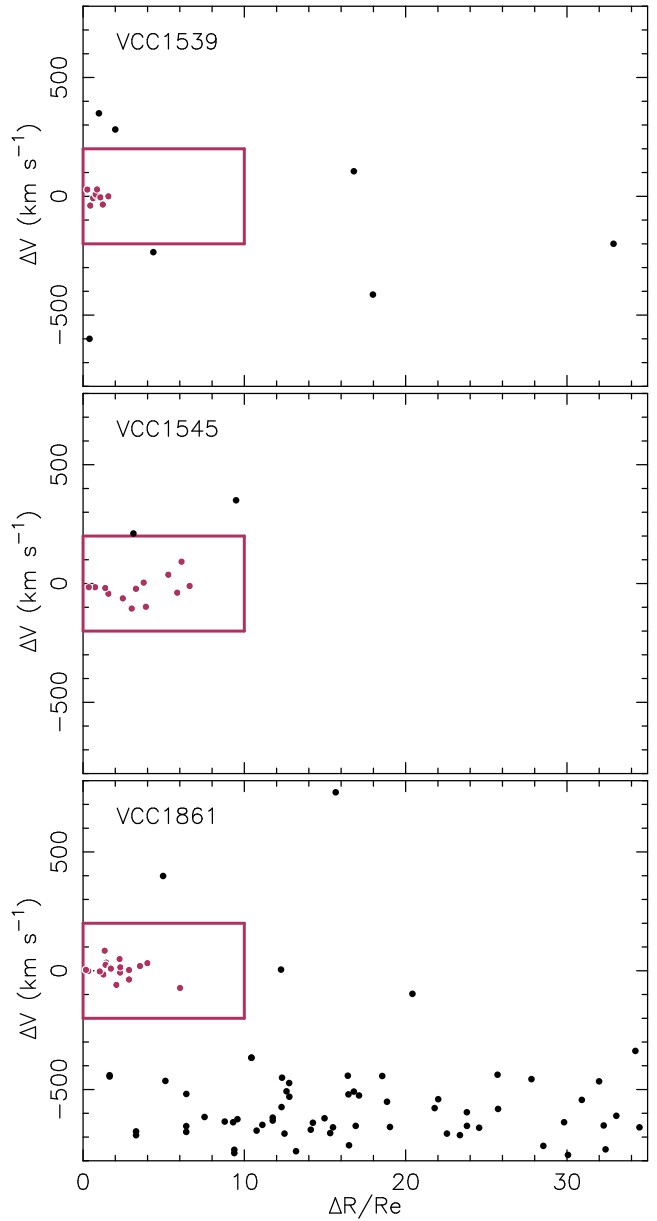

\centering
\includegraphics[angle=-90,width=8.5cm]{fig4a.ps}
\includegraphics[angle=-90,width=8.5cm]{fig4b.ps}
\includegraphics[angle=-90,width=8.5cm]{fig4c.ps}
\caption{Membership diagram. The velocity vs. projected angular position space is used to determine which GC candidates are satellites of the dEs. From top to bottom the membership criteria is applied to VCC~1539, VCC~1545, and VCC~1861. The parameter $\Delta V$ is the difference between the radial velocity of the candidate and the radial velocity of the dE. The parameter $\Delta R/$\Reff\ is the projected angular difference between the candidate and the dE normalized by the \Reff\ of the dE. All the GCs that are within the red box ($\Delta V < 200$~\kms\ and $\Delta R/$\Reff$< 10$) are considered to be satellites of that dE. The black dots outside the red box are GC satellites of a different galaxy or foreground stars.}\label{members}
\end{figure}

To determine which of the observed GC candidates are GC satellites of the three dEs studied in this paper we use the two dimensional space shown in Figure \ref{members}.
We define as satellites those globular clusters that lie within the region $|\Delta V| < 200$~\kms, which represents $\sim 8$ times the central velocity dispersion of the dE,  and $\Delta R/$\Reff$<10$. This region is chosen because it has the lowest fraction of contaminants by chance superposition. 

To measure the fraction of contaminants we scramble the systemic velocities of all 21 dEs preserving their position in the sky, thus, these artificial dEs have a sky position and systemic velocity consistent with being members of the Virgo cluster.  However, the randomly assigned velocity must be at least 800~\kms\ different from its true systemic velocity to avoid artificial dEs that are very close to the true dEs.
We make this scramble exercise 100 times and count how many objects are found within the satellite region defined by $|\Delta V| < 200$~\kms and $\Delta R/$\Reff$<10$. On average, $3.2\pm 0.7$~GCs appear in the GC satellite region combining all 21 galaxies. Thus the contamination rate in this region is $3.7\%$. Making this GC satellite region larger rapidly increases the number of contaminants, however, making it smaller does not make this number decrease. Therefore, $|\Delta V| < 200$~\kms and $\Delta R/$\Reff$<10$ represents the best compromise.

For each object targeted in our observations, which include all GC candidates and foreground stars, we combine the projected angular distance and the relative velocity with respect to the full sample of 21 dEs in the following way:

\begin{equation}\label{dist}
D=\sqrt{(\Delta R/R_e)^2+(\Delta V/\sigma_0)^2}
\end{equation}

\noindent where $\Delta R$ is the projected angular distance between the target object and a dE, \Reff\ is the half light radius of the dE, $\Delta V=$~\VGC$-$\Vsys\ is the radial velocity difference between the target object and the dE, and $\bar{\sigma}$ is the average velocity dispersion of the dE, we use a value of 30~\kms\ for all dEs. However, the number of GC satellites does not change if $\bar{\sigma}$ is $\sim 15$~\kms\ larger or smaller.

For each observed target we have 21 values of $D$ (obtained following Equation \ref{dist}), one with respect to each dE in our full sample of dEs.
We find the minimum $D$ for each target and plot it in the membership diagram shown in Figure \ref{members}. If the candidate lies within the region defined by $|\Delta V| < 200$~\kms\ and $\Delta R/$\Reff$<10$ the candidate is considered a GC satellite. All the objects that are classified as GC satellites have sizes and colors expected for GCs bound to dEs \citep{Peng06}, i.e. none of the targeted foreground stars ended up in the membership region.

This method of minimizing simultaneously the position and the velocity allows us to find the most likely host amongst the 21 dEs in our parent sample. 
Our dEs are separated by a few arcmin, since our target selection includes as many galaxies as possible in a single slitmask, and are not close in projection to any other large galaxy. This means there are minimal chances that a GC satellite of one of our dEs, is in fact part of the GC system of a galaxy that is not in our sample. We will, however, remark on the fact that there are 64 GCs in our slitmasks that do not belong to any of the 21 dEs that comprise our parent sample. These are likely to be members of Virgo cluster massive galaxies at $\Delta R/$\Reff$\gg 10$. This analysis is the focus of a future paper.

The properties of the GC satellites of VCC~1539, VCC~1545, and VCC~1861 are presented in Table \ref{GCsat_table}. VCC~1539 has a GC system that is more spatially concentrated than that of VCC~1545 and VCC~1861 (Figures \ref{candidates} and \ref{members}). This may be related with its lower luminosity and S\'ersic index (see Table \ref{dEs_properties}), which could be an indication of a lower mass and less extended halo. See Li et al. (in prep.) for a discussion of the halo masses of these galaxies.

\input{table4.tex}

\section{Measurement of Rotation Speed, Position Angle, and Velocity Dispersion of the Globular Cluster Systems}\label{measurements}

We apply two different methods to measure the rotation speed, position angle, and velocity dispersion of the GC system.
The first method, rotation fit (RF), consists on fitting a rotation model to obtain the rotation speed and the position angle, and then estimate the velocity dispersion of the GCs as their mean deviation from the model \citep[e.g.;][]{Cote01,Cote03,Beasley06,Beasley09}.

The second method, rotation and dispersion simultaneous fit (RDSF), consists on simultaneously fitting the rotation speed, the position angle, and the velocity dispersion of the GC system \citep[e.g.;][]{Strader11,Foster11,Pota13}.

Regardless of the method used, we assume that the GC system is spherical. The ellipticity of the GC system cannot be constrained with the very small number of GCs observed in each dE. We also assume that the projected velocity field of a galaxy can be described in a first order approximation by a cosine function:

\begin{equation}\label{cosine_eqn}
V_{\rm mod} = V_{\rm sys} + V_{\rm rot} \cos(PA - PA_{\rm max})
\end{equation}

\noindent where $V_{\rm mod}$ is the velocity model that best fits the GC system, \Vsys\ is the velocity of the main body of the dE, \Vmax\ is the amplitude of the cosine function which indicates the maximum measured rotation speed, $PA$ is the position angle of each GC satellite with respect to the center of the dE, and \PAmax\ is the phase of the cosine function and indicates the position angle of the maximum rotation speed. This model fits an average rotation curve to the whole system and it assumes that this curve has sinusoidal symmetry, i.e. it is maximum along an axis characterized by \PAmax\ and minimum in the orthogonal axis. 

\subsection{Rotation Fit Method (RF)}\label{RF}

We use a non-linear least squares method to find the best fit cosine function for the GC system of each dE.
The inverse squared of the velocity uncertainties are used as weights in the fitting procedure following the mathematical expression:

 \begin{equation}\label{leastsq_eqn}
\chi^2 \propto \sum_{i=1}^{i=N_{\rm GC}} \bigg\{ \frac{(V_i-V_{\rm mod})^2}{\delta V_i^2}\bigg\}
\end{equation}

\noindent where the subindex $i$ indicates the $i-$th GC, $N_{\rm GC}$ is the observed number of GCs for the target galaxy, $V_{\rm mod}$ is defined in Equation \ref{cosine_eqn}, and $\delta V$ is the uncertainty in the line-of-sight radial velocity $V$. The velocity dispersion of the GC system, \sigGC, is measured as the mean departure of the GC satellites with respect to the best fit cosine function for each dE.

The uncertainties in \Vmax, \PAmax, and \sigGC\ are estimated using a numerical bootstrap procedure as suggested by \citet{Cote01,Cote03}. This method does not make any assumption about the size and shape of the uncertainties and about the parent distribution. We generate 1000 artificial data sets by choosing at random globular clusters from the actual sample under consideration allowing repetition. The sizes of the artificial data sets are the same as the size of the actual sample of globular clusters in the target galaxy. Each artificial data set is analyzed using the same procedure as described above. The estimated uncertainties are the $68\%$ confidence intervals ($1\sigma_G$).


\subsection{Rotation and Dispersion Simultaneous Fit Method (RDSF)}\label{RDSF}

The parameters \Vmax, \PAmax, and \sigGC\ can be simultaneously obtained by minimizing the following function using a maximum likelihood statistical approach:

\begin{equation}\label{maxlik_eqn}
\chi^2 \propto \sum_{i=1}^{i=N_{\rm GC}} \bigg\{ \frac{(V_i-V_{\rm mod})^2}{\sigma_{\rm GC}^2+\delta V_i^2} + \ln[\sigma_{\rm GC}^2+\delta V_i^2] \bigg\}
\end{equation}

\noindent the subindex $i$, $N_{\rm GC}$, $V_{\rm mod}$, and $\delta V$ are the same as in Equation \ref{leastsq_eqn}. This method has been extensively used to estimate the rotation and dispersion of the GC systems bound to massive early-type galaxies \citep[e.g.][]{Foster11,Strader11,Pota13}. Note that, in those cases, the analysis is done in annuli centered in the galaxy. For dwarf galaxies, the number of GC satellites is not large enough to split it in annuli of different radius.

The uncertainties in \Vmax, \PAmax, and \sigGC\ are estimated applying the same bootstrap procedure as the one described in Section \ref{RF}.

The RDSF method is based on a kinematical model that includes both rotation and dispersion simultaneously and uses a likelihood function to evaluate the probability of a measument ($V_i \pm \delta V_i$) being drawn from a Gaussian distribution of model velocities. Contrary to the RF method, where the kinematical model only includes rotation and the dispersion is obtained as a second step in the analysis, assuming that the mean scatter of the measured velocities with respect to the best fit rotating model is a good estimation of the dispersion.

\subsection{Statistical Significance of the Measured Rotation}\label{stat_sig}

The fitting methods RF and RDSF in combination with the bootstrap procedure provide a measurement and an uncertainty of the rotation amplitude of the system. However, this measurement needs to be further tested to check whether that value is consistent or not with a spurious measurement of a non-rotating system. For example, in a extreme case where a dE only has one GC satellite, the best fit model will perfectly fit it providing a \sigGC\ of $0$~\kms and a \Vmax\ of infinity. If instead of one GC, the dE has two GC satellites, the best fit model will always go through both GCs, thus providing again a \sigGC\ of $0$~\kms\ and, in this case, \Vmax\ will depend on the PAs of the GCs. If the PAs are very similar, then \Vmax\ will tend to infinity. In these two cases, it is not possible to know whether the GC system is rotating or it is not, however, the methods will provide a very high rotation amplitude.

To address the statistical significance of the the measured rotation, we test whether the null hypothesis of the measured rotation being consistent with a non-rotating system can be rejected. We perform two types of simulations to test this null hypothesis: velocity constrained simulations and velocity unconstrained simulations.

The {\it velocity constrained simulations} (VCS) consist of simulating artificial non-rotating GC systems based on the observed line-of-sight GC velocity measurements and their uncertainties for each dE. We make these simulations using three independent approaches, but all of them keep coupled the line-of-sight velocity measurements and their associated uncertainties. The first approach generates a uniform distribution of PAs. Then a random PA between $0^{\circ}$ and $360^{\circ}$ is assigned to each GC velocity. We repeat this exercise 10000 times and each time we apply the RDSF method to obtain the best fit \Vmax, \sigGC, and \PAmax. The second and third approaches consist of randomly assigning a measured PA to each GC velocity and its associated uncertainty. In the second method we bootstrap the PAs with repetition, i.e. we assign a new PA to each velocity measurement and its uncertainty by randomly choosing  a new value from the measured PAs, repeated values are allowed, and in the third method we bootstrap the PAs without repetition, i.e. the PAs are scrambled. As done for the first approach, we repeat this exercise 10000 times and each time we apply the RDSF method to obtain the best fit \Vmax, \sigGC, and \PAmax.

The {\it velocity unconstrained simulations} (VUS) consist of simulating artificial non-rotating GC systems based on the velocity uncertainties of the full sample of 82 GC satellites observed in all 21 Virgo cluster dEs. We assume \Vmax~$=0$~\kms\ for these models but we test different \sigGC\ given that the intrinsic value is unknown. These artificial GC systems are generated following these steps: (1) we randomly choose \NGC\ velocity uncertainties from our sample of 82 GC satellites which has a median velocity uncertainty of $\delta V_{\rm GC}=19$~\kms; (2) we assign a random PA between $0^{\circ}$ and $360^{\circ}$ to each GC; (3) we assign a velocity of 0~\kms\ to each GC; (4) to make a realistic model, we perturb the velocity of each GC within a Gaussian function whose width is the assumed velocity dispersion of the GC system and within another Gaussian function whose width is the velocity uncertainty chosen in step (1). As done for the VCS, we repeat this exercise 10000 times, apply the RDSF method to each model, and measure \Vmax, \sigGC, and \PAmax.

\section{Finite Sampling and Velocity Uncertainty Effects on the Kinematics of GC Systems}\label{finite_sampling_effects}

We explore the effects due to discrete samples of velocities on modeled GC systems with different \Vmax$/$\sigGC\ values, different GC sample sizes, and different velocity uncertainties.
To create realistic GC systems we base our models on real data. For this reason, we use the full sample of 82 GC satellites observed in all 21 Virgo cluster dEs.

We explore these effects by running the VUS simulations following the steps described in Section \ref{stat_sig}. As these simulations model rotating and non-rotating systems, the velocity assigned to each GC is calculated following Equation \ref{cosine_eqn}. We assume that each modelled GC system has \PAmax~$=45^{\circ}$ and \sigGC~$=30$~\kms. The assumed \Vmax\ is the value that corresponds to the fixed \Vmax/\sigGC\ in the model. We repeat this exercise 10000 times, apply the RDSF method to each model, and measure \Vmax\ and \sigGC.

For each assumed \Vmax$/$\sigGC\ we select subsamples of GCs of different sizes. The largest subsample consists of all 82 GCs, the remaining subsamples consist of 30, 20, 11, 7, and 5 GCs.
There are two ways of selecting a subsample of GCs: (1) by creating a parent sample of GCs assigning a random PA to each one of the 82 GCs. In each realization, the GCs are selected from the parent sample; (2) by randomly choosing a subsample of GCs in each realization. Both methods get the same quantitative results and, thus, the same conclusions. We show the results obtained by using the second method.

\subsection{Finite sampling effects on rotating and non-rotating GC systems}\label{vs}

\begin{figure*}
\centering
\includegraphics[angle=0,width=5.9cm]{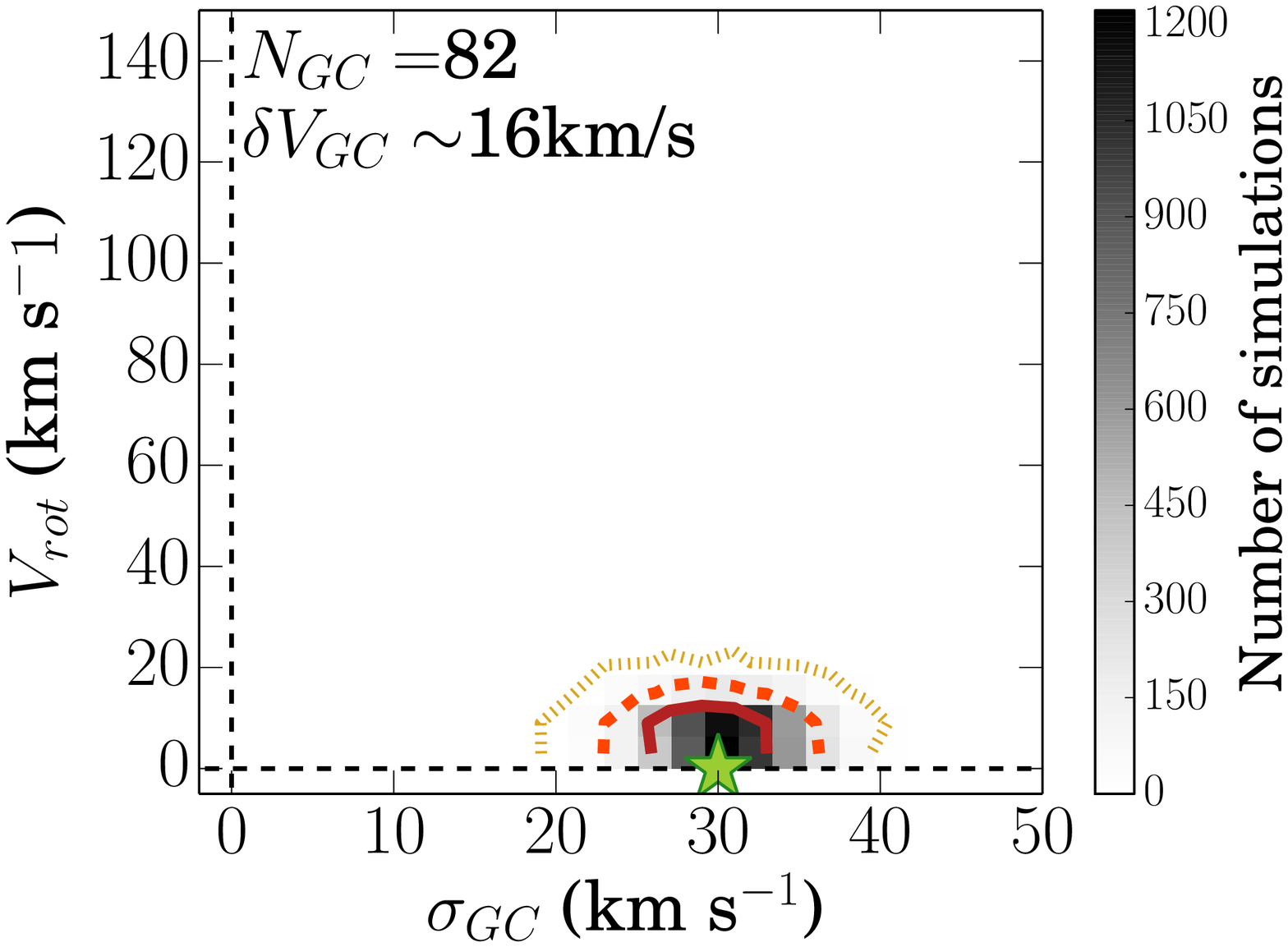}
\includegraphics[angle=0,width=5.9cm]{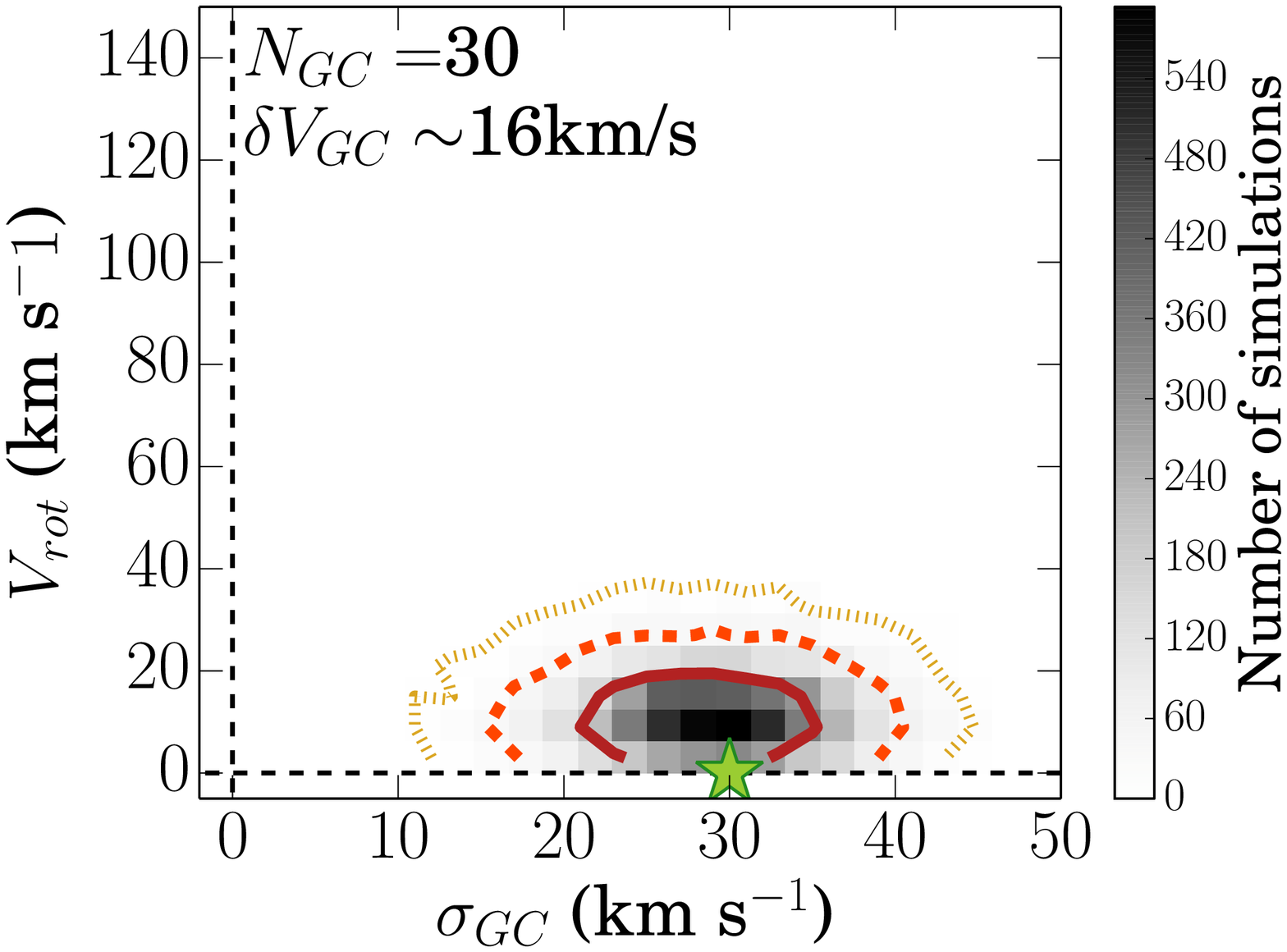}
\includegraphics[angle=0,width=5.9cm]{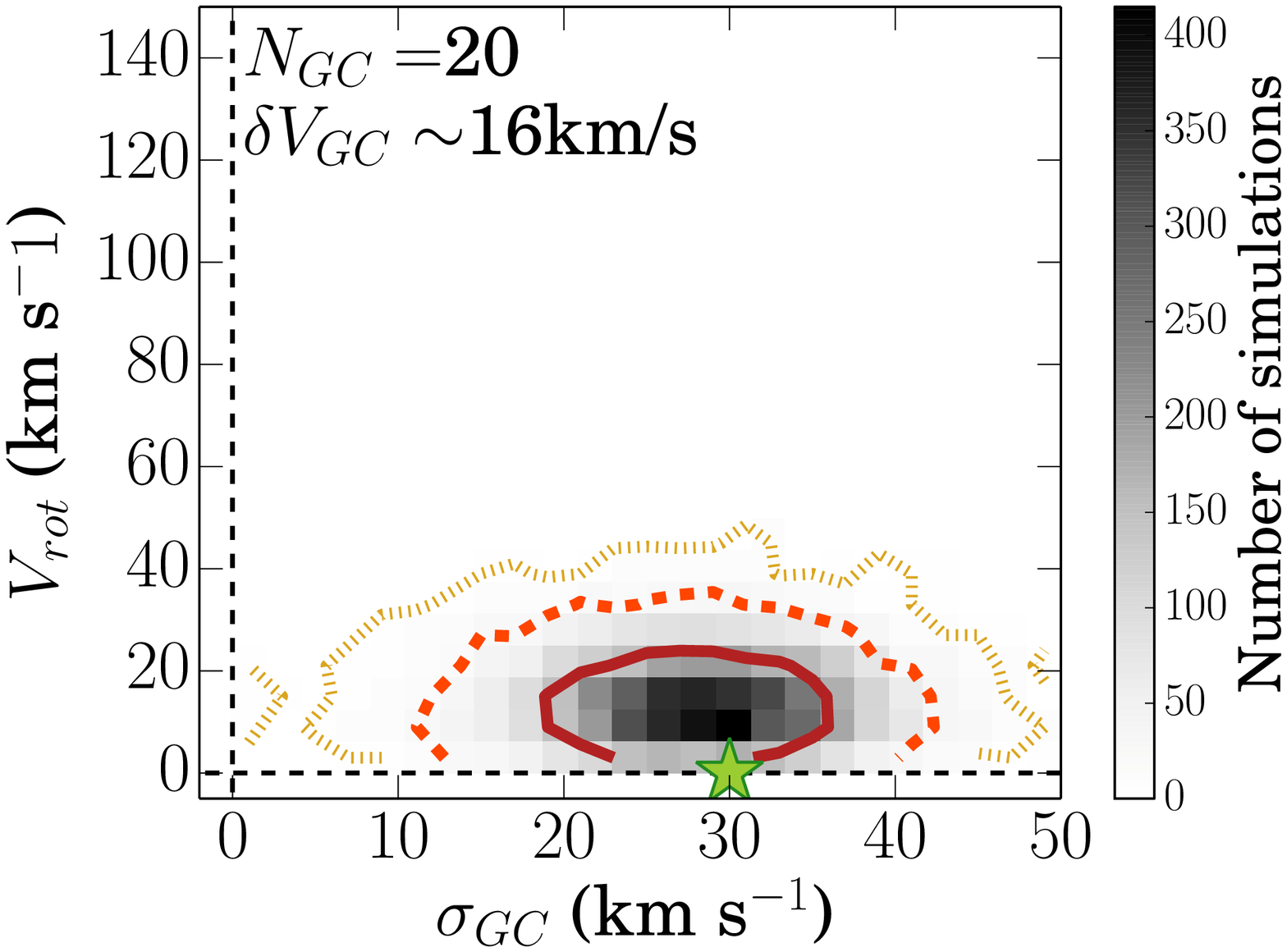}
\includegraphics[angle=0,width=5.9cm]{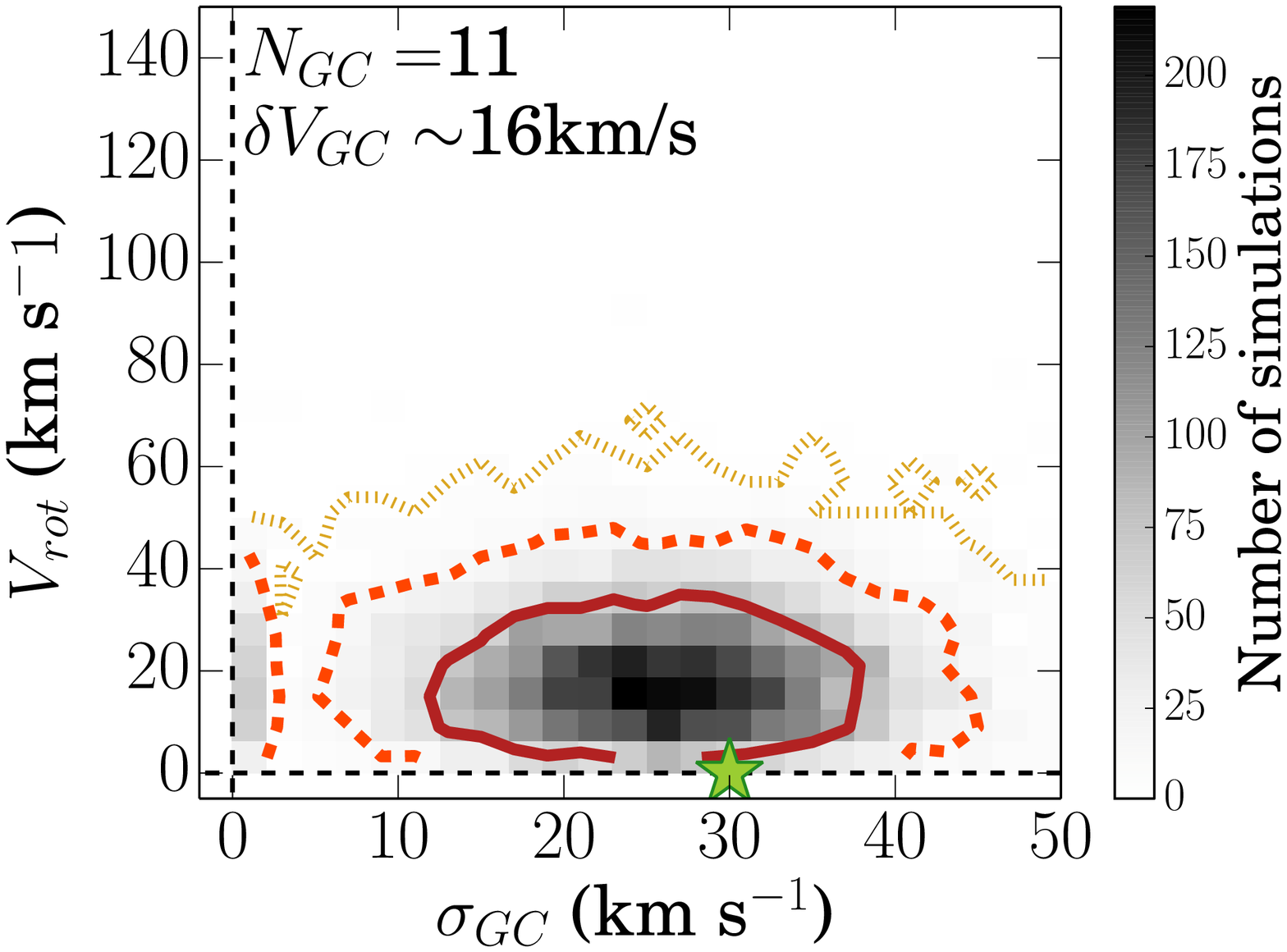}
\includegraphics[angle=0,width=5.9cm]{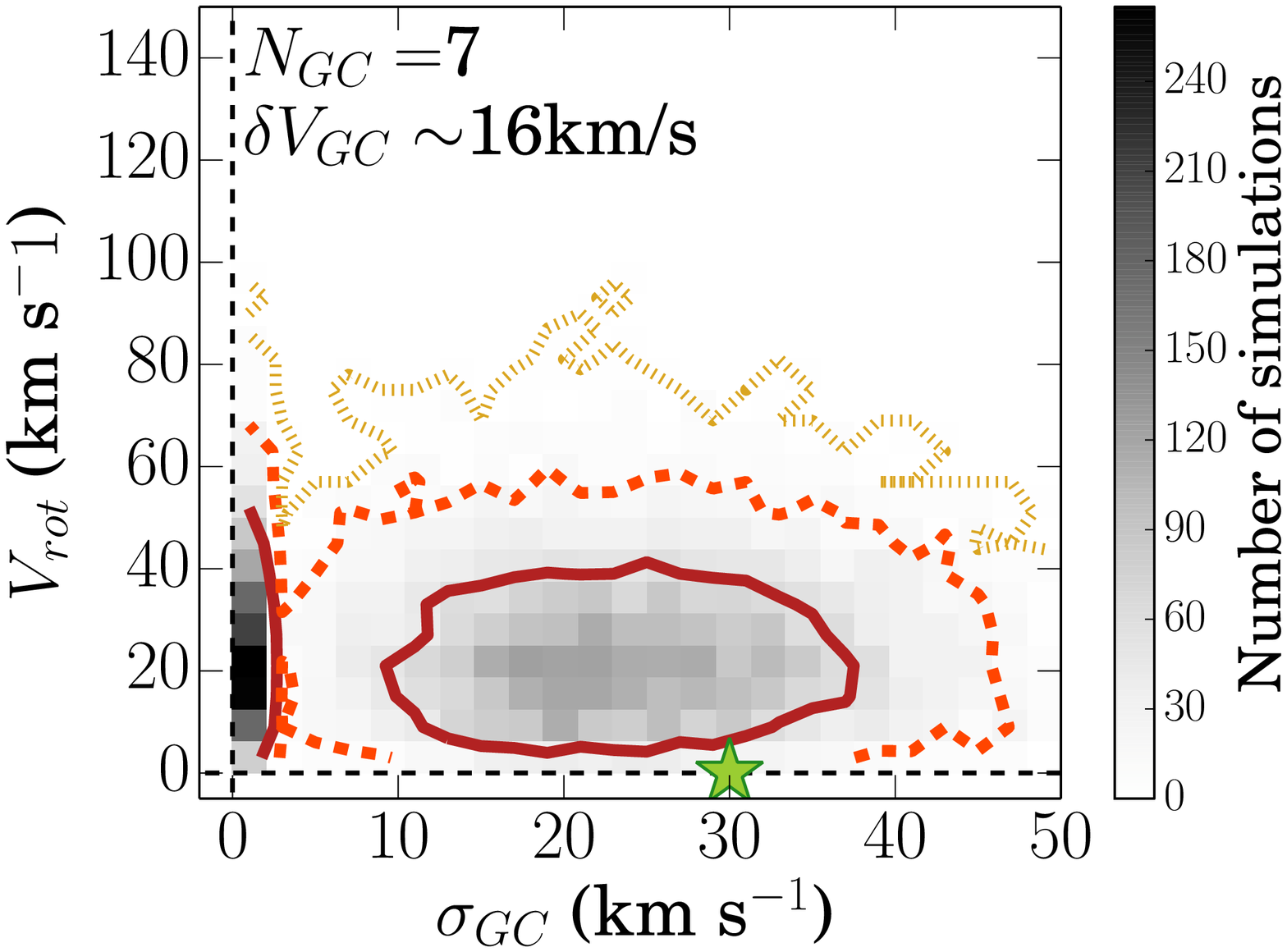}
\includegraphics[angle=0,width=5.9cm]{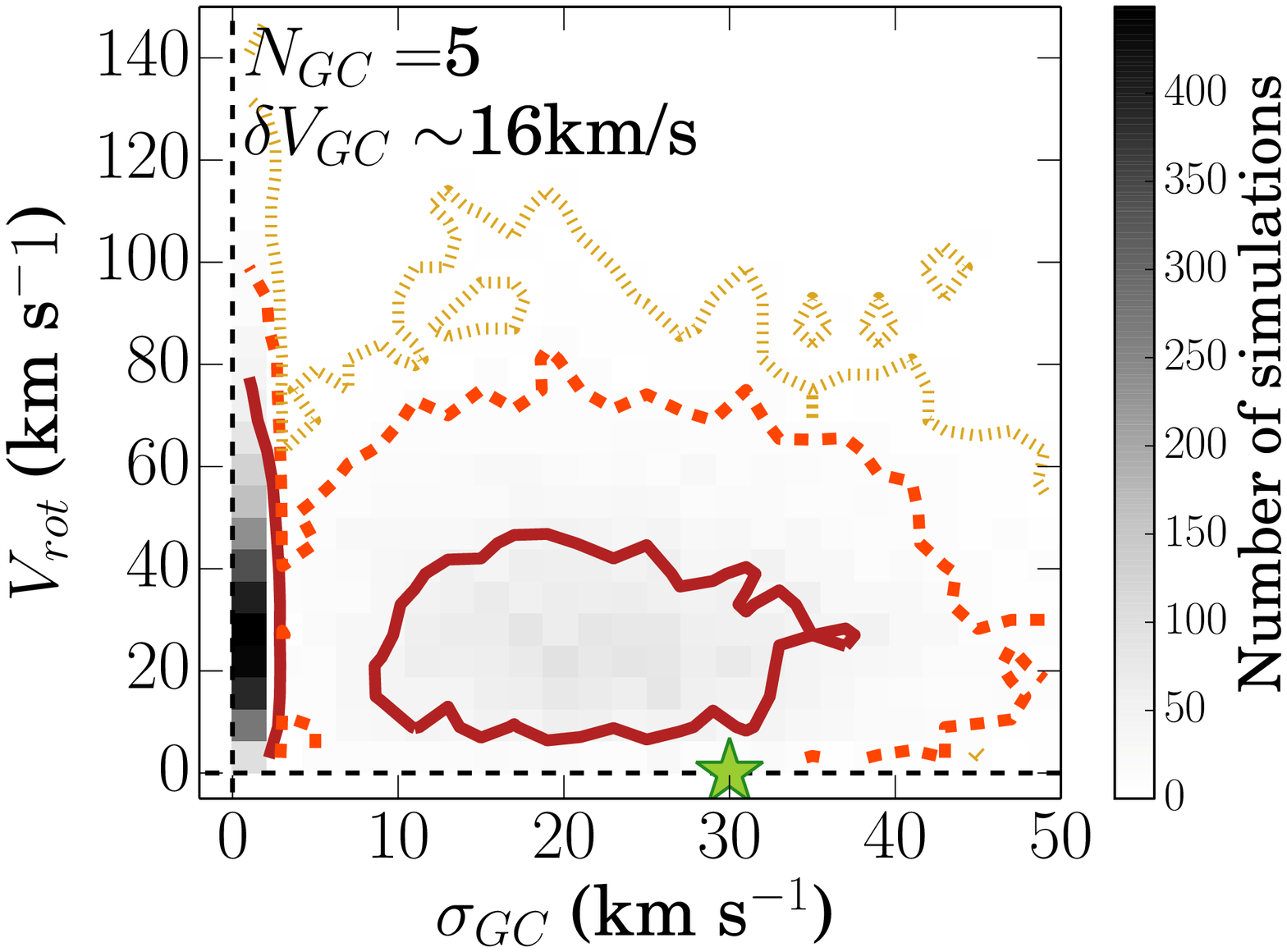}
\caption{Density map of VUS simulations (see Section \ref{stat_sig}) for a non-rotating GC system with $\sigma_{GC}=30$~\kms. The green star indicates the \Vmax\ and \sigGC\ values of the input model. Each panel is a set of VUS simulations with a different number of GCs (\NGC) but with the same median velocity uncertainty ($\delta V_{\rm GC}$). The solid red contour encloses $68\%$ of the simulations. The dashed orange contour encloses $95.5\%$ of the simulations. The dotted yellow contour encloses $99.7\%$ of the simulations.}\label{vs0}
\end{figure*}

\begin{figure*}
\centering
\includegraphics[angle=0,width=5.9cm]{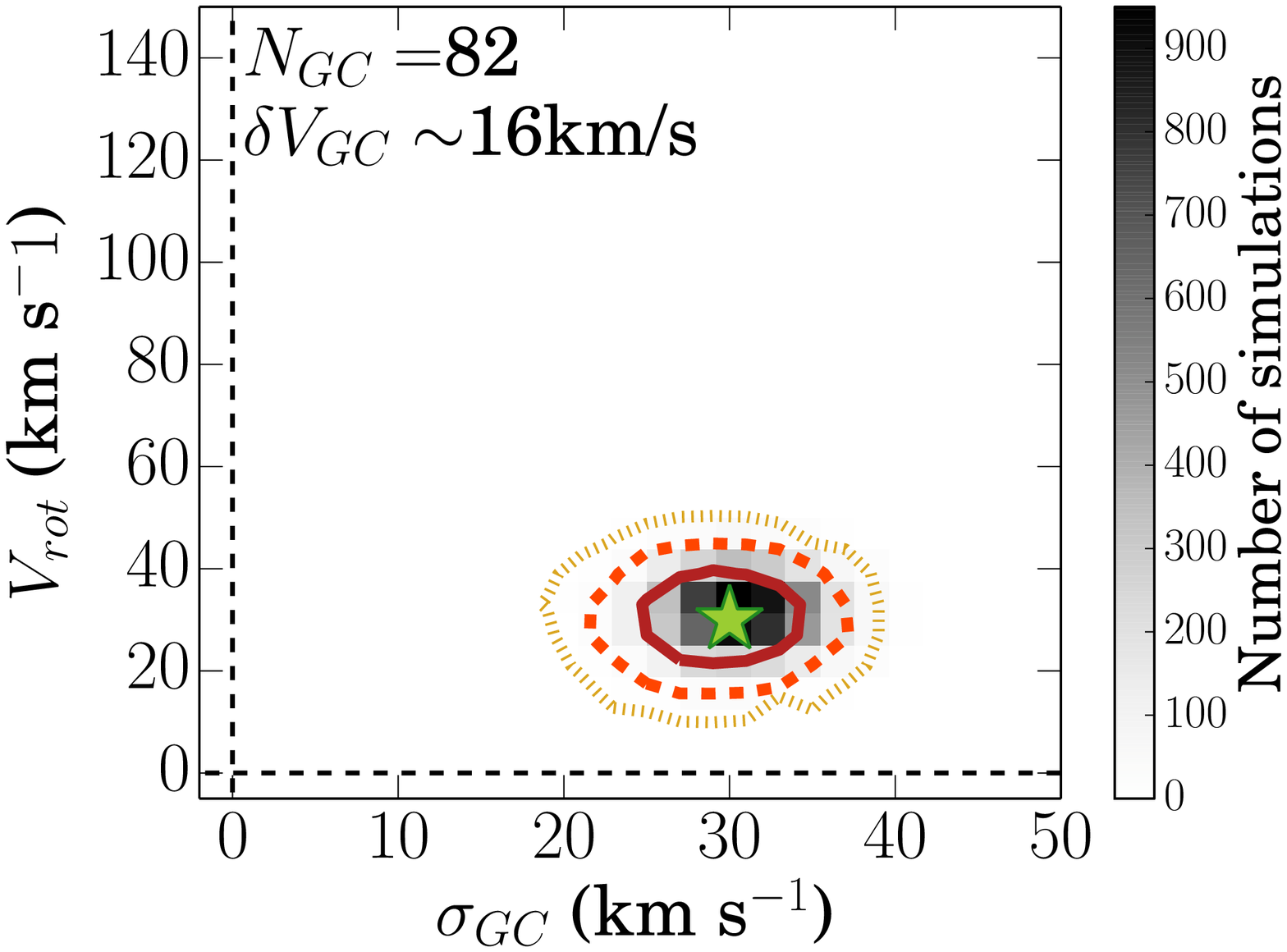}
\includegraphics[angle=0,width=5.9cm]{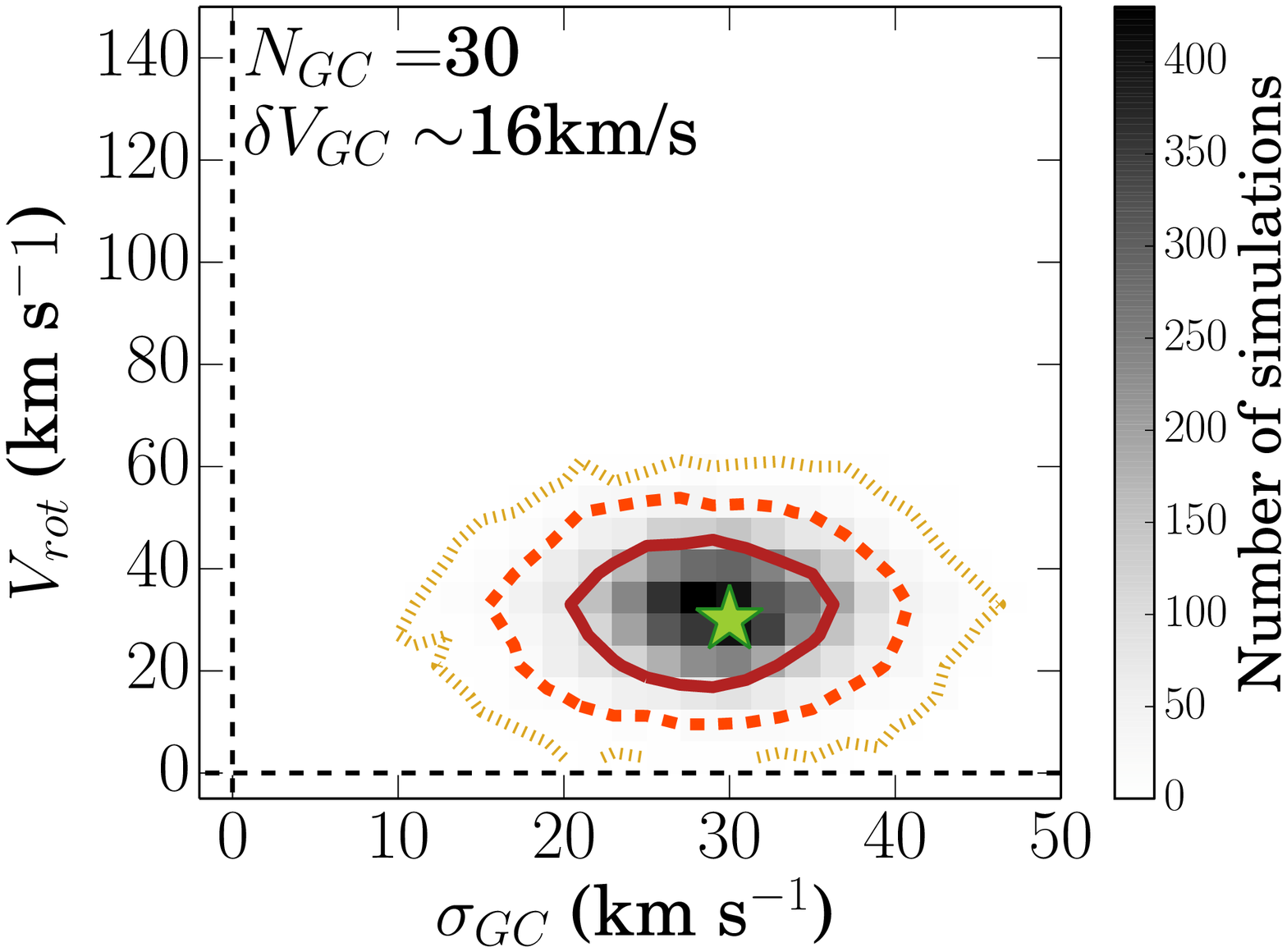}
\includegraphics[angle=0,width=5.9cm]{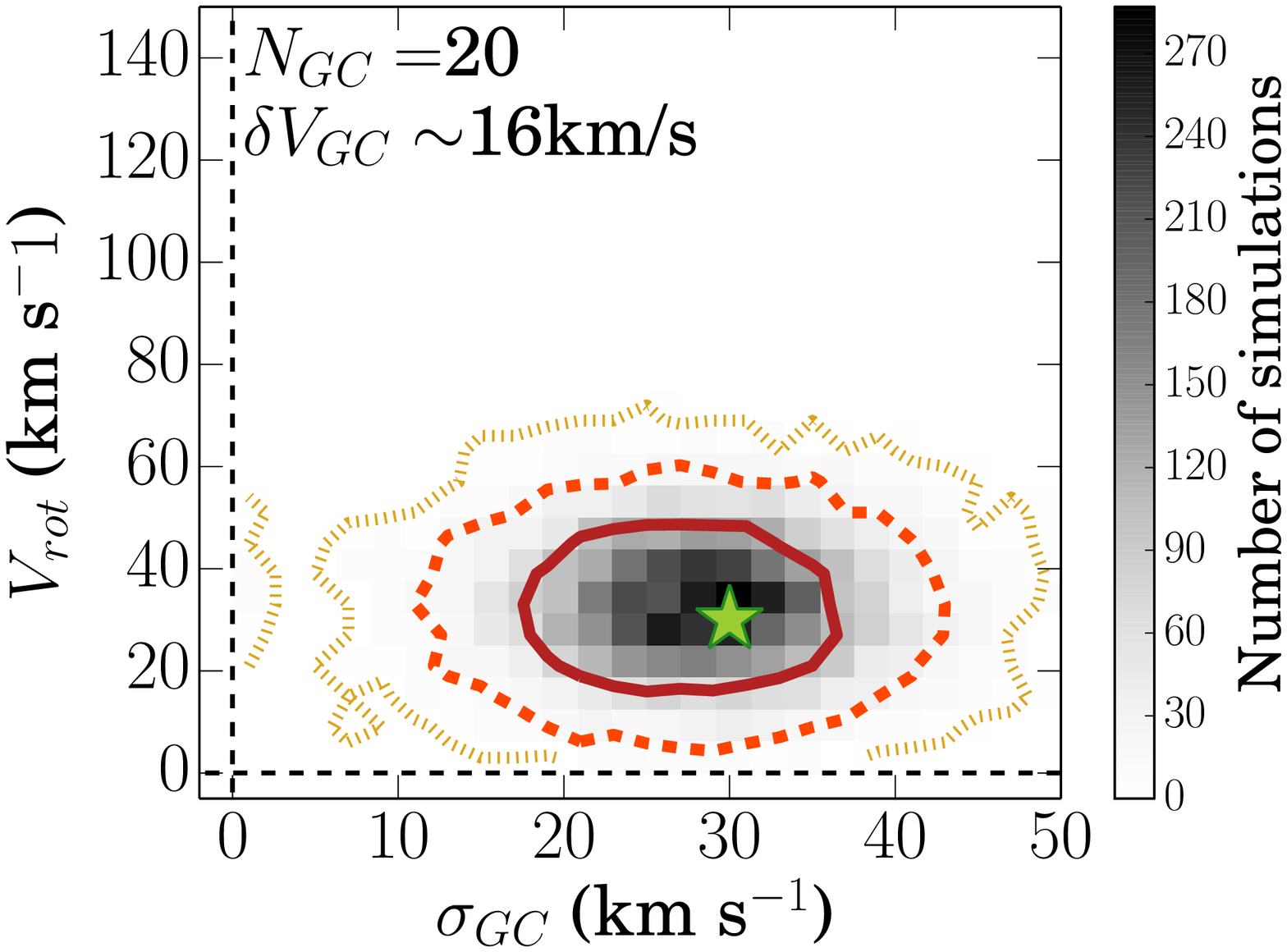}
\includegraphics[angle=0,width=5.9cm]{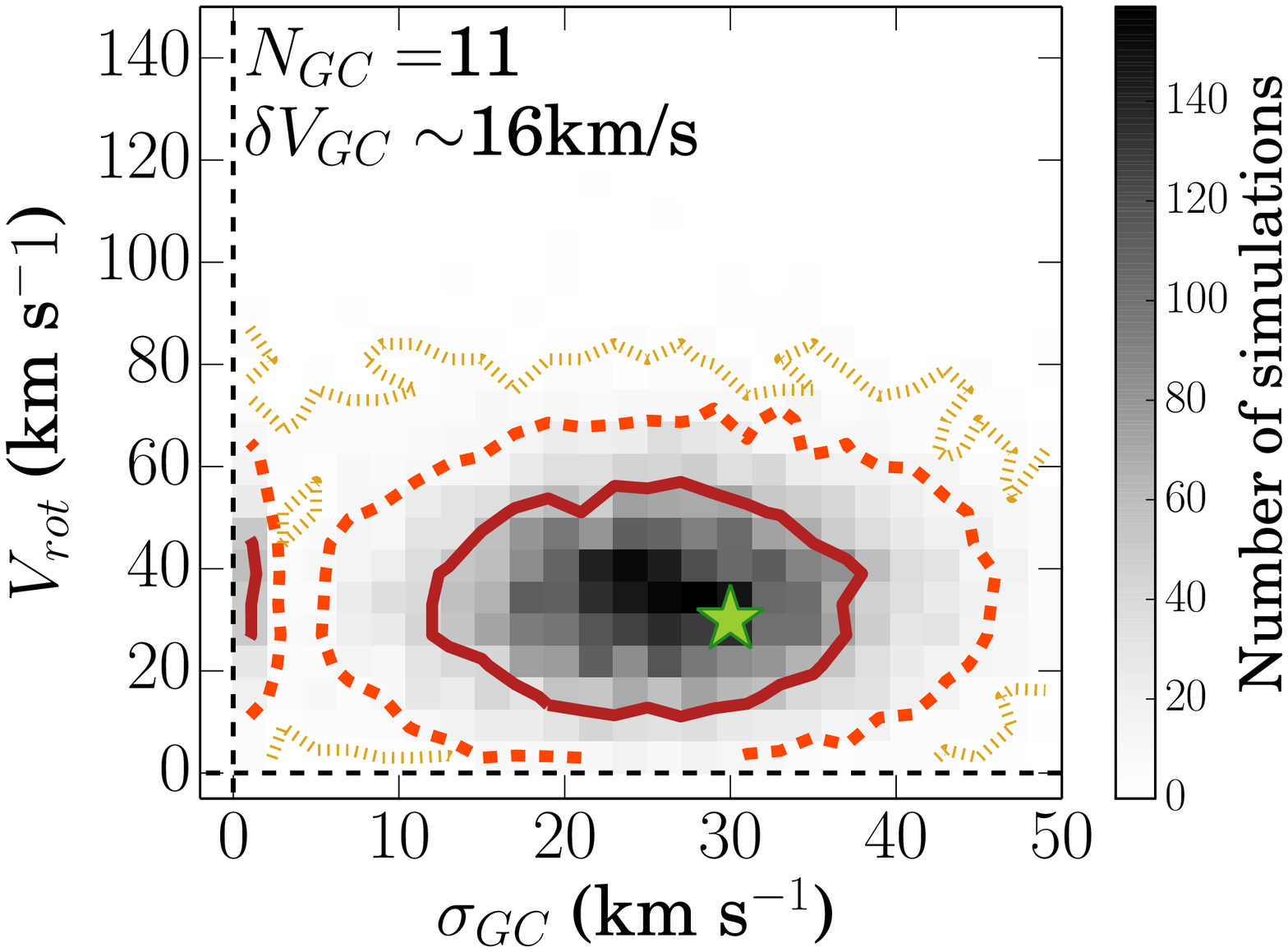}
\includegraphics[angle=0,width=5.9cm]{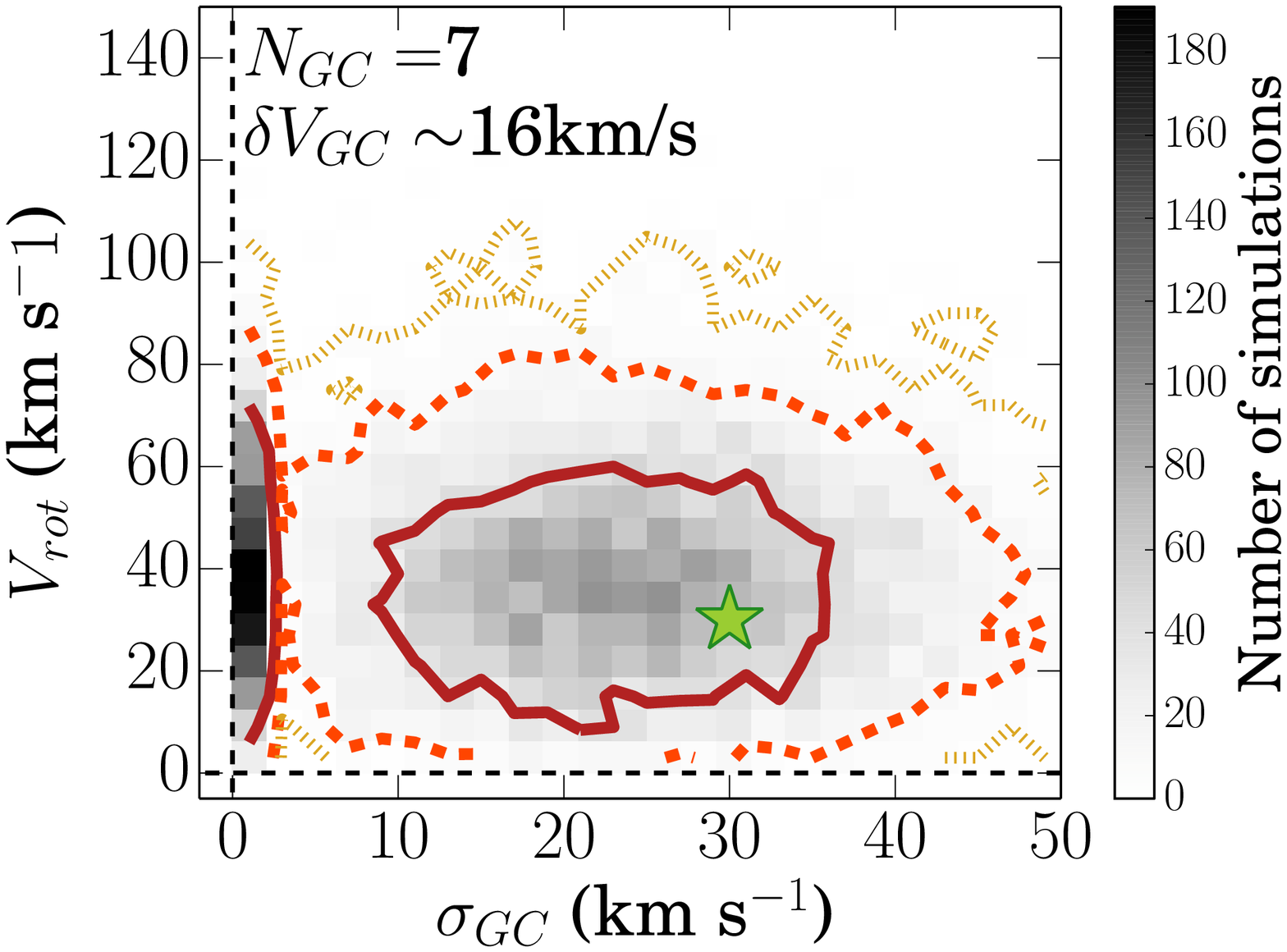}
\includegraphics[angle=0,width=5.9cm]{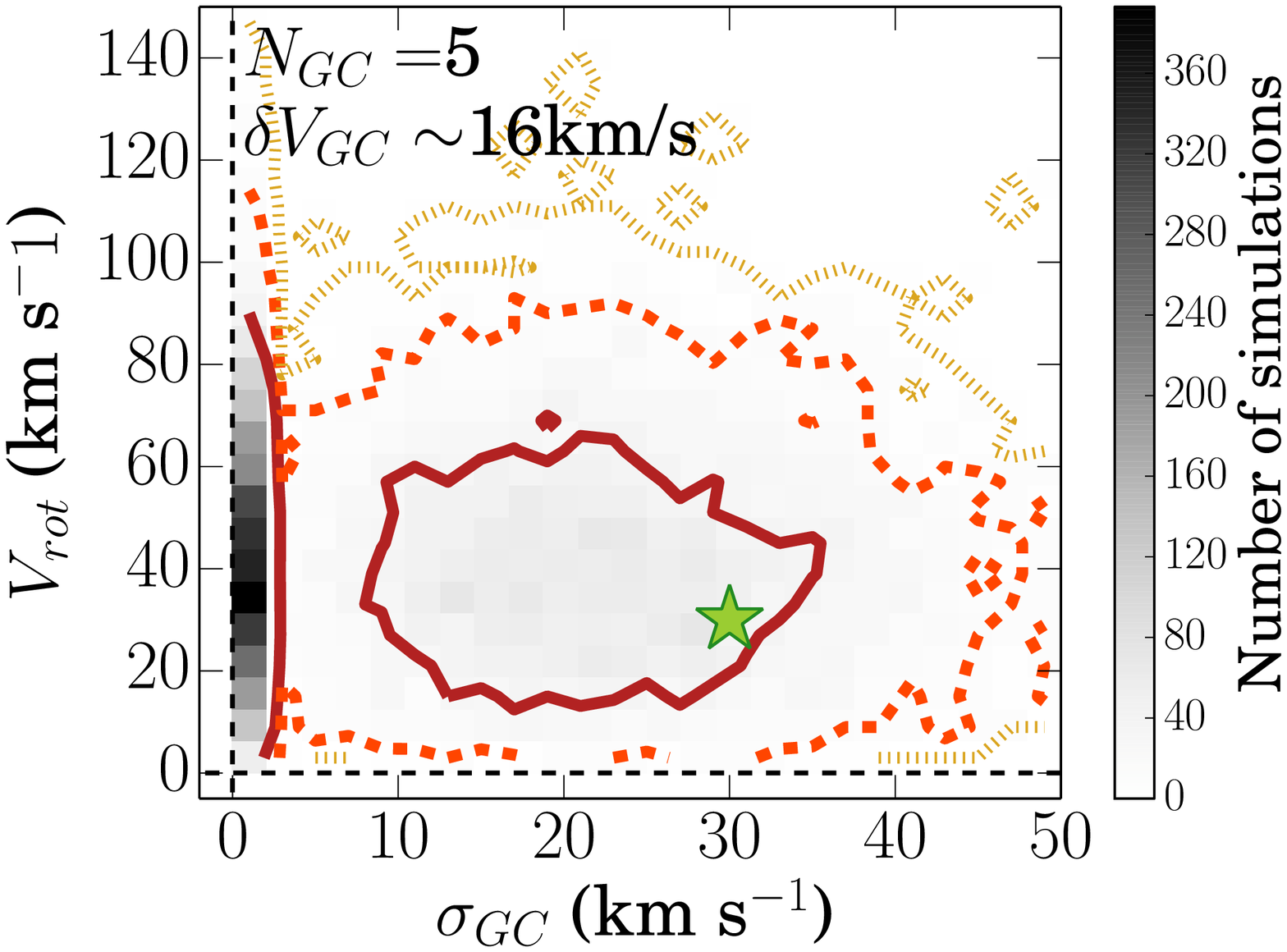}
\caption{Same as Figure \ref{vs0} for a rotating GC system with \Vmax$/$\sigGC~$=1$.}\label{vs1}
\end{figure*}

We explore the finite sampling effects on GC systems with \Vmax$/$\sigGC~$=0$, 0.5, 1, and 2. 
Figures \ref{vs0} and \ref{vs1} show the density maps that result from running 10000 VUS simulations for modelled GC systems with \Vmax$/$\sigGC~$=0$ and 1, respectively, and a different number of GC satellites. The results for modelled GC systems with \Vmax$/$\sigGC~$=0.5$ and 2 are not shown because the results for \Vmax$/$\sigGC~$=0.5$ are in between the two examples shown and the results for \Vmax$/$\sigGC~$=2$ are very similar those of \Vmax$/$\sigGC~$=1$. 

Regardless of the \Vmax/\sigGC\ of the GC model, the distribution of simulations becomes bimodal for \sigGC\ if \NGC~$< 20$. The number of simulations clustered in each peak of the distribution depends on \NGC. As \NGC\ decreases, the peak that contains the largest number of simulations moves towards \sigGC~$=0$~\kms. This is an artifact produced by the fitting process as discussed in Section \ref{stat_sig}.

The velocity dispersion \sigGC\ is underestimated in all the modelled GC systems if \NGC~$\leq 20$. This underestimation increases when \NGC\ decreases. In contrast, the rotation amplitude \Vmax\ tends to be overestimated. This overestimation is larger for GC systems with lower \Vmax/\sigGC, indicating that GC systems that rotate very slowly or do not rotate at all are significantly more difficult to measure than GC systems that rotate fast. This \Vmax\ overestimation also increases when \NGC\ decreases. 

In summary, a data set with \NGC~$\gtrsim20$ and a median velocity uncertainty of $\sim 16$~\kms\ is enough to estimate whether the GC system is rotating or not. If the GC system is rotating, the \Vmax\ estimated is not affected by large biases if the GC system has \Vmax$/$\sigGC~$\gtrsim 1$, or, in other words, \Vmax~$\gtrsim30$~\kms\ for an assumed \sigGC$\sim30$~\kms. If the GC system is not rotating, the \Vmax\ obtained will be affected by a systematic bias of the order of the median velocity uncertainty, i.e. $\sim 16$~\kms\ in this case. If the data set consists of \NGC~$\lesssim10$, the systematic bias, both in the rotation speed and the velocity dispersion, is in between the median velocity uncertainty and the intrinsic velocity dispersion of the dE. This bias depends on the intrinsic \Vmax/\sigGC\ which is unknown, thus, this bias cannot be corrected. However, the measured \Vmax\ can be considered as an upper limit and the measured \sigGC\ as a lower limit.

\subsection{Velocity uncertainty effects on rotating and non-rotating GC systems}

We explore the effects that the velocity uncertainties ($\delta V_{\rm GC}$) have on the measured \Vmax\ and \sigGC. We repeat the VUS simulations run in Section \ref{vs} but assuming velocity uncertainties that are $40\%$, $70\%$, and $90\%$ smaller than our measured values. These correspond to median uncertainties of 10, 4, and 2~\kms. 

\begin{figure}
\centering
\includegraphics[angle=-90,width=8.5cm]{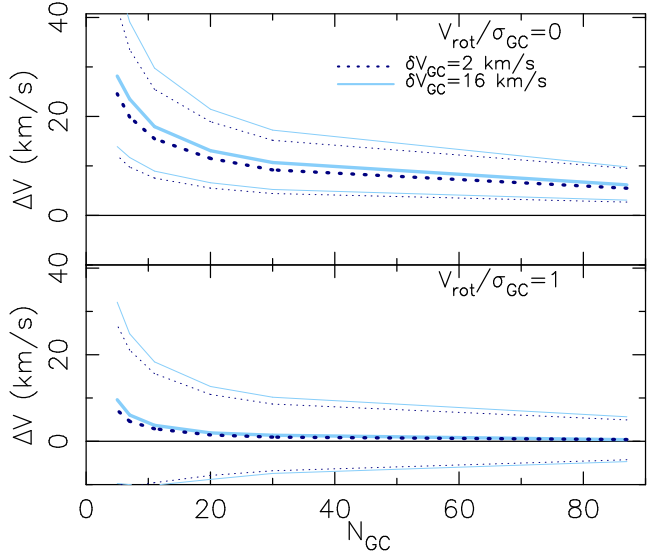}
\caption{Rotation amplitude bias as a function of the number of observed GCs satellites for different \Vmax/sigGC\ values. The parameter $\Delta V$ measures the difference between the median rotation amplitude obtained in the 10000 VUS and the rotation amplitude of the input model. For this particular case, the models have \sigGC$=30$~\kms.
The thick lines indicate the rotation amplitude bias and the thin lines indicate the percentiles 84 and 16 of the distribution of the VUS. The light blue solid lines indicate the results obtained for a median line-of-sight GC velocity uncertainty of 19~\kms, the same as in our data. The dark blue dotted lines indicate the results for an 11 times smaller median uncertainty. As \Vmax/\sigGC\ is an unknown parameter of the GC system the bias cannot be subtracted from the data but it can be significantly reduced by increasing the number of observed GC satellites.}\label{vel_bias}
\end{figure}

\begin{figure}
\centering
\includegraphics[angle=-90,width=8.5cm]{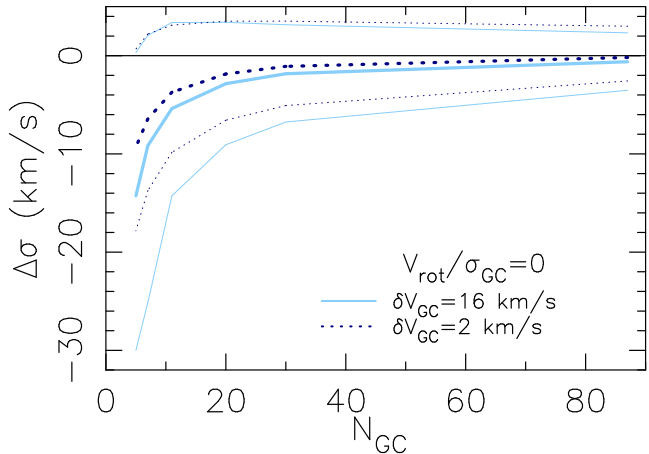}
\caption{Same as Figure \ref{vel_bias} but for the velocity dispersion. This trend looks nearly identical for all the \Vmax/\sigGC\ values analyzed.}\label{sig_bias}
\end{figure}

\begin{figure*}
\centering
\includegraphics[angle=0,width=5.9cm]{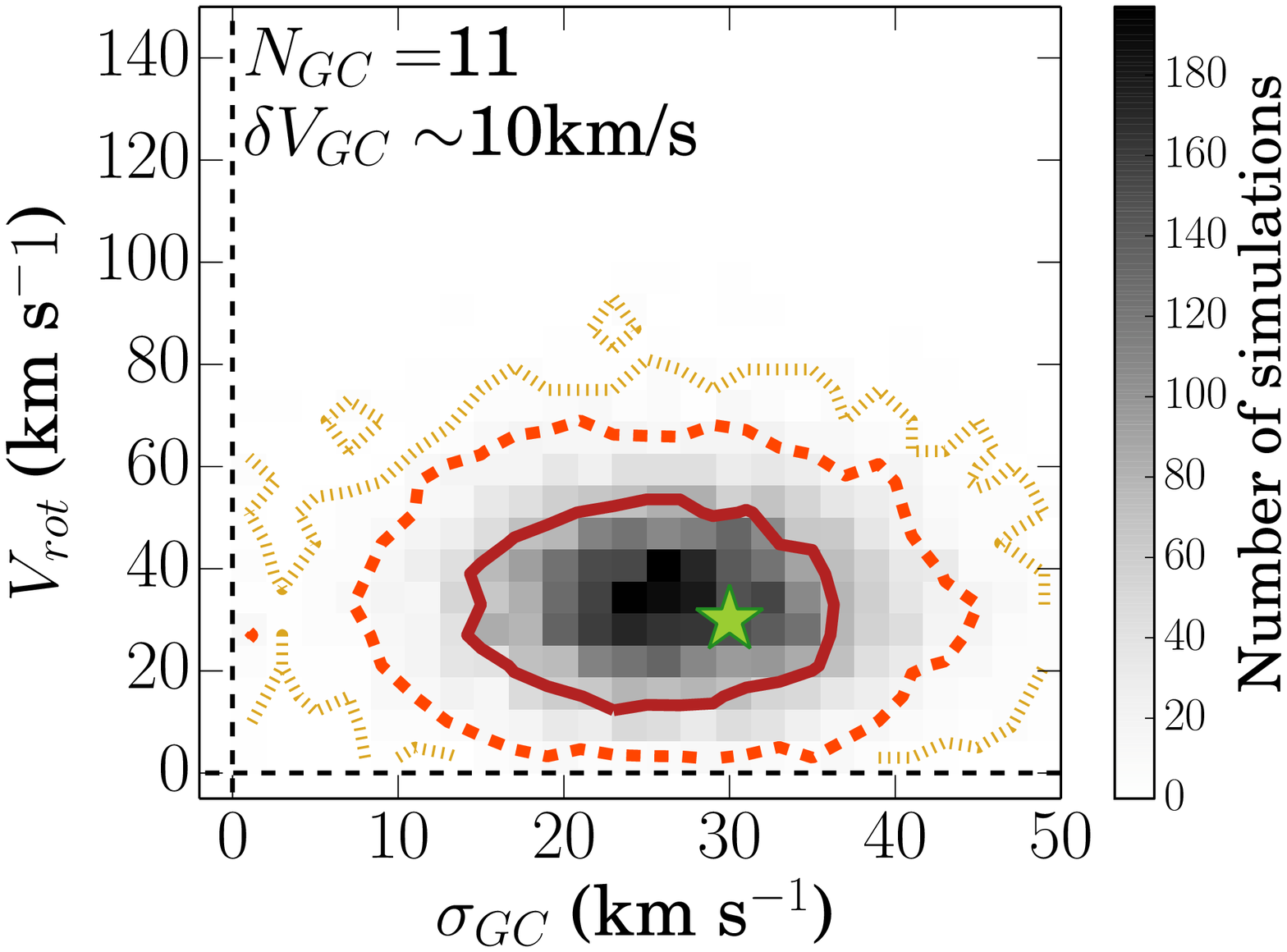}
\includegraphics[angle=0,width=5.9cm]{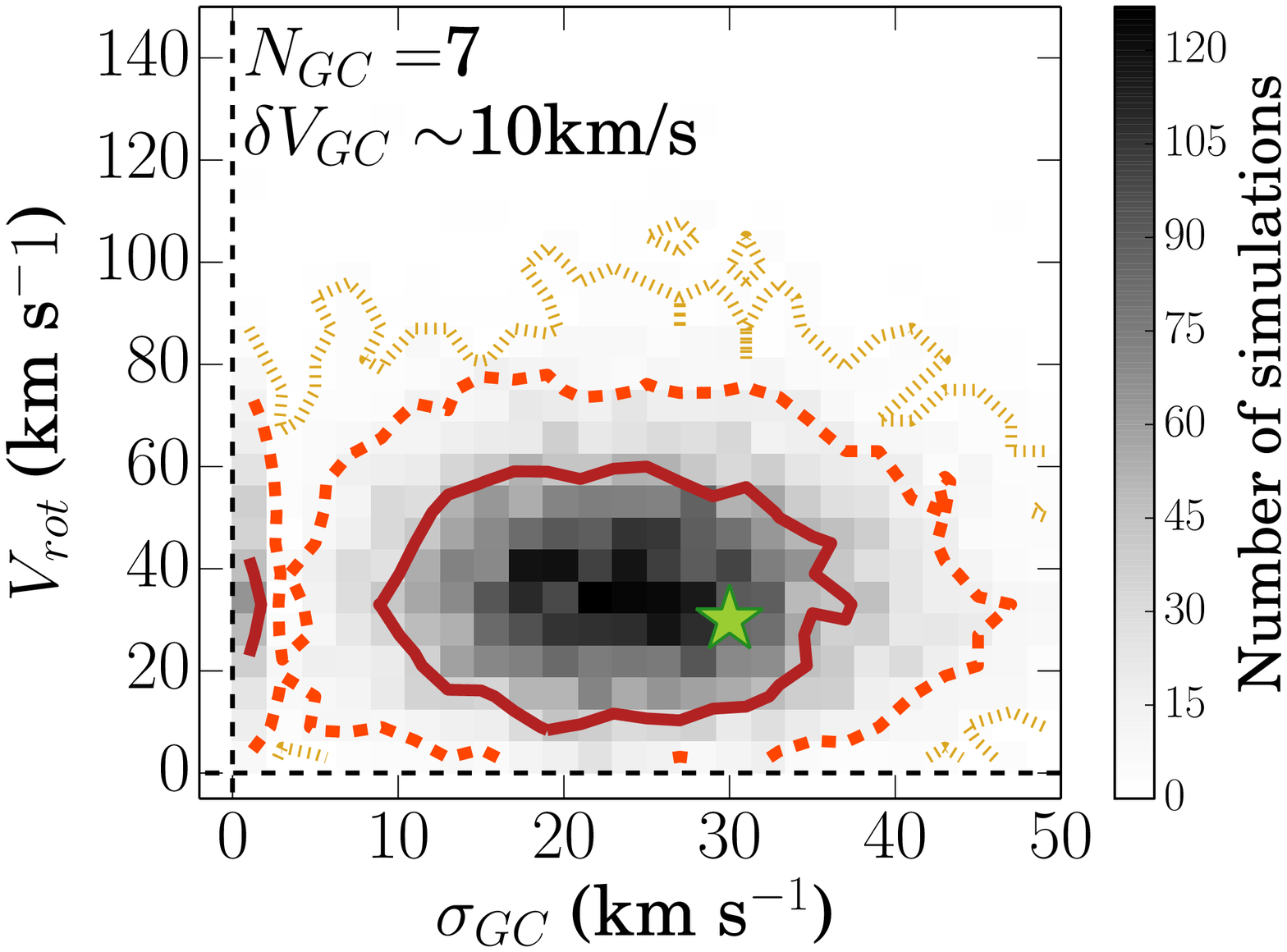}
\includegraphics[angle=0,width=5.9cm]{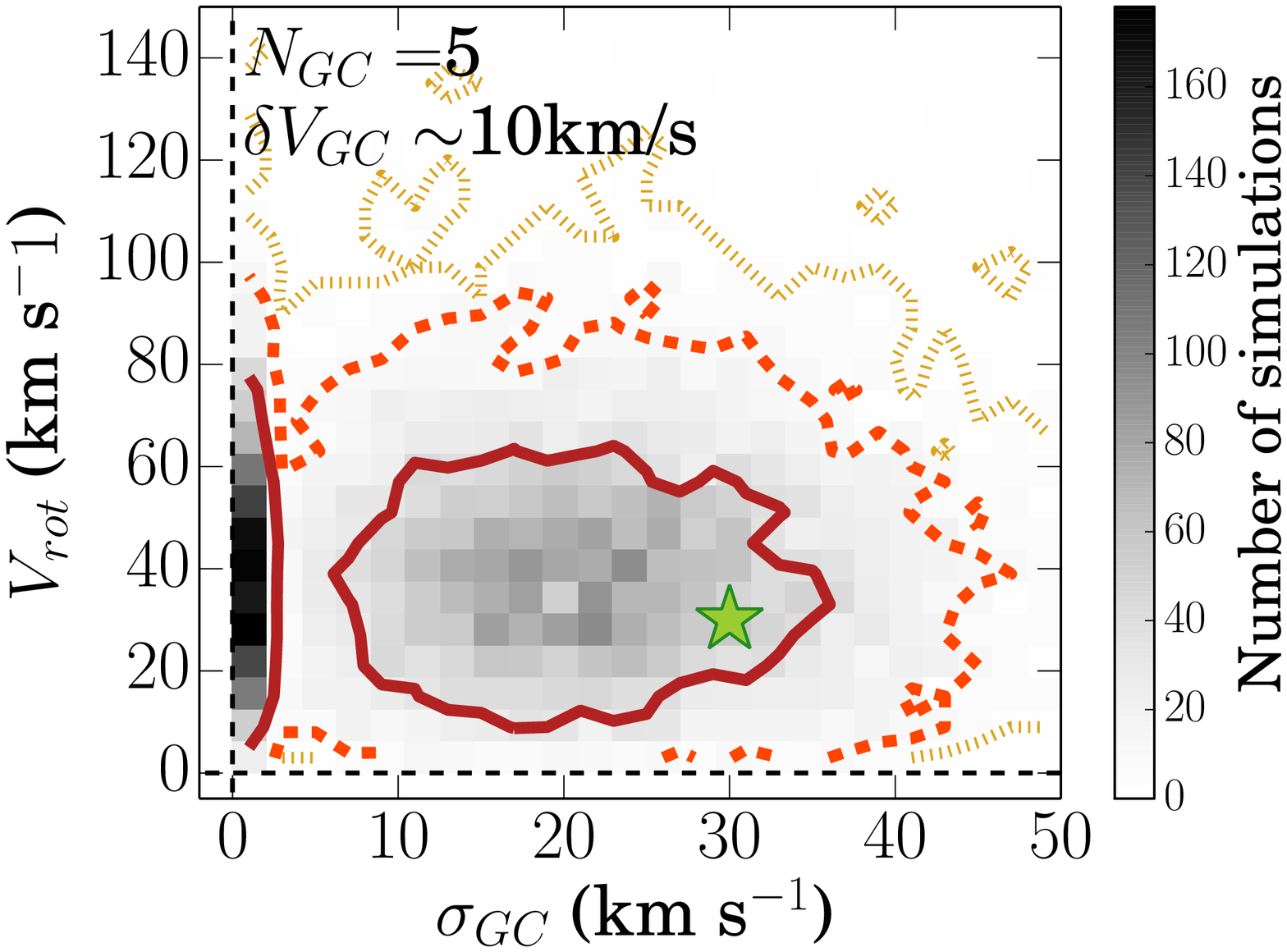}
\includegraphics[angle=0,width=5.9cm]{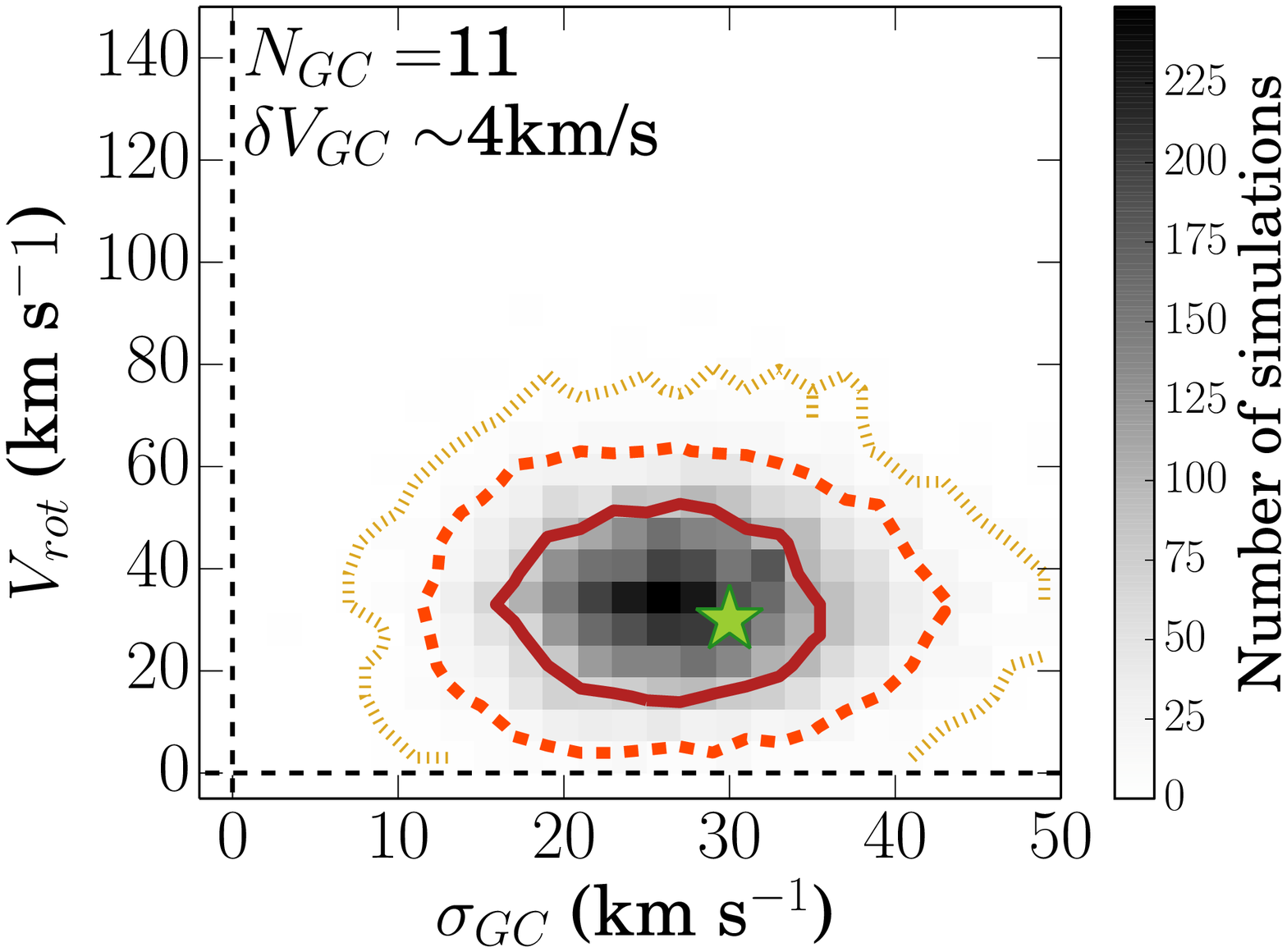}
\includegraphics[angle=0,width=5.9cm]{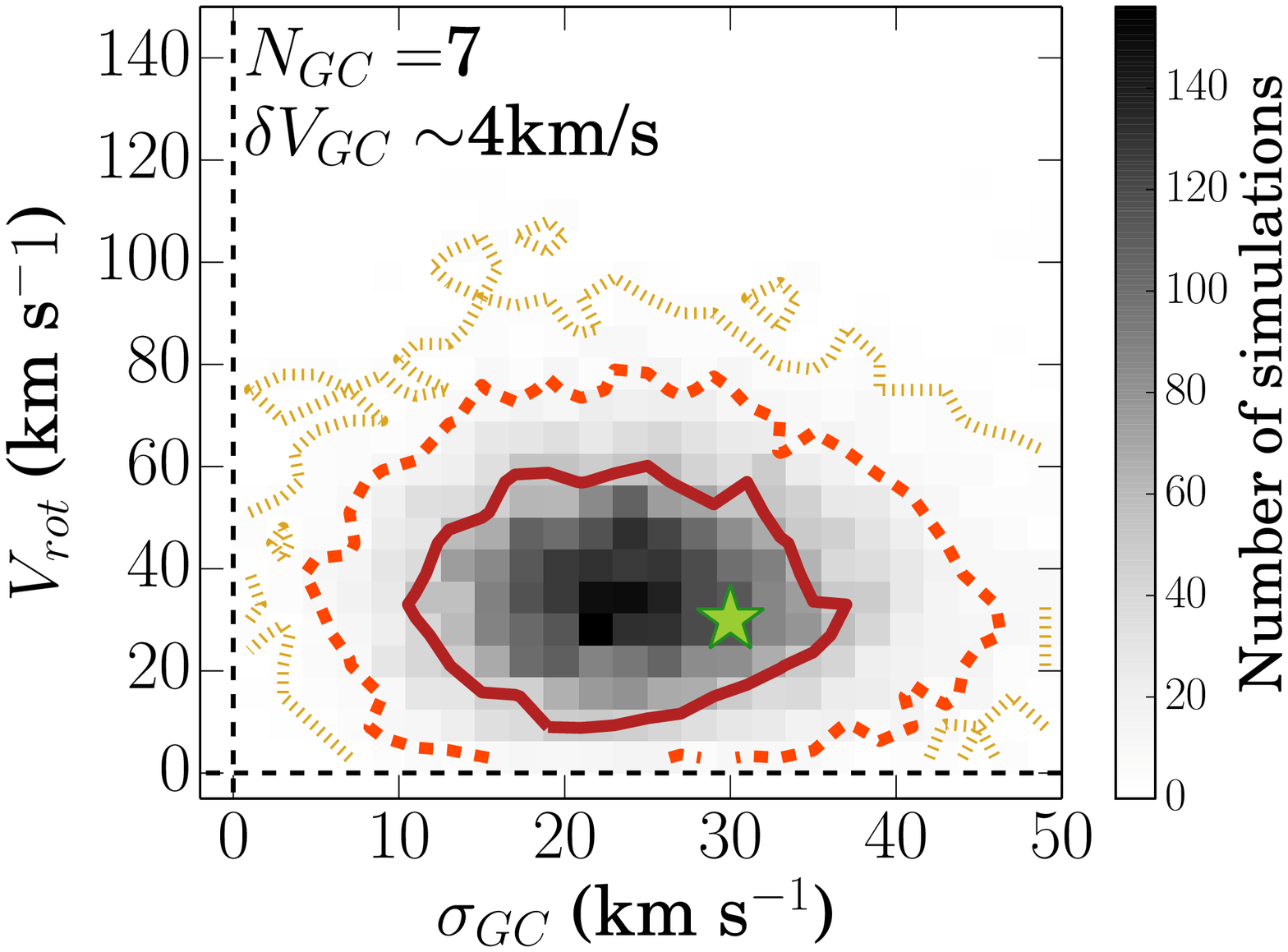}
\includegraphics[angle=0,width=5.9cm]{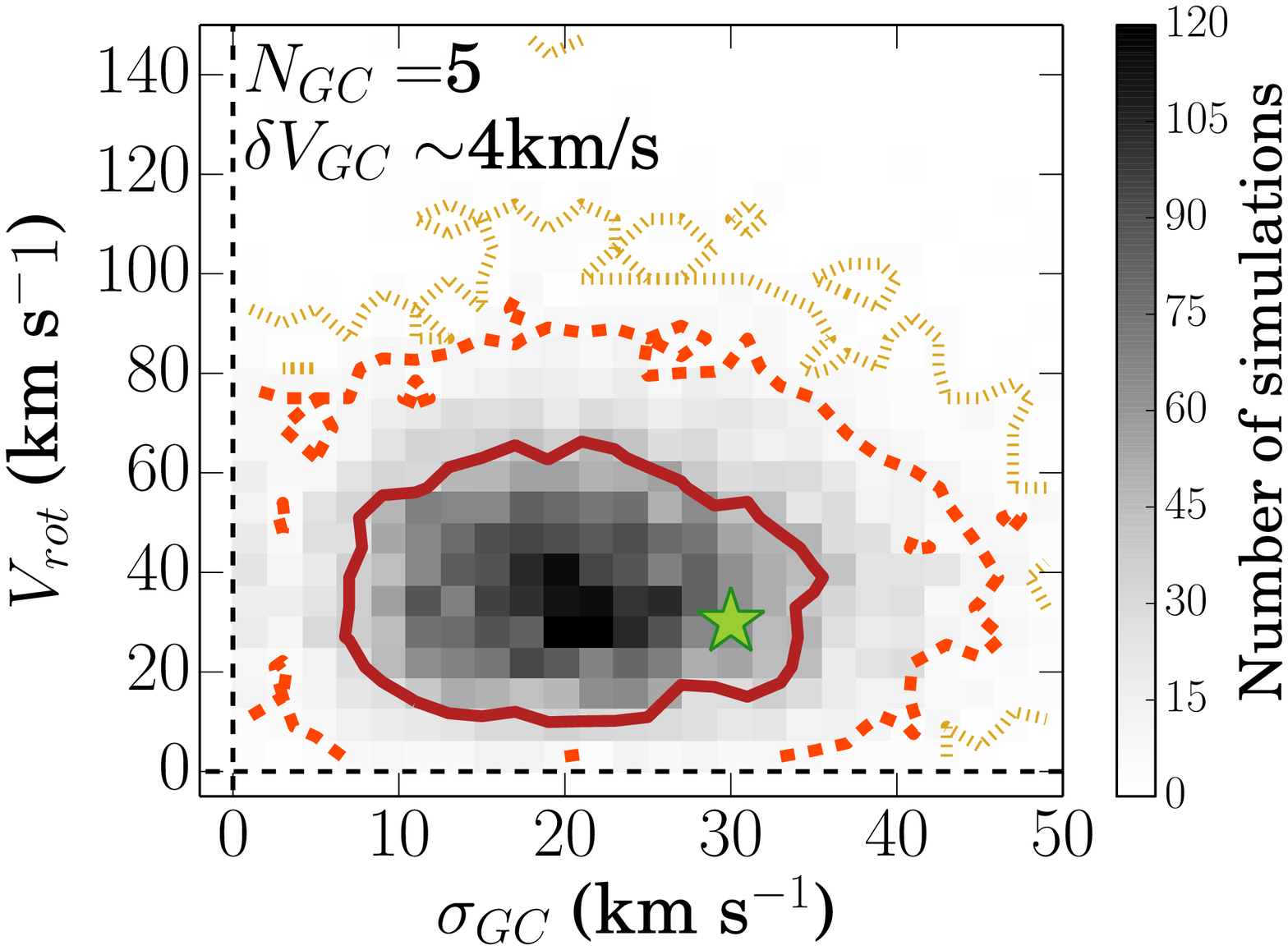}
\includegraphics[angle=0,width=5.9cm]{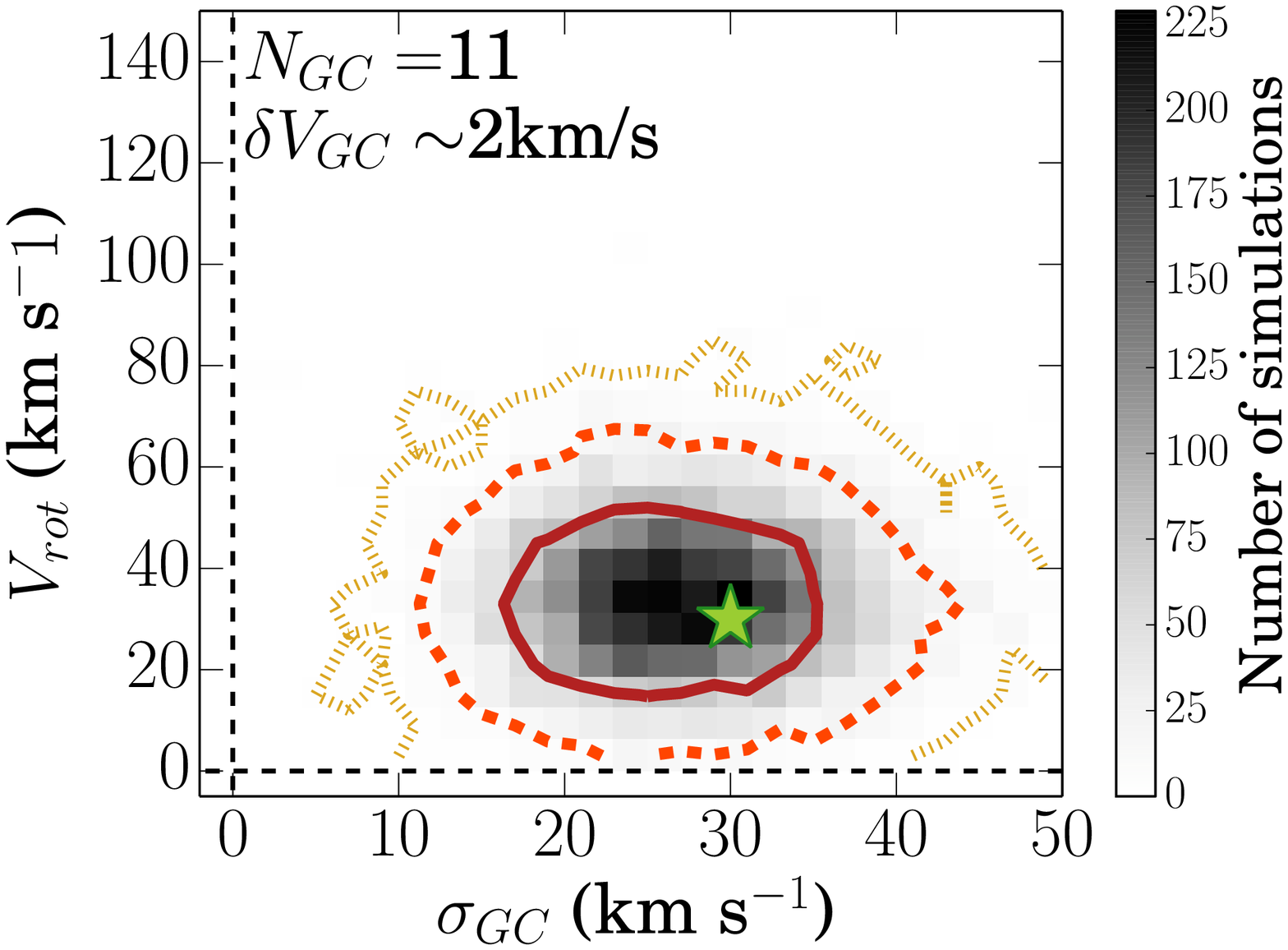}
\includegraphics[angle=0,width=5.9cm]{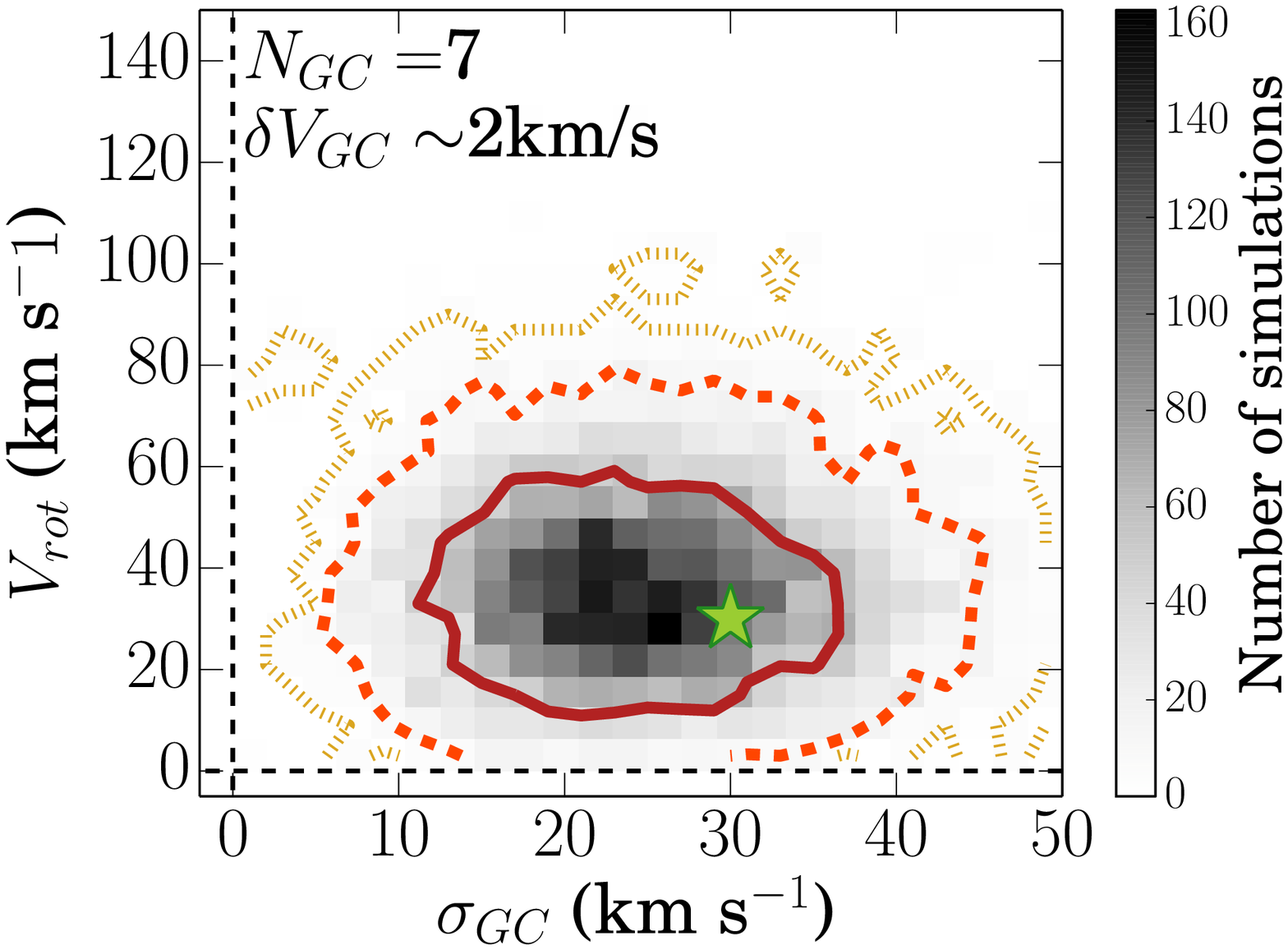}
\includegraphics[angle=0,width=5.9cm]{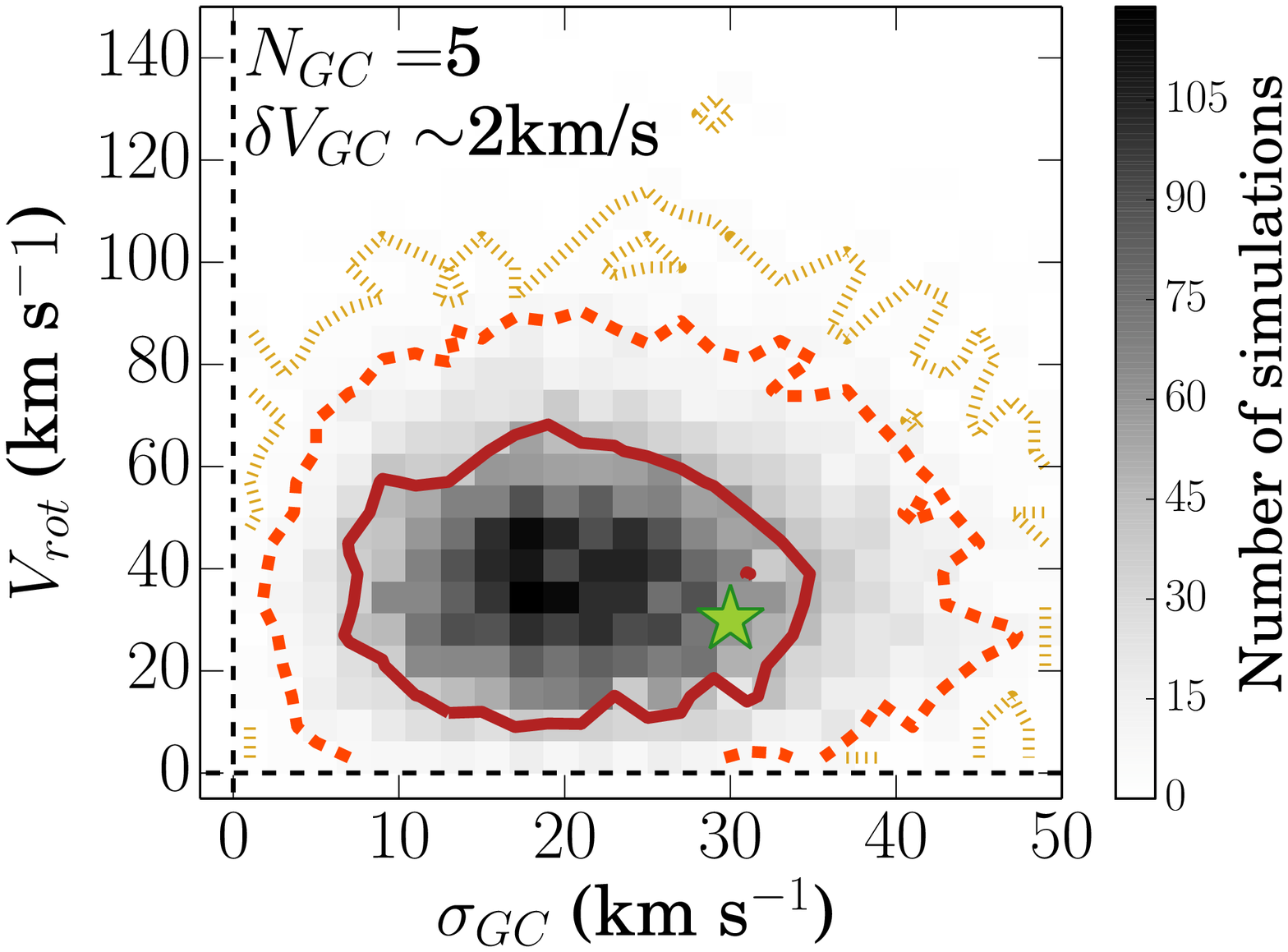}
\caption{Same as Figure \ref{vs1} for \NGC~$=11$, 7, 5, and $\delta V_{\rm GC}=10$, 4, 2~\kms. Decreasing the velocity uncertainties makes the bimodality in the measured \sigGC\ disappear.}\label{figstats}
\end{figure*}

Figures \ref{vel_bias} and \ref{sig_bias} show the \Vmax\ and \sigGC\ biases as a function of \NGC\ for different median velocity uncertainties and \Vmax/\sigGC\ values. The velocity uncertainties have a larger effect in systems with small numbers of GCs. 
Figure \ref{figstats} shows these effects for a GC system with \sigGC~$=30$~\kms, \Vmax/\sigGC~$=1$, and \NGC~$\leq 11$.
The bimodal distribution of \sigGC\ seen in Figures \ref{vs0} and \ref{vs1} for \NGC~$\leq 11$ disappears when $\delta V_{\rm GC} < 10$~\kms. In addition, the underestimation of \sigGC\ and overestimation of \Vmax\ described in Section \ref{vs} decrease for smaller velocity uncertainties.

\section{Analysis the Globular Cluster Systems}\label{our_dEs}

In this Section we analyze the three new GC systems presented in this paper, those bound to VCC~1539, VCC~1545, and VCC~1861, and we reanalyze the GC systems of VCC~1087, VCC~1261, and VCC~1528 from \citet{Beasley06,Beasley09}.

\subsection{Globular Cluster Systems of VCC~1539, VCC~1545, and VCC~1861}

\begin{figure*}
\centering
\includegraphics[angle=0,width=2.9cm]{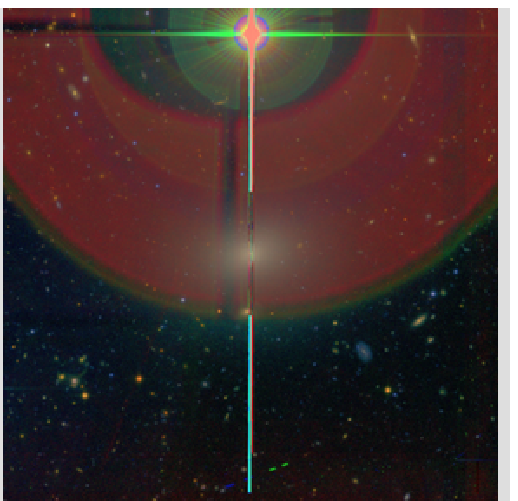}
\includegraphics[angle=0,width=2.9cm]{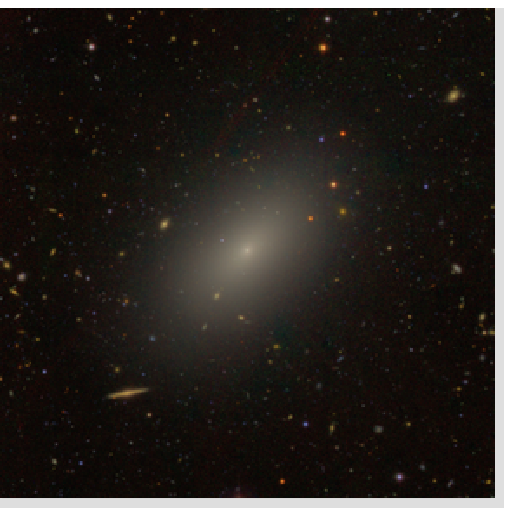}
\includegraphics[angle=0,width=2.9cm]{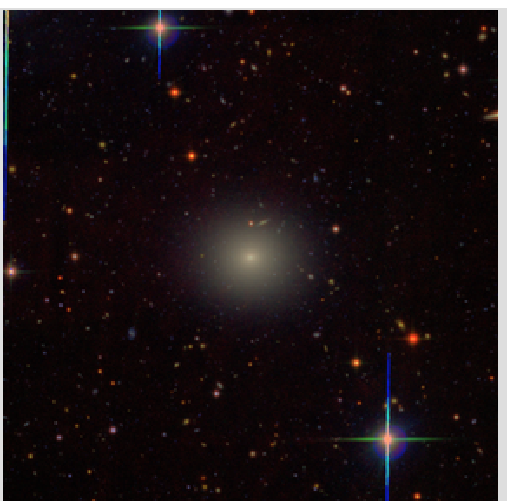}
\includegraphics[angle=0,width=2.9cm]{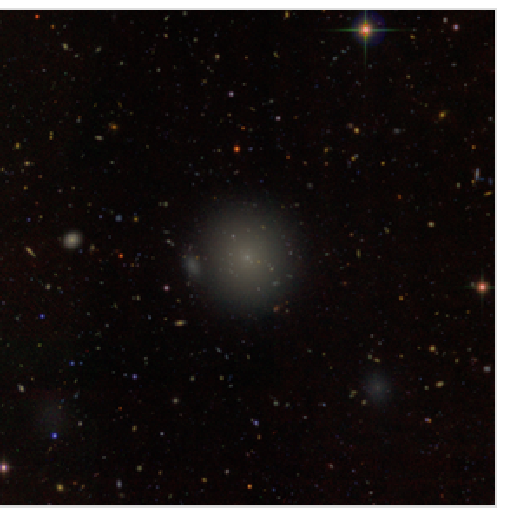}
\includegraphics[angle=0,width=2.9cm]{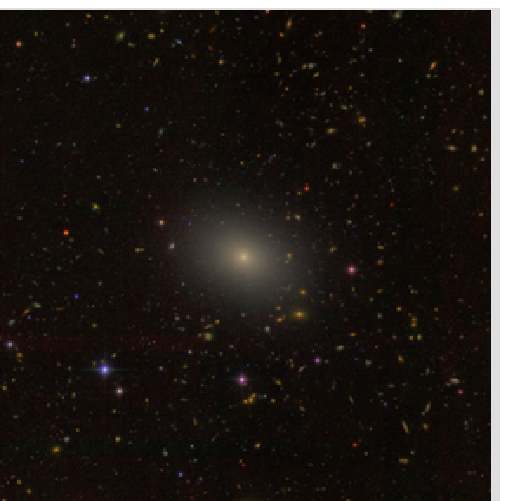}
\includegraphics[angle=0,width=2.9cm]{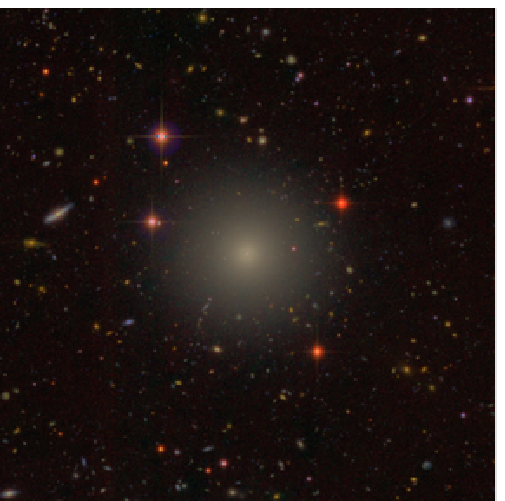}
\includegraphics[angle=0,width=2.9cm]{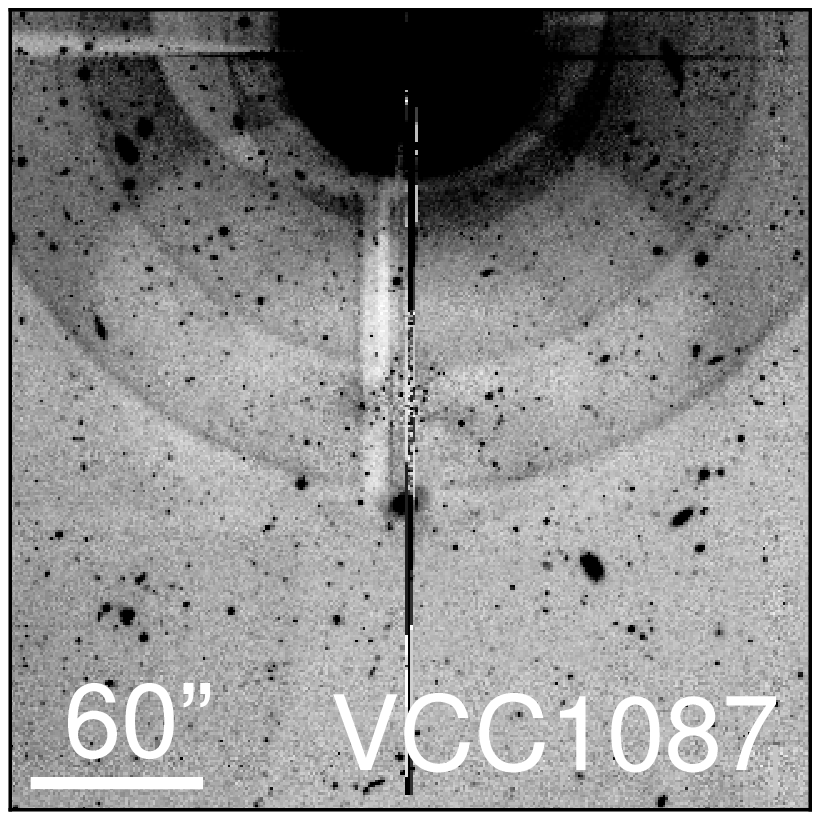}
\includegraphics[angle=0,width=2.9cm]{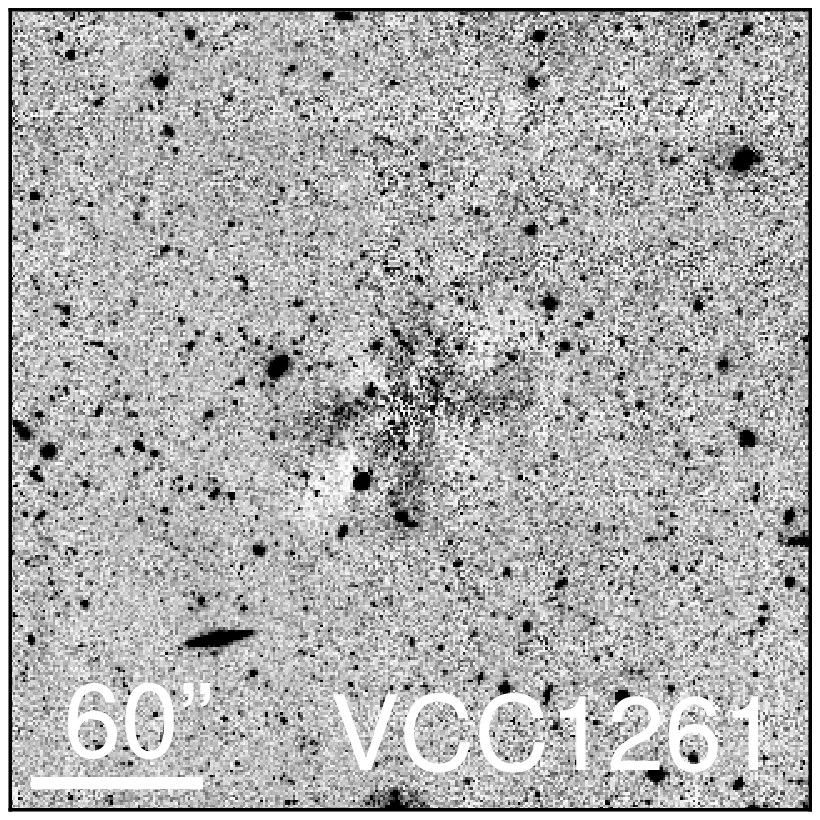}
\includegraphics[angle=0,width=2.9cm]{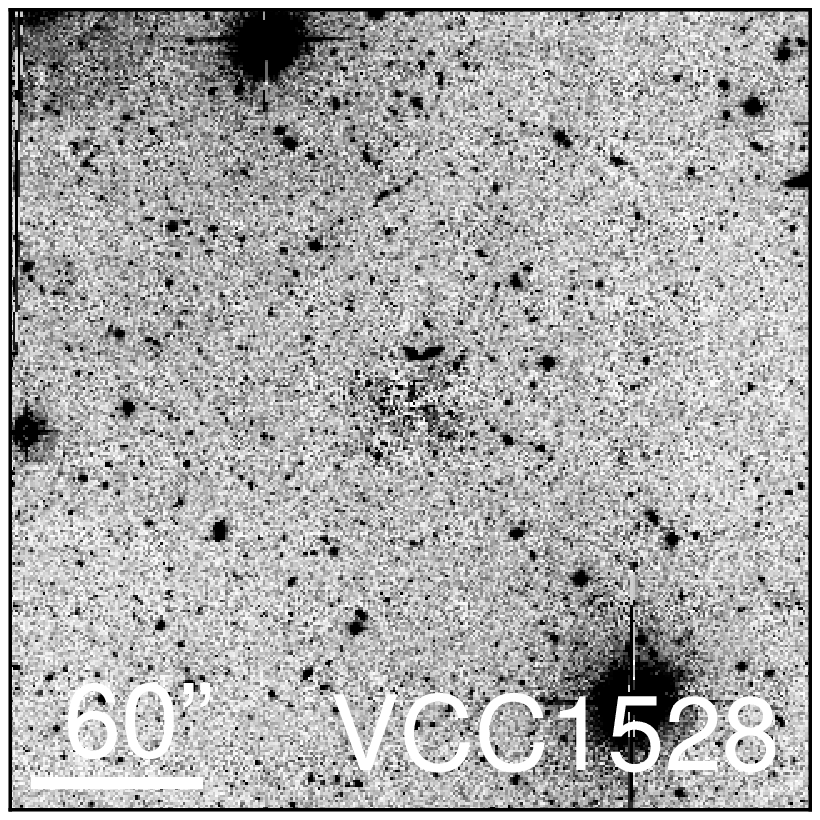}
\includegraphics[angle=0,width=2.9cm]{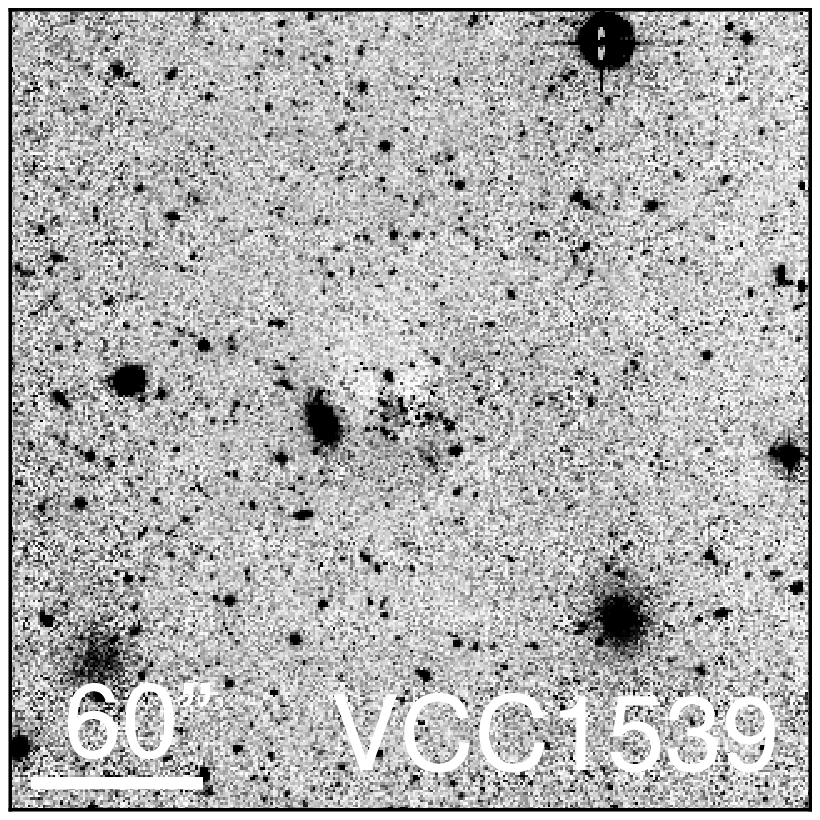}
\includegraphics[angle=0,width=2.9cm]{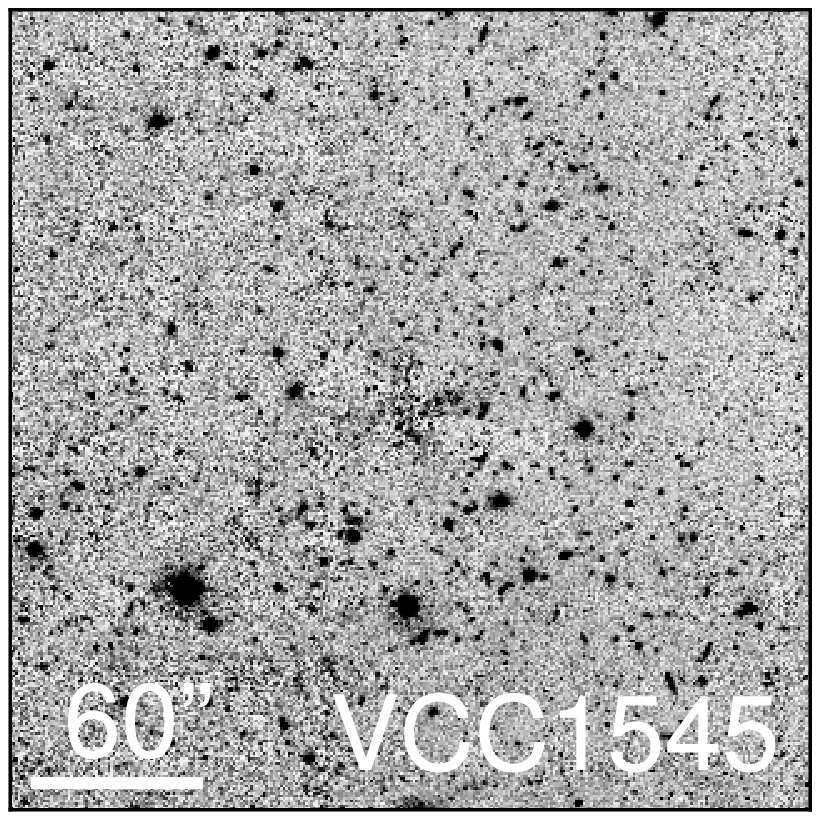}
\includegraphics[angle=0,width=2.9cm]{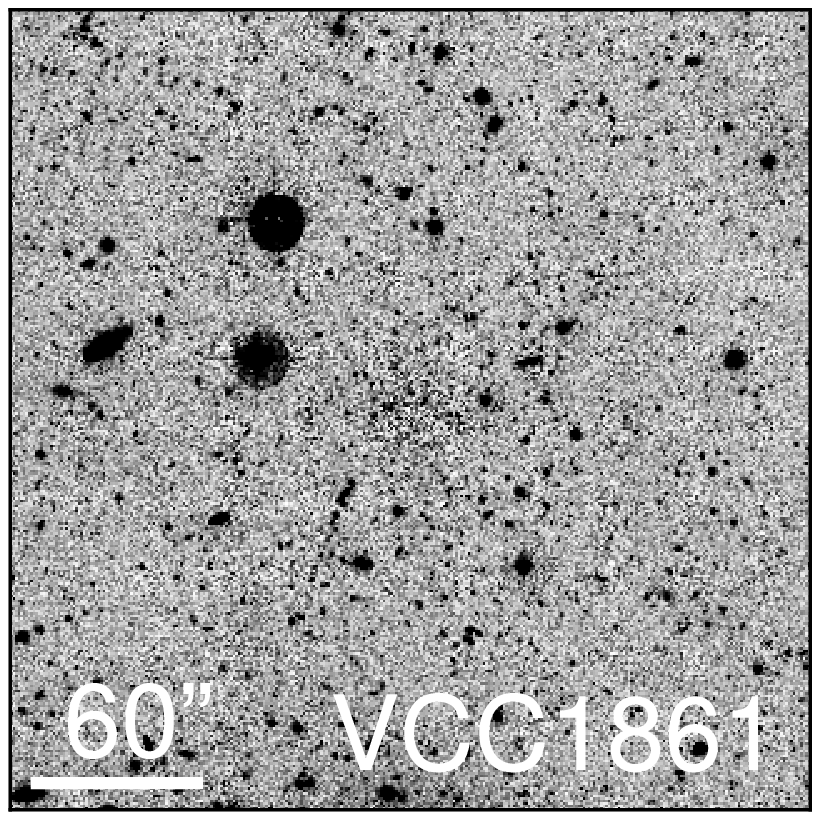}
\caption{Upper panel: composite $giz$ color images from the NGVS. The images are oriented with the North up and the East on the left. Lower panel: $g$ band residual images obtained by subtracting a smooth model based on the ellipse-fitting of the isophotes. The red circle and the corresponding black circle in the images for VCC~1087 are an artifact due to a saturated bright star. None of the six dEs have disk-like structures, such as bars or spiral arms. The x-shapes seen for VCC~1261, VCC~1528, and VCC~1545 are the typical residuals of boxy and disky isophotes.}\label{colorimages}
\end{figure*}

VCC~1539, VCC~1545, and VCC~1861 are located in the inner core of the Virgo cluster. Their projected distance to M87 are between $0.88^{\circ}$, $0.89^{\circ}$, and $2.76^{\circ}$, respectively \citep[the virial radius of the Virgo cluster, considering M87 at its center, is $R_{200}=5.38^{\circ}=1.55$~Mpc;][]{Ferrarese12}. 

These three dEs are $\sim 1$~mag fainter than the brightest dEs, which have M$_V \sim -18$, \citep[e.g.;][]{FB94}. They are very round, with ellipticities of $\epsilon \lesssim 0.1$, and they do not show any disk-like structures, such as bars, spiral arms, or irregular features, in the NGVS deep images (see Figure \ref{colorimages}), confirming the lack of substructure found by \citet{Lisk06a,Lisk07} based on the much swallower Sloan Digital Sky Survey images \citep[SDSS;][]{SDSS_DR6}. The stellar kinematics is available for VCC~1545 and VCC~1861. While VCC~1545 rotates at a speed of $\sim 25$~\kms\  \citep{Chil09}, VCC~1861 is not rotating  \citep[its stellar rotation is $5.3\pm2.5$~\kms;][]{etj11,etj14b,etj14c}. See Table \ref{dEs_properties} for a summary of the main properties of these dEs.

\begin{figure}
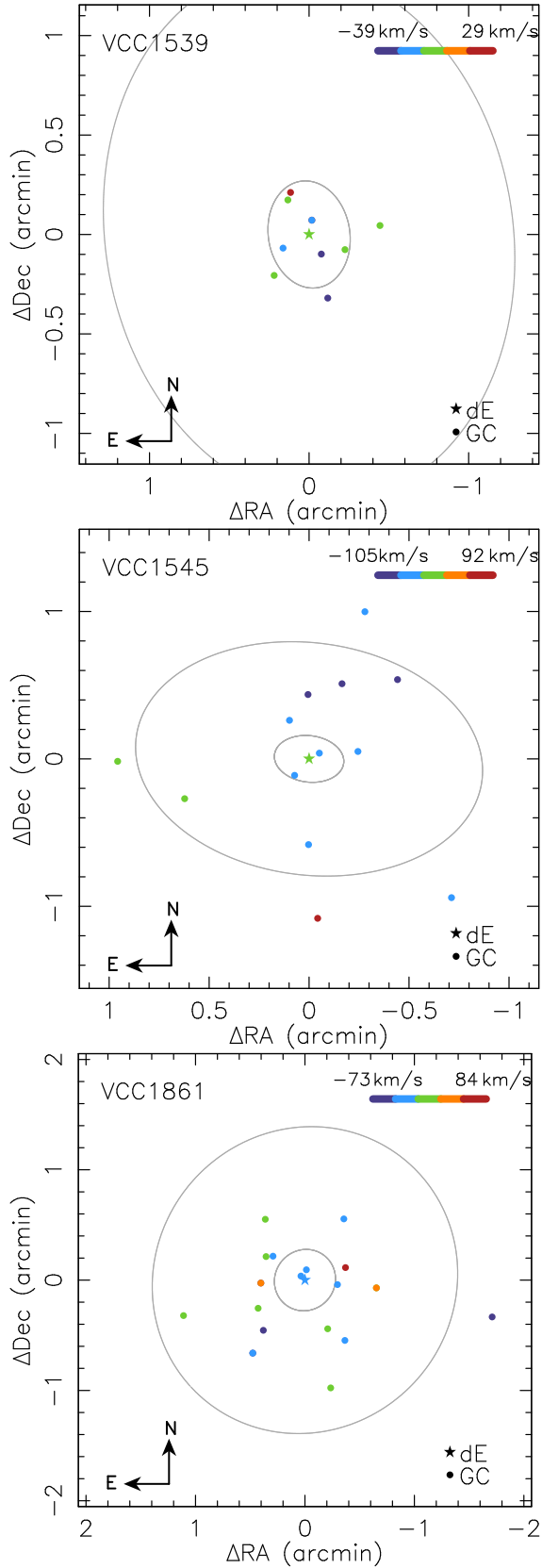

\centering
\includegraphics[angle=-90,width=7.5cm]{fig11a.ps}
\includegraphics[angle=-90,width=7.5cm]{fig11b.ps}
\includegraphics[angle=-90,width=7.5cm]{fig11c.ps}
\caption{Velocity maps in the North-East direction for, from top to bottom, VCC~1539, VCC~1545, and VCC~1861. The asterisk in the center of each panel represents the centre of the dE. The dots indicate the position of the GC satellites. The colors indicate the velocity difference between the GC and the dE (\VGC$-$\Vsys). The velocity scale is indicated in the top right color bar. The ellipses centered on the dE indicate the 1\Reff\ and 5\Reff\ isophotes measured in the $i$ band from the NGVS.}\label{velmaps}
\end{figure}

The number of spectroscopically observed and confirmed GC satellites for VCC~1539, VCC~1545, and VCC~1861 is similar to the dEs presented in the literature \citep{Beasley06,Beasley09}. 
Figure \ref{velmaps} shows the velocity maps of their observed GC satellites. For VCC~1539, the 9 observed GC satellites are within 2\Reff. For VCC~1545, the 14 observed GC satellites are are within 7\Reff. Finally, for VCC~1861, the 18 observed GC satellites are within 7\Reff. This very different spatial distribution of the GC satellites may be related to the intrinsic properties of these galaxies. For example, VCC~1539 has a very compact GC system. This galaxy is the faintest of the three and has the lowest S\'ersic index, however, its light concentration, measured by the ratio $R_{80}/R_{20}$ (see Table \ref{dEs_properties}), is not the highest. Li et al. (in prep.) analyze the total mass and radial extent of the dark matter halos of these galaxies, which may shed more light on this.

\begin{figure}
\centering
\includegraphics[angle=0,width=8.5cm]{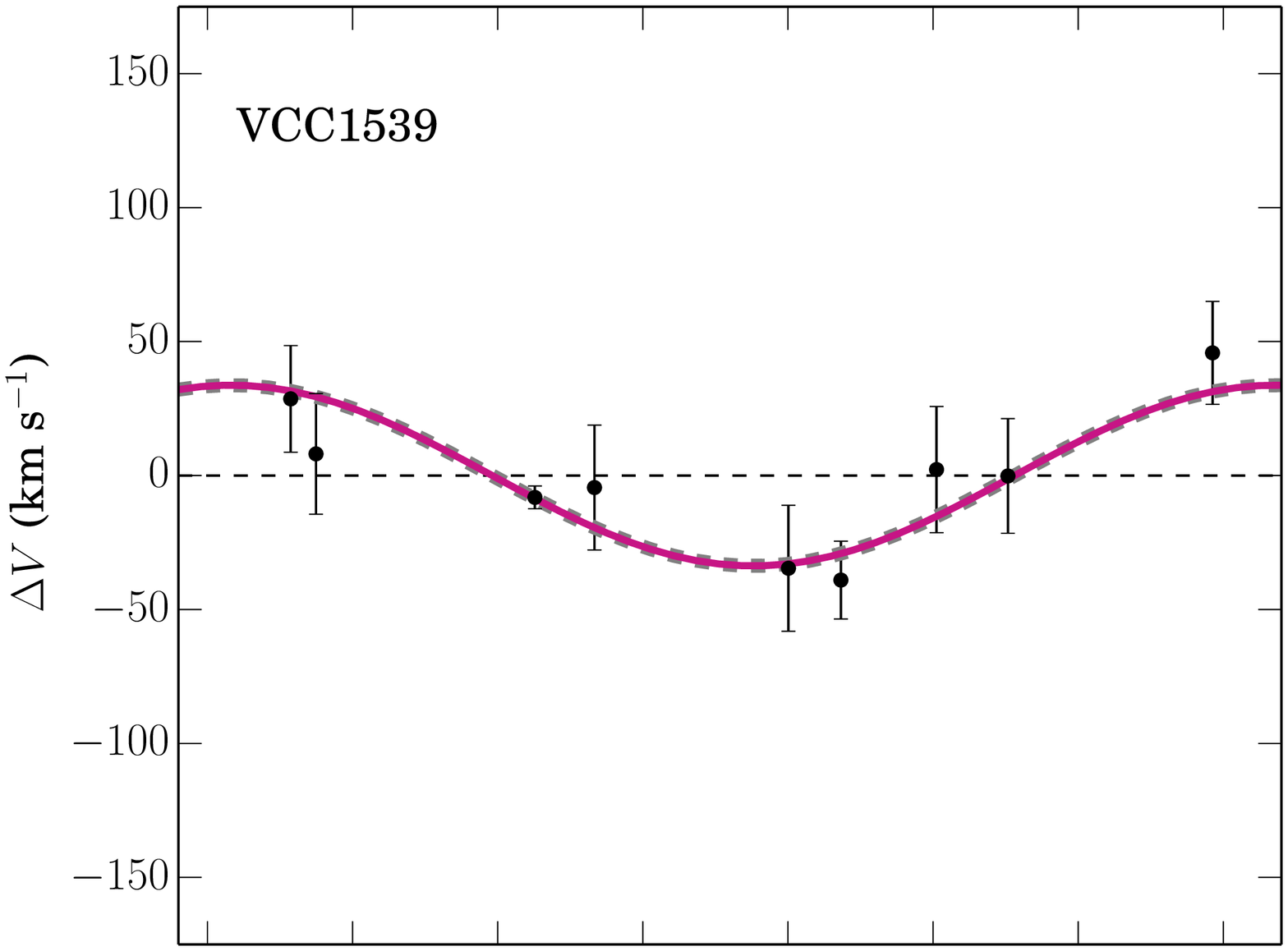}
\includegraphics[angle=0,width=8.5cm]{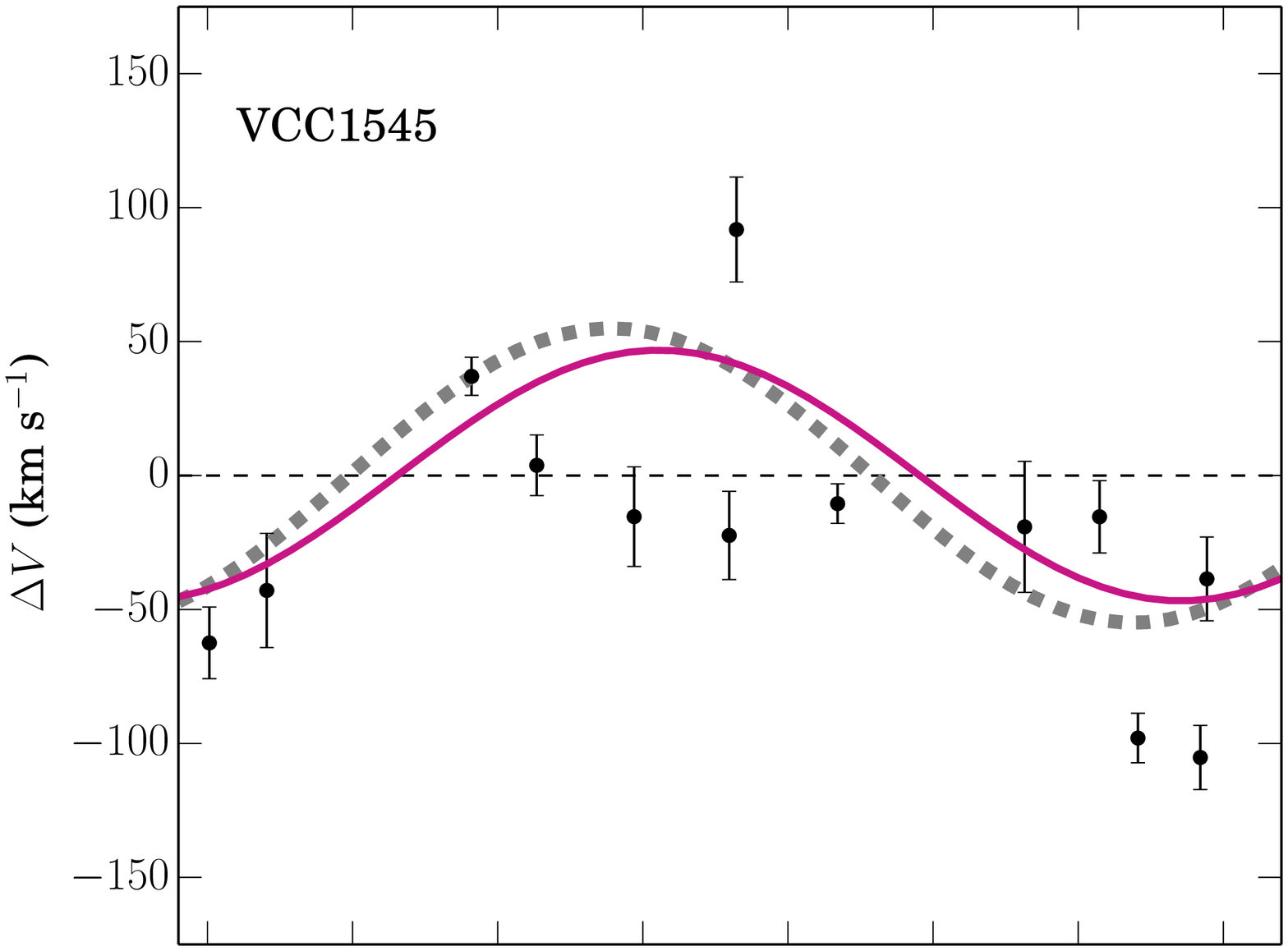}
\includegraphics[angle=0,width=8.5cm]{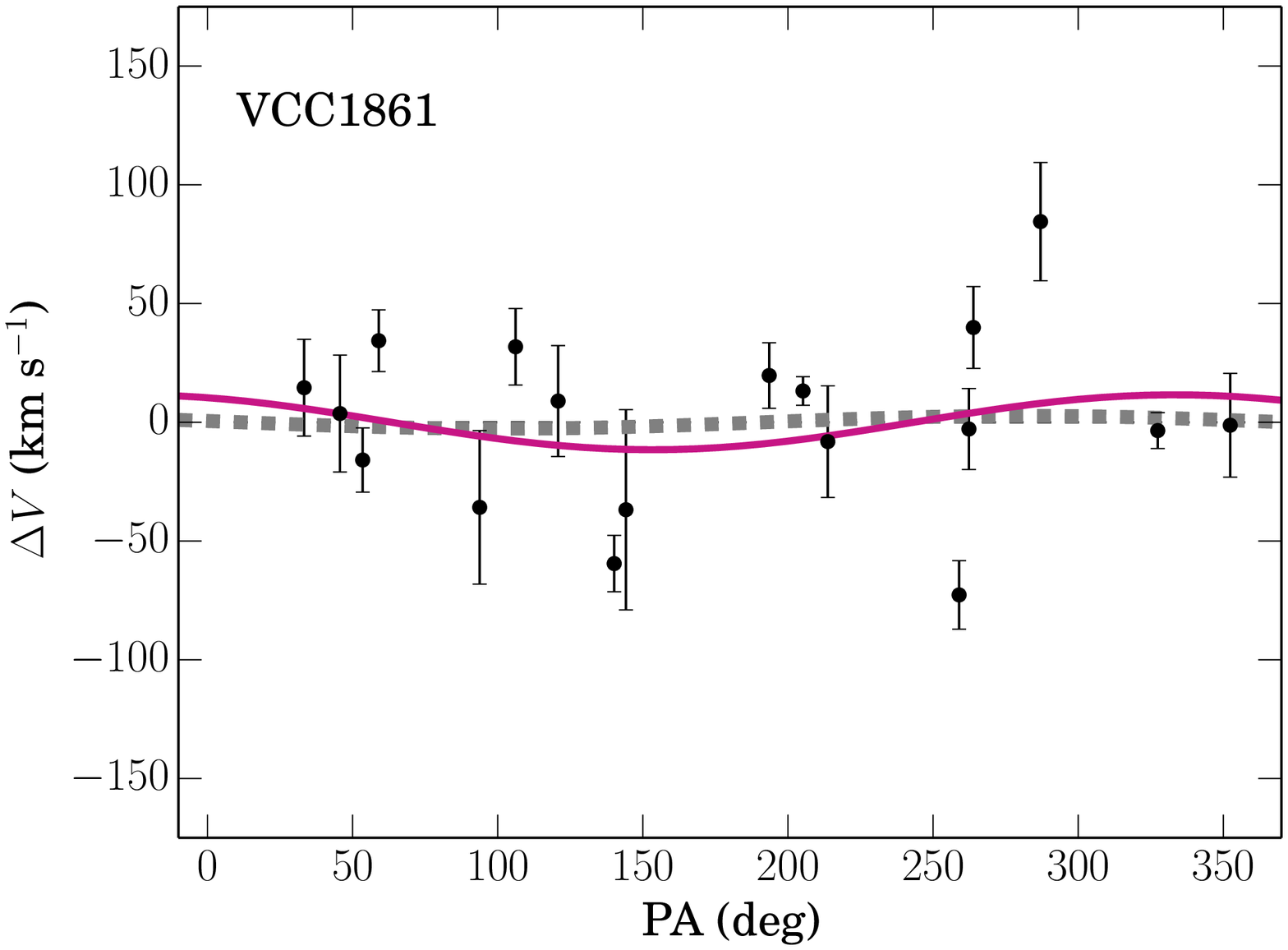}
\caption{Velocity of the GC satellites with respect to the main body of the dE as a function of position angle measured North-East. The black dots indicate the GC satellites. The dotted red line is the best fit cosine function obtained by using the RF method. The red line is the best fit cosine function obtained by using the RDSF method. }\label{cosine}
\end{figure}

We apply the RF and RDSF methods to the GC systems of these three dEs. The best fit cosine function for each method and each galaxy is shown in Figure \ref{cosine}. The best fit \Vmax, \sigGC, and \PAmax\ can be found in Table \ref{rotation}. The resulting kinematic parameters obtained in the RF and RDSF methods agree well within the $1\sigma_G$ uncertainties. Fixing the rotation axis, \PAmax, to reduce the number of free parameters in the fitting procedure does not change the estimated kinematics.


\begin{figure*}
\centering
\includegraphics[angle=0,width=5.9cm]{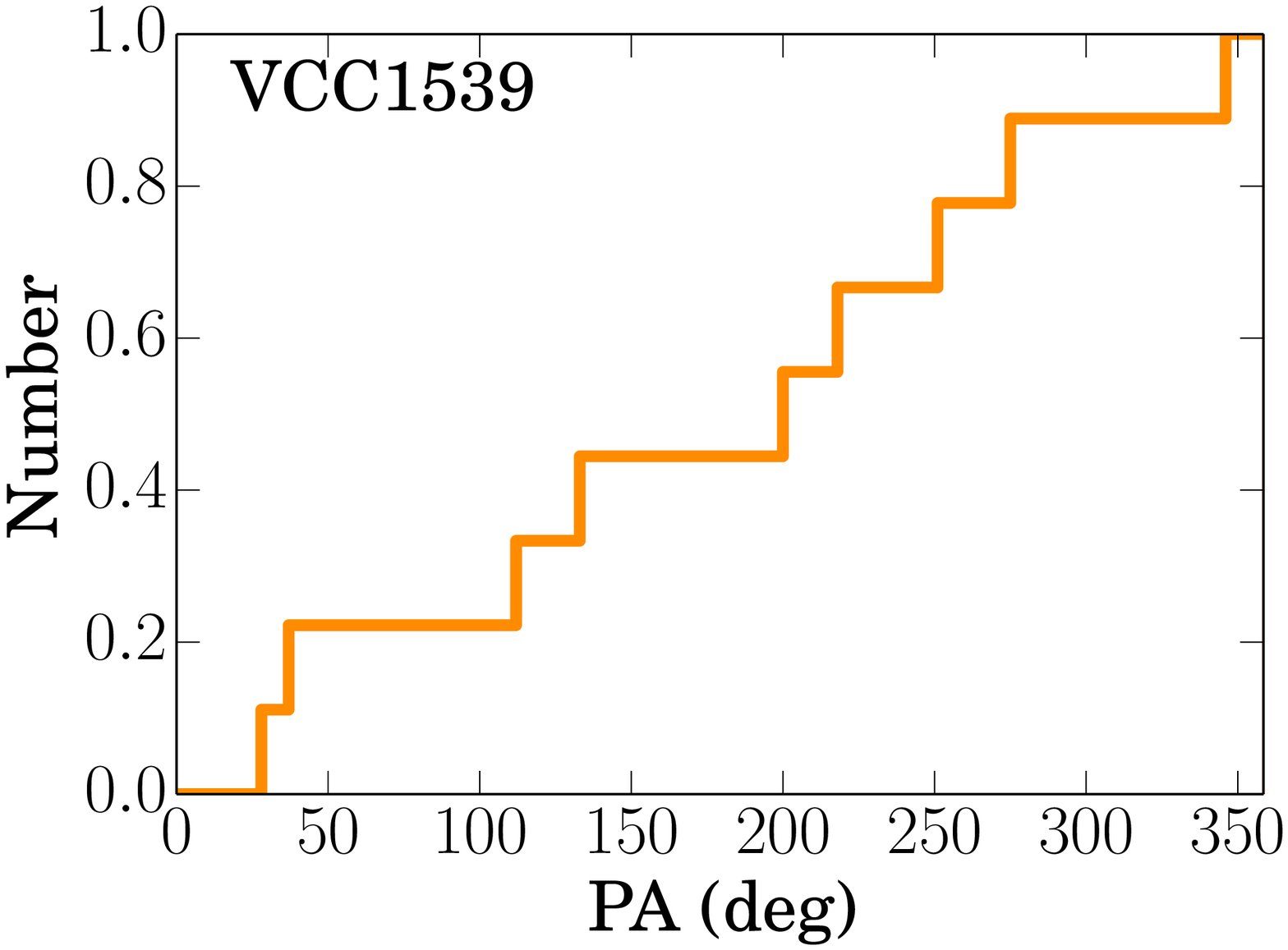}
\includegraphics[angle=0,width=5.9cm]{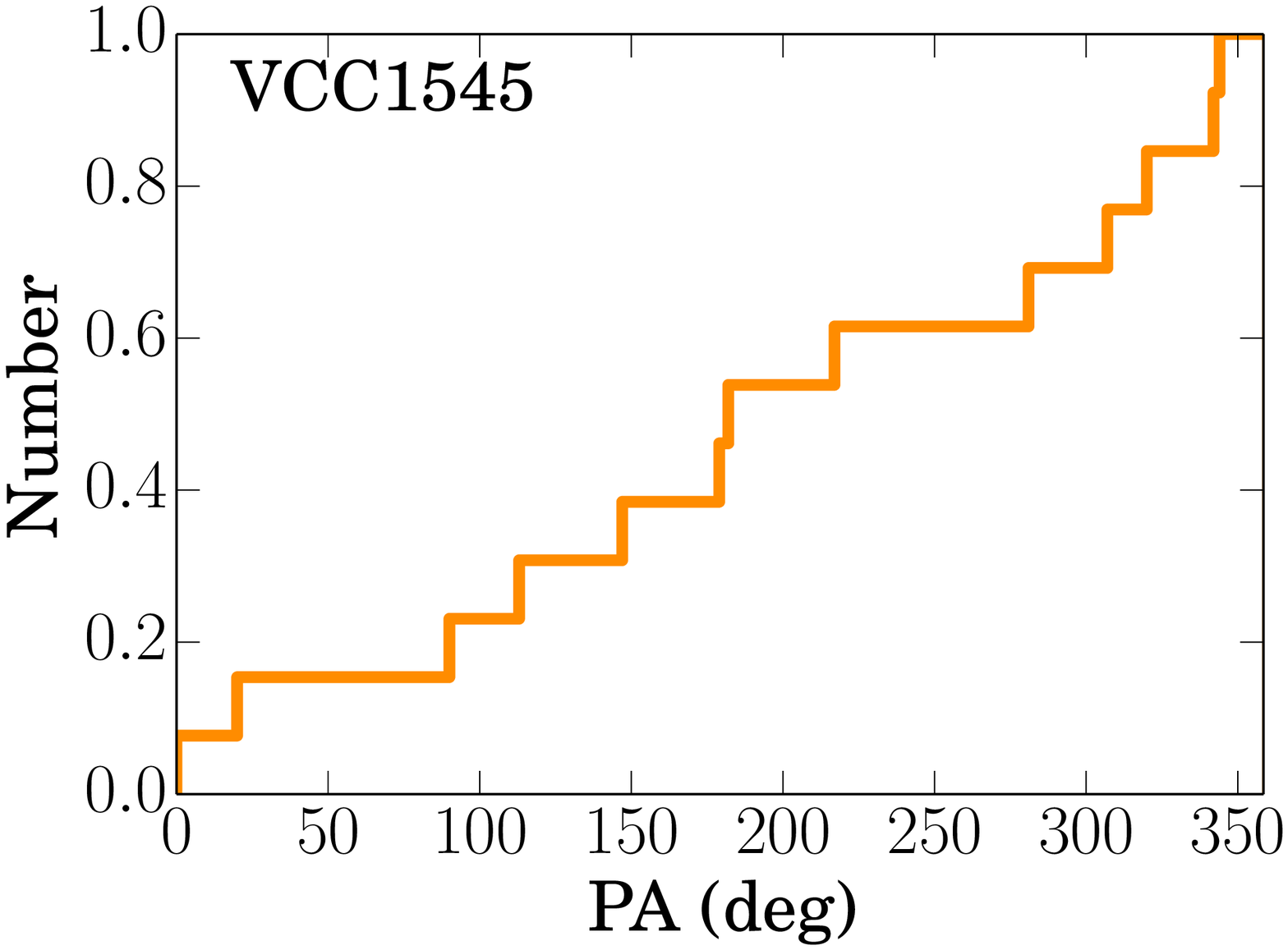}
\includegraphics[angle=0,width=5.9cm]{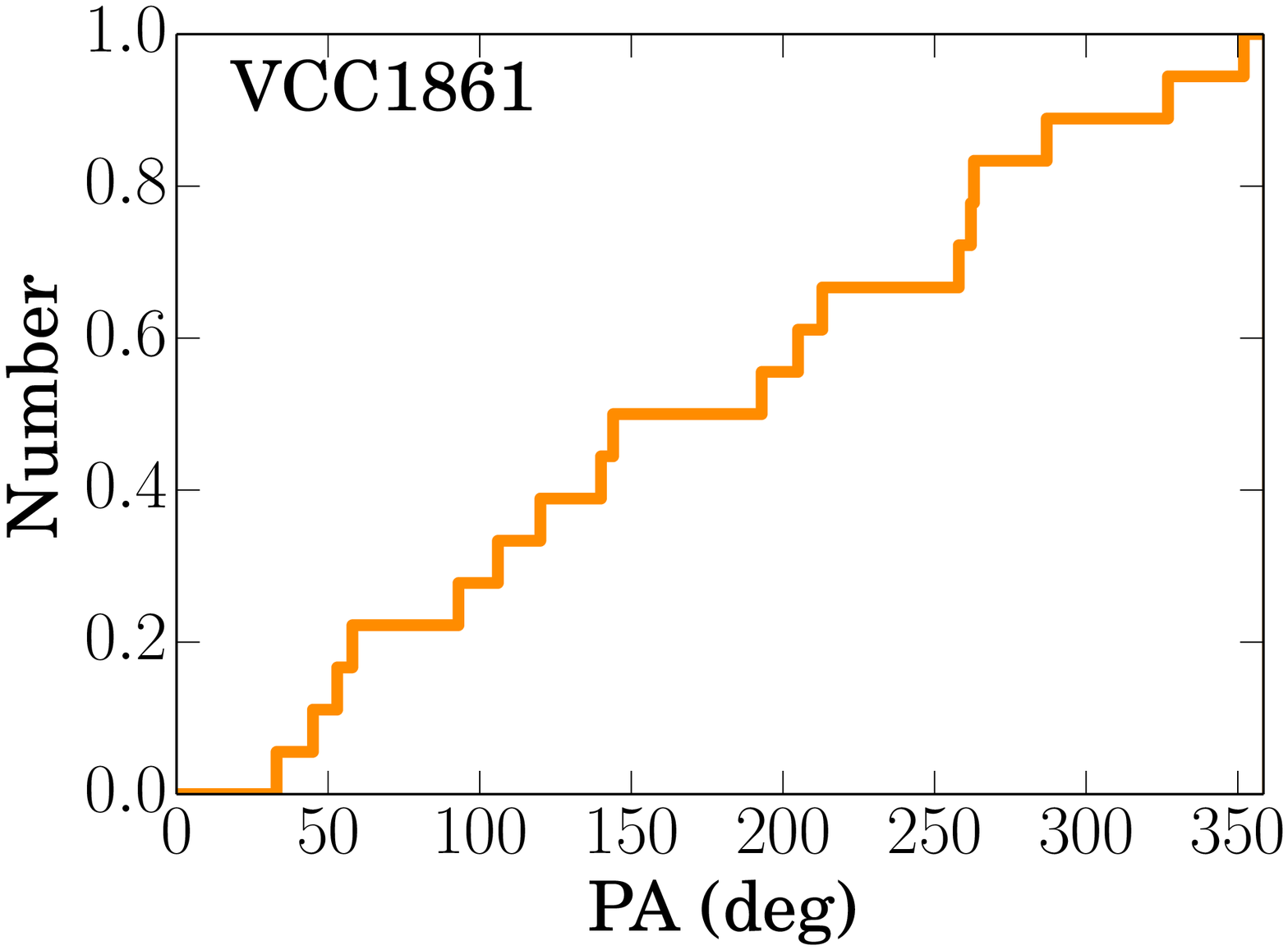}
\caption{Cumulative distribution of PAs for the observed GC satellites of, from left to right, VCC~1539, VCC~1545, and VCC~1861. In all cases, the distribution of PAs is uniform between $0^{\circ}$ and $360^{\circ}$.}\label{hists}
\end{figure*}

\begin{figure*}
\centering
\includegraphics[angle=0,width=5.9cm]{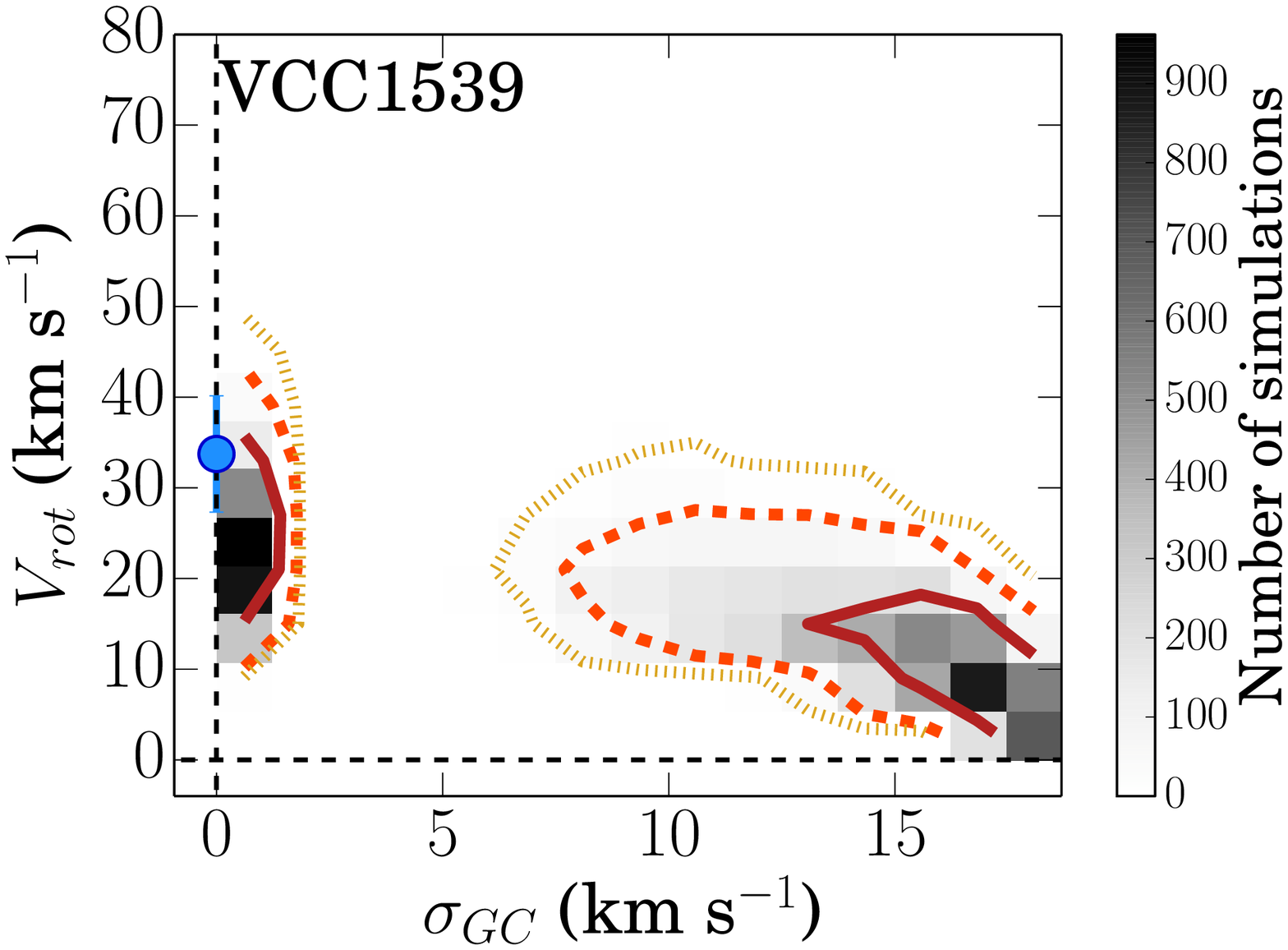}
\includegraphics[angle=0,width=5.9cm]{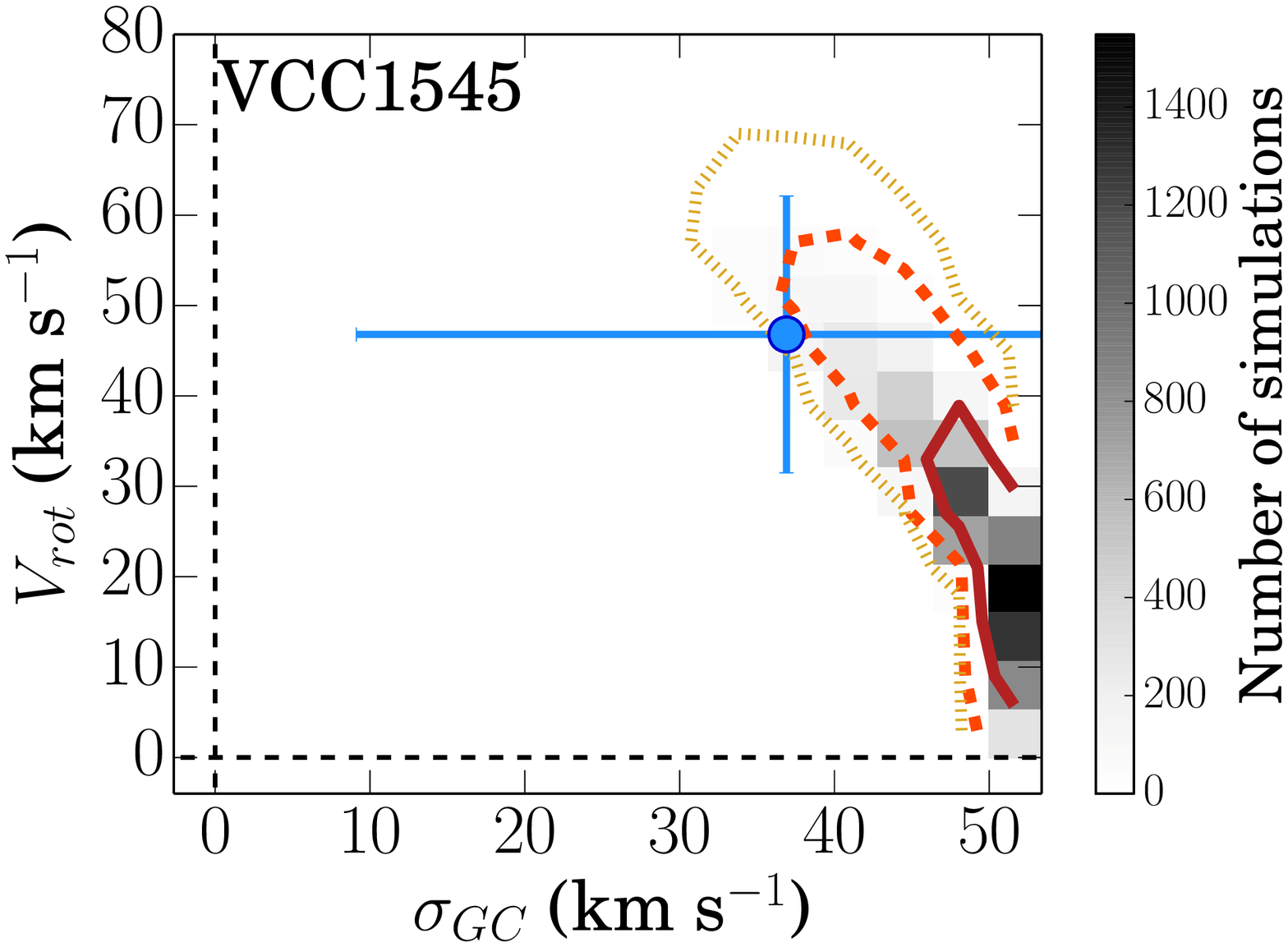}
\includegraphics[angle=0,width=5.9cm]{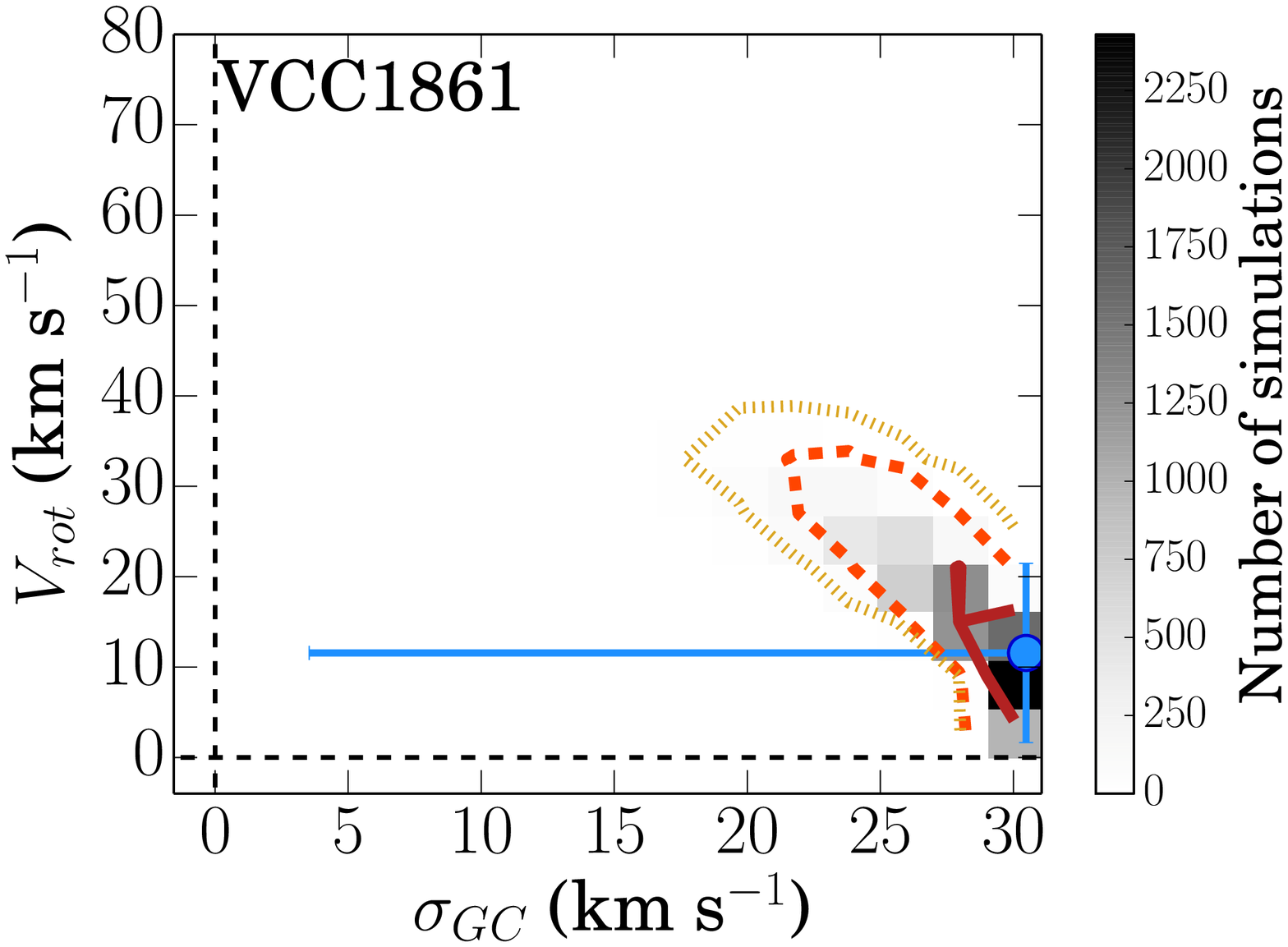}
\includegraphics[angle=0,width=5.9cm]{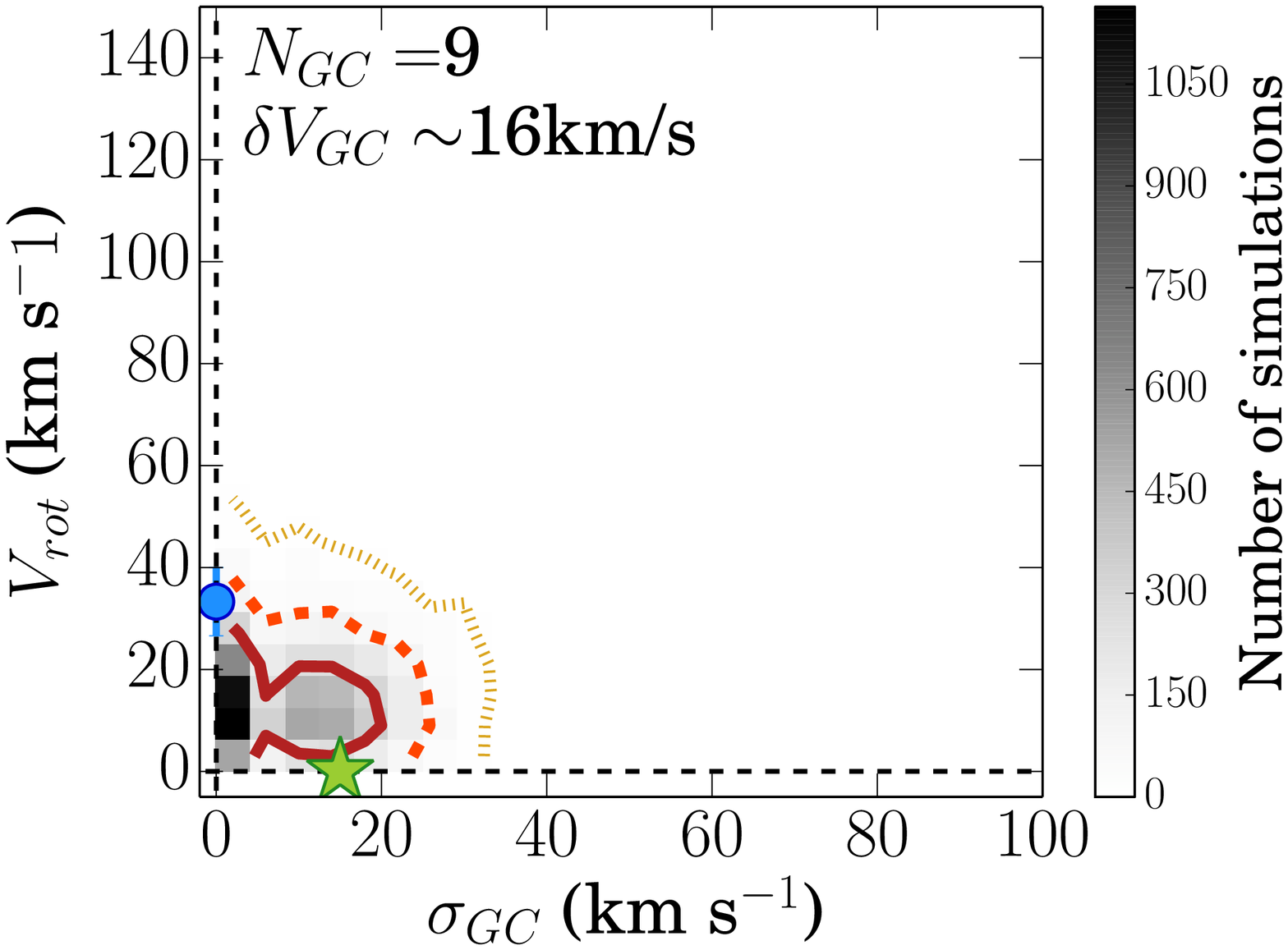}
\includegraphics[angle=0,width=5.9cm]{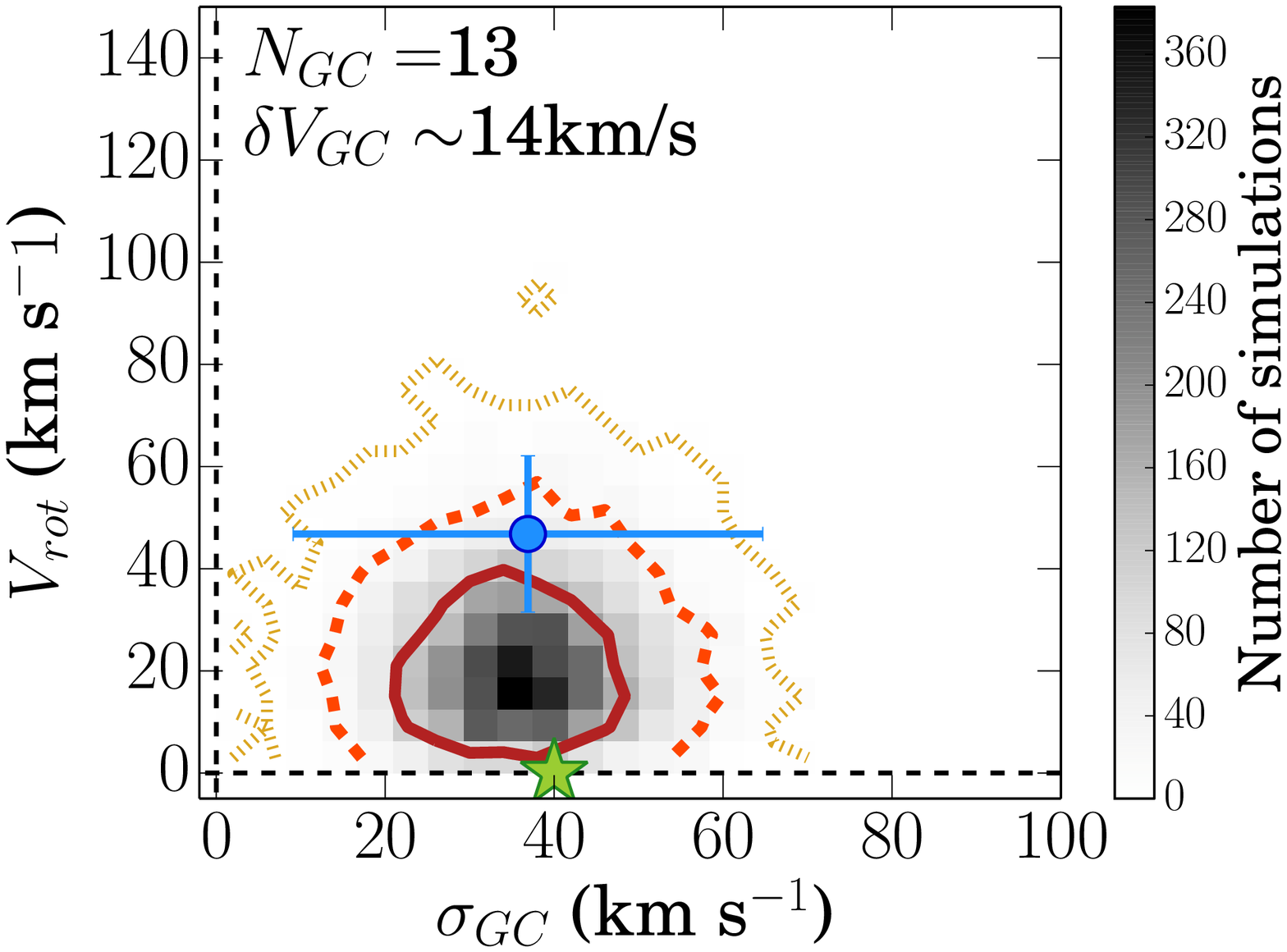}
\includegraphics[angle=0,width=5.9cm]{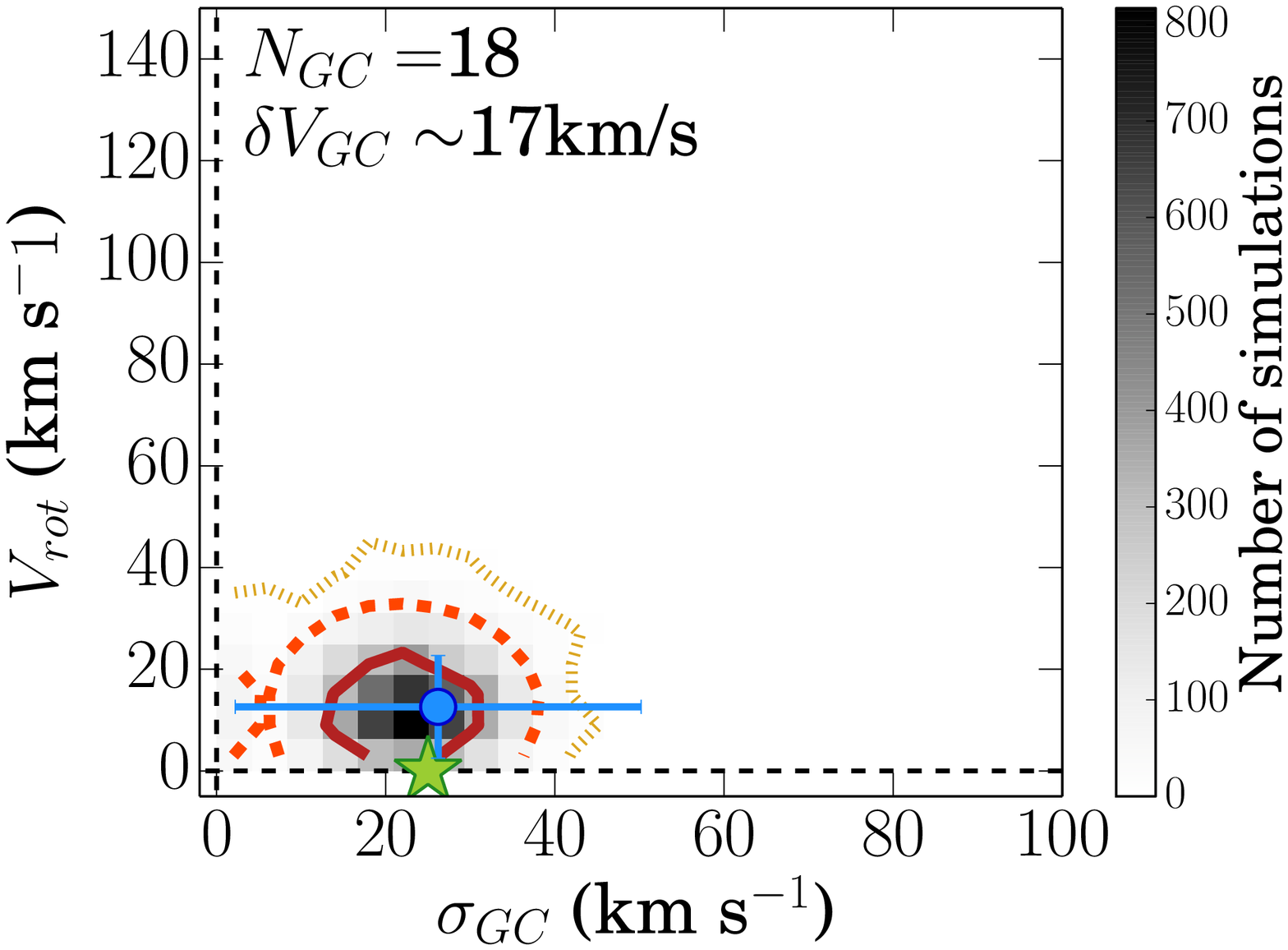}
\caption{Density map that results from creating 10000 non-rotating GC systems using the VCS simulations in the upper row and the VUS simulations in the lower row (see Section \ref{stat_sig} for a description of the VCS and VUS simulations). The blue dot indicates the best fit \Vmax\ and \sigGC\ for each GC system using the RDSF method. The green dot indicates the input \Vmax/\sigGC\ for the VUS simulations. The solid red line contour encloses $68\%$ of the simulations. The dashed orange contour encloses $95.5\%$ of the simulations. The dotted yellow contour encloses $99.7\%$ of the simulations. \NGC\ indicates the number of GCs in the system, and thus in the simulations, and $\delta V_{\rm GC}$ indicates the median velocity uncertainty in our data. The best fit \Vmax\ and \sigGC\ are always within the $2\sigma_G$ contours for non-rotating systems. This indicates that the measured kinematics are not statistically significant because they can be obtained by chance in a non-rotating GC system.}\label{Vsdens}
\end{figure*}

\begin{table}
\begin{center}
\caption{Kinematics of the GC systems \label{rotation}}
{\renewcommand{\arraystretch}{1.}
\resizebox{9cm}{!} {
\begin{tabular}{|c|c|c|c|c|c|}
\hline \hline
Galaxy  &  $N_{\rm GC}$ &  \PAmax  & \Vmax  &  \sigGC  &   \Vmax/\sigGC  \\
            &                      &    (deg)     & (\kms)    &    (\kms)   &                            \\
  (1)       &      (2)           &    (3)        &    (4)      &     (5)       &          (6)              \\
\hline
                   \multicolumn{6}{|c|}{ RF Method}    \\
\hline
VCC~1087 & 12           & 125.7$\pm$  26.5& 74.9 $\pm$ 21.7 & 43.1 $\pm$ 21.1 & 1.7 $\pm$ 1.0 \\
VCC~1261  & 12           &122.6 $\pm$ 22.0 & 48.5 $\pm$ 14.4 & 37.6 $\pm$ 21.1 & 1.3 $\pm$ 0.8\\
VCC~1528 & 10           &226.0 $\pm$ 12.4 & 67.9 $\pm$ 12.0 & 19.8 $\pm$ 26.1 & 3.4 $\pm$ 4.6\\
VCC~1539 & 9             &8.0  $\pm$ 130.8 & 33.2 $\pm$ 6.5    &   9.8 $\pm$ 14.2 & 3.4 $\pm$ 5.0\\
VCC~1545 & 13            & 138.8  $\pm$ 18.8 & 55.0 $\pm$ 17.4 & 35.6 $\pm$ 23.8 & 1.5 $\pm$ 1.1\\
VCC~1861 & 18            & 272.5 $\pm$ 84.1 &   9.4 $\pm$ 10.1 & 27.3 $\pm$ 7.4 & 0.3 $\pm$ 0.4\\
\hline
                   \multicolumn{6}{|c|}{ RDSF Method}  \\
\hline
VCC~1087 & 12           &104.2 $\pm$ 37.8&  40.2 $\pm$ 18.8 & 35.4 $\pm$ 9.2 & 1.1 $\pm$ 0.6\\
VCC~1261 &12            &94.7  $\pm$ 26.0  & 52.0 $\pm$ 19.2 & 35.8 $\pm$ 12.1 & 1.5 $\pm$ 0.7\\
VCC~1528 &10            &226.0 $\pm$ 12.6 & 67.9 $\pm$ 11.9 & 0.0 $\pm$ 4.0 & $--$\\
VCC~1539 & 9             &  8.0  $\pm$ 130.8 & 33.2 $\pm$ 6.5 &  0.0 $\pm$ 0.0 & $--$  \\
VCC~1545 & 13            & 155.4  $\pm$ 20.1 & 46.8 $\pm$ 15.8 & 36.9 $\pm$ 28.2 & 1.3 $\pm$ 1.1\\
VCC~1861 & 18            & 318.3 $\pm$ 108.2 & 12.6 $\pm$ 10.1 & 26.2 $\pm$ 24.0 & 0.5 $\pm$ 0.6\\
\hline
\end{tabular}
}}
\end{center}
\tablecomments{Column 1: galaxy name. Column 2: number of observed and spectroscopically confirmed GC satellites. Column 3: position angle, measured N-E, of the maximum rotation of the GC system. Column 4: maximum rotation speed of the GC system. Column 5: velocity dispersion of the GC system. Column 6: ratio between the rotation speed and the velocity dispersion of the GC system. This ratio is obtained from the best fit parameters. For details on how we measure these parameters see Section \ref{measurements}. The GC data for VCC~1087, VCC~1261, and VCC~1528 are from \citet{Beasley06,Beasley09} as discussed in Section \ref{Beasleys_dEs}.}
\end{table}

We test the statistical significance of our measurements by running the VCS and VUS simulations described in Section \ref{stat_sig}.
Figure \ref{hists} shows that the cumulative distribution of the PAs for the observed GC systems are uniform, thus the first approach for the VCS simulations is justified. The upper row of Figure \ref{Vsdens} shows the resulting density map of applying the first approach of the VCS simulations. The second approach shows an even broader and more elongated density maps because the PA and the velocity are coupled, and the third approach gives density maps that look nearly identical to the first approach. Because of the similarities between the three approaches, we choose to show only the first approach for the VCS simulations.

To run the VUS simulations, we need to choose a \sigGC. We estimate \sigGC\ by measuring the standard deviation of the line-of-sight radial velocities of the target GC system and subtracting the squared uncertainties. We run the VUS simulations for different \sigGC\ in the range \sigGC~$\pm 15$~\kms. The results are the same for all the tried velocity dispersions, the lower row of Figure \ref{Vsdens} shows the resulting density map for the estimated \sigGC.

Figure \ref{Vsdens} shows that the measured \Vmax\ and \sigGC\ for each GC system are between the $1\sigma_G$ and the $2\sigma_G$ contours of the VCS and VUS simulations for non-rotating systems. Therefore, the null hypothesis of the measured \Vmax\ being consistent with a non-rotating GC system cannot be rejected. In the case of VCC~1539 and VCC~1545, where the RF and RDSF methods estimate some rotation, this indicates that the measured rotation amplitude is not significant. However, in the case of VCC~1861, where the rotation speed estimated by using the RF and RDSF methods is consistent with zero, this confirms that the GC system bound to this dE is not rotating.

\subsection{Analysis of Globular Cluster Systems from the Literature}\label{Beasleys_dEs}

VCC~1087, VCC~1261, and VCC~1528 are also located in the inner core of the Virgo cluster. Their projected distance to M87 are $0.88^{\circ}$, $1.62^{\circ}$, and $1.20^{\circ}$, respectively. They are brighter and more elongated than the three dEs analyzed here for the first time. They do not show any disk-like structures, such as bars, spiral arms, or irregular features, in the NGVS deep images (see Figure \ref{colorimages}), also confirming the previous analysis by \citet{Lisk06a,Lisk07} based on the SDSS images. The x-shapes found for VCC~1261, VCC~1528, and VCC~1545 in the $g$ band residual images are typical of disky and boxy isophotes (see Figure \ref{colorimages}). These three dEs are slow rotators based on the stellar rotation speed estimation by \citet{etj14b,etj14c}. See Table  \ref{dEs_properties} for a summary of the main properties of these dEs.

\begin{figure}
\centering
\includegraphics[angle=0,width=8.5cm]{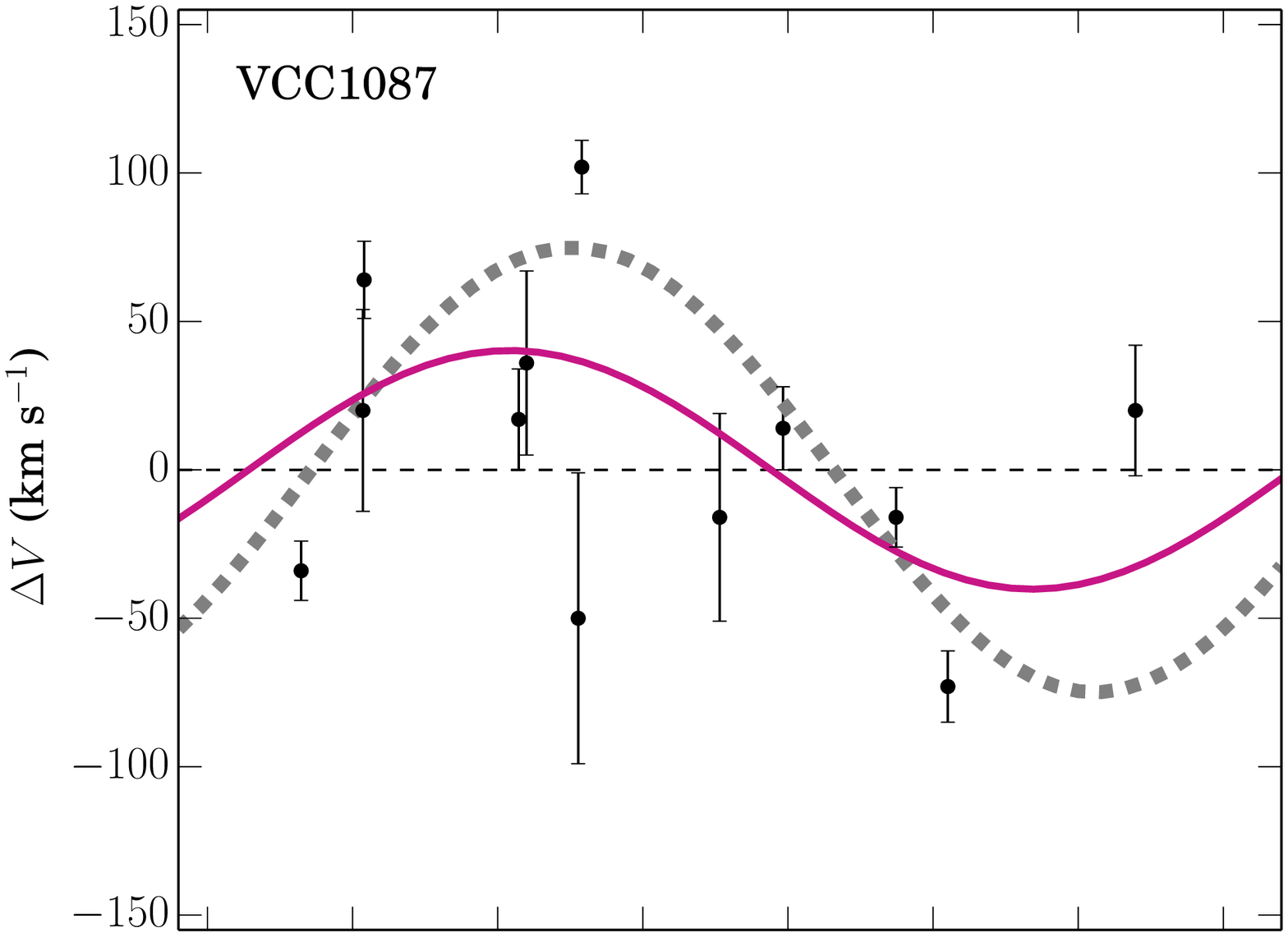}
\includegraphics[angle=0,width=8.5cm]{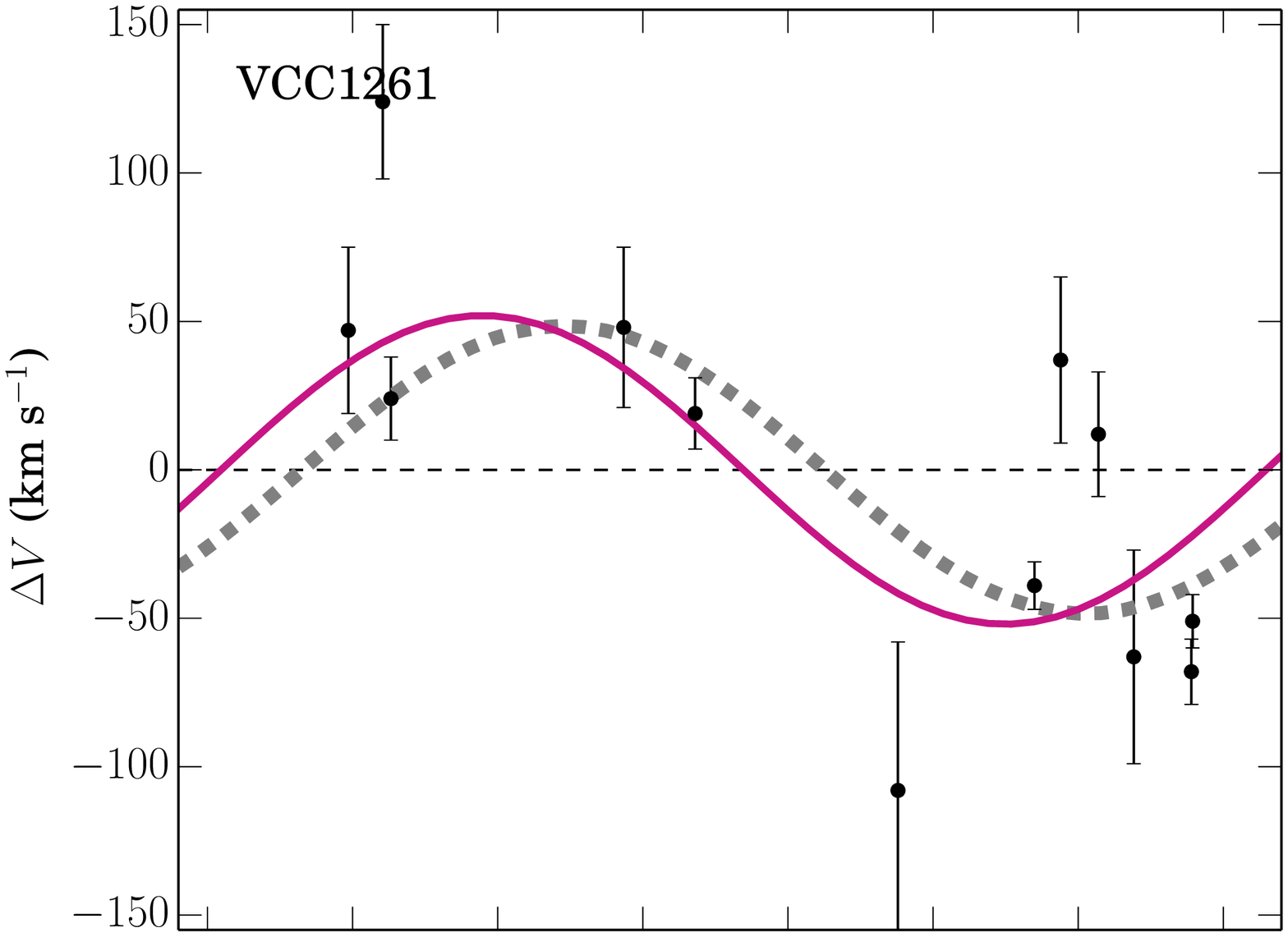}
\includegraphics[angle=0,width=8.5cm]{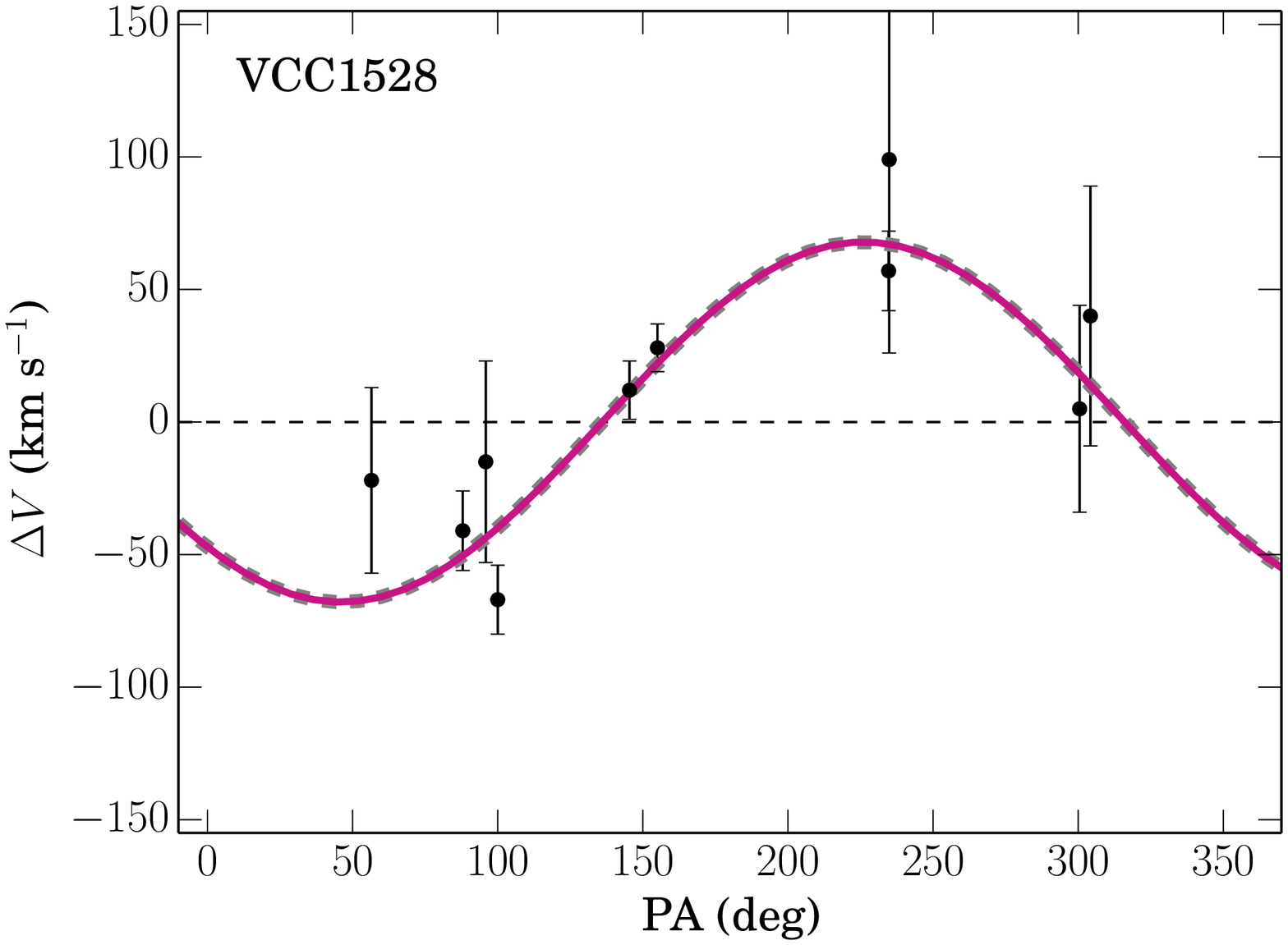}
\caption{Same as Figure \ref{cosine} for VCC~1087, VCC~1261, and VCC~1528. The data for these galaxies are from the literature \citep{Beasley06,Beasley09}, see Section \ref{Beasleys_dEs} for details.}\label{Beasley_cosine}
\end{figure}

We apply the RF method to these three dEs, as it was previously done by Beasley et al., and we find a good agreement with their published values within the $1\sigma_G$ uncertainties. We also apply the RDSF method and find a good agreement with the RF method within the $1\sigma_G$ uncertainties. The estimated kinematics obtained by fixing the rotation axis, thus, with one less free parameter in the fitting process, is also in good agreement within the $1\sigma_G$ uncertainties. Figure \ref{Beasley_cosine} shows the best fit cosine functions using the RF and RDSF methods for each dE. The best fit \Vmax, \sigGC, and \PAmax\ can be found in Table \ref{rotation}.

\begin{figure*}
\centering
\includegraphics[angle=0,width=5.9cm]{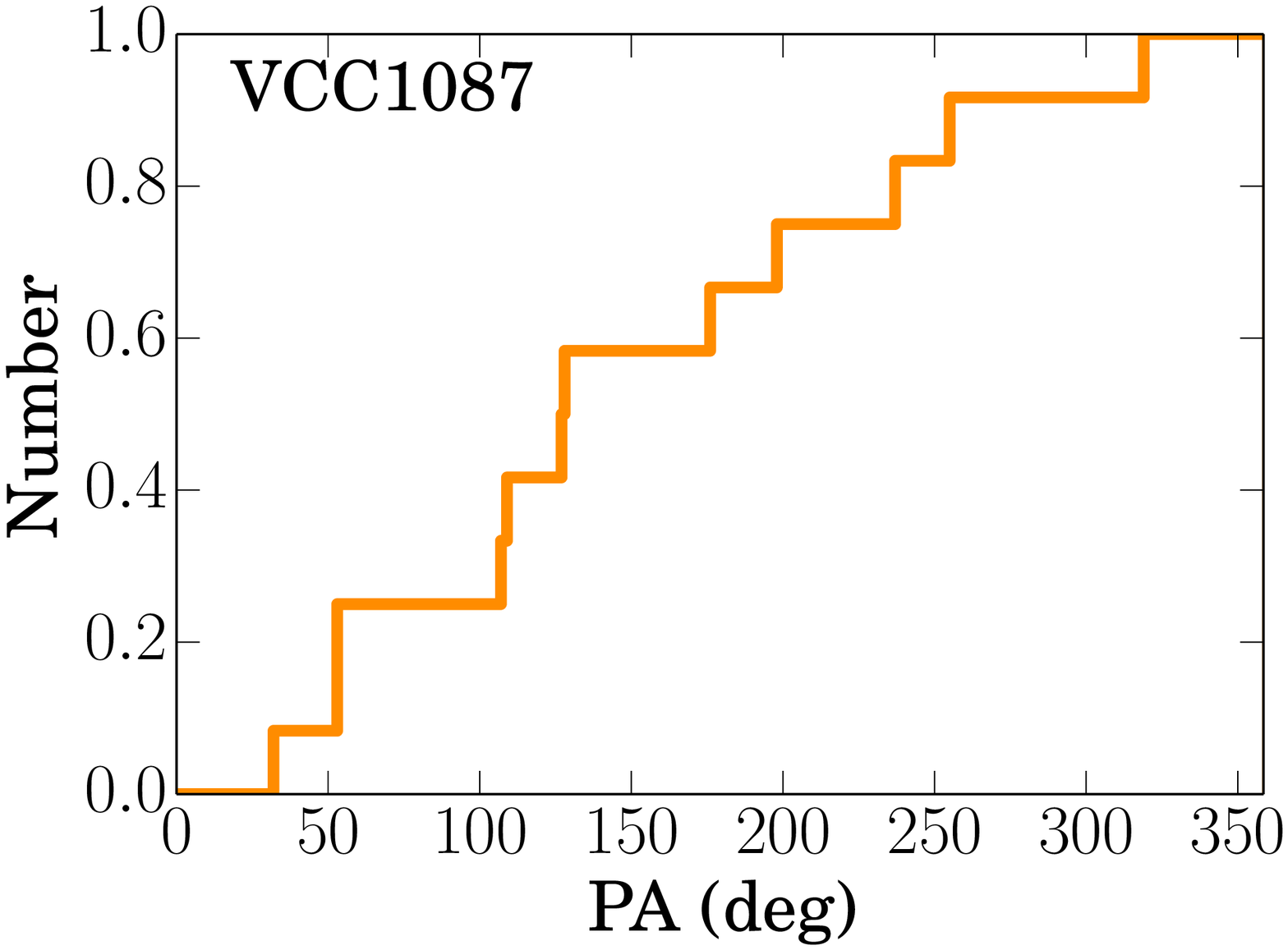}
\includegraphics[angle=0,width=5.9cm]{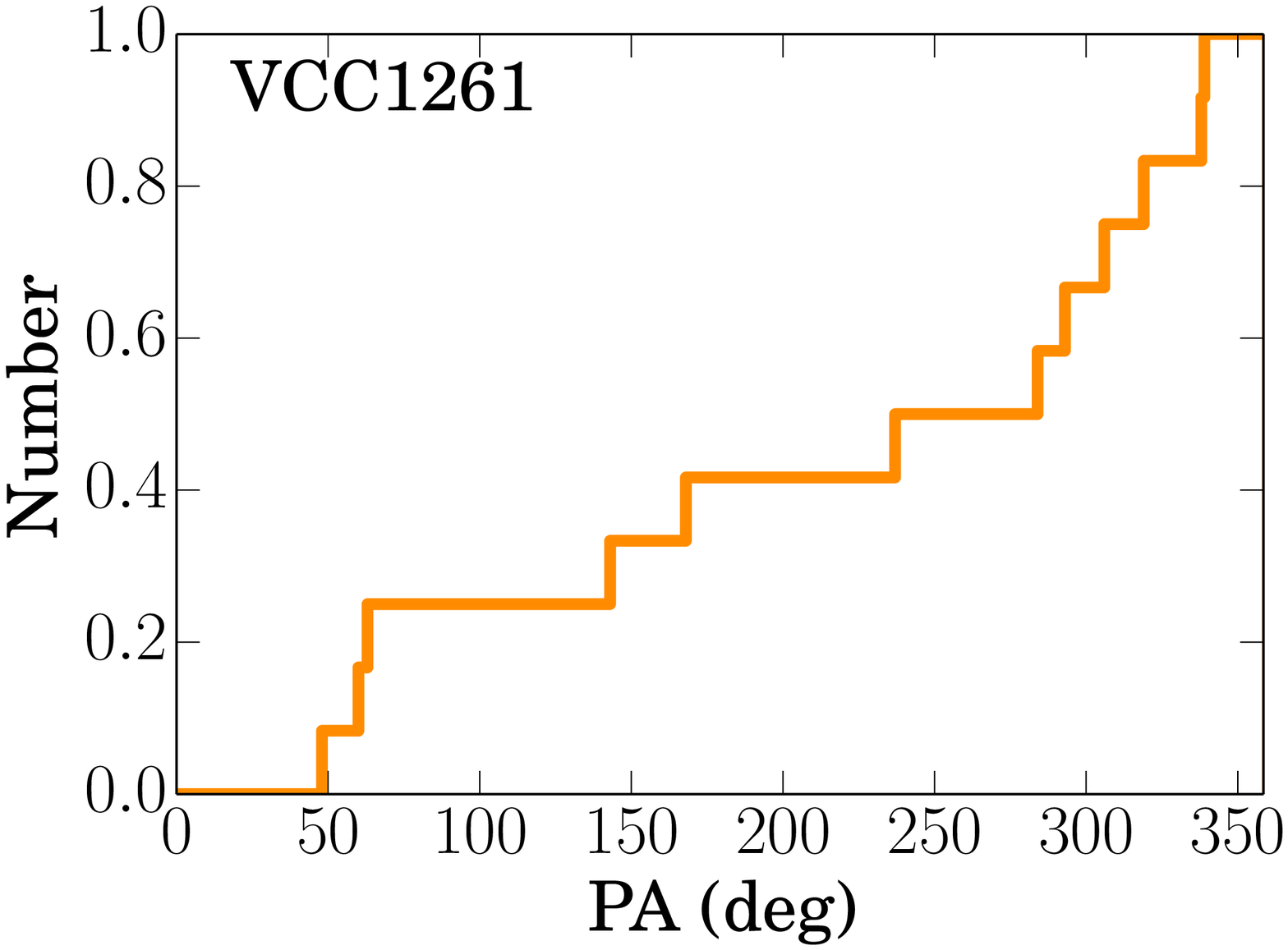}
\includegraphics[angle=0,width=5.9cm]{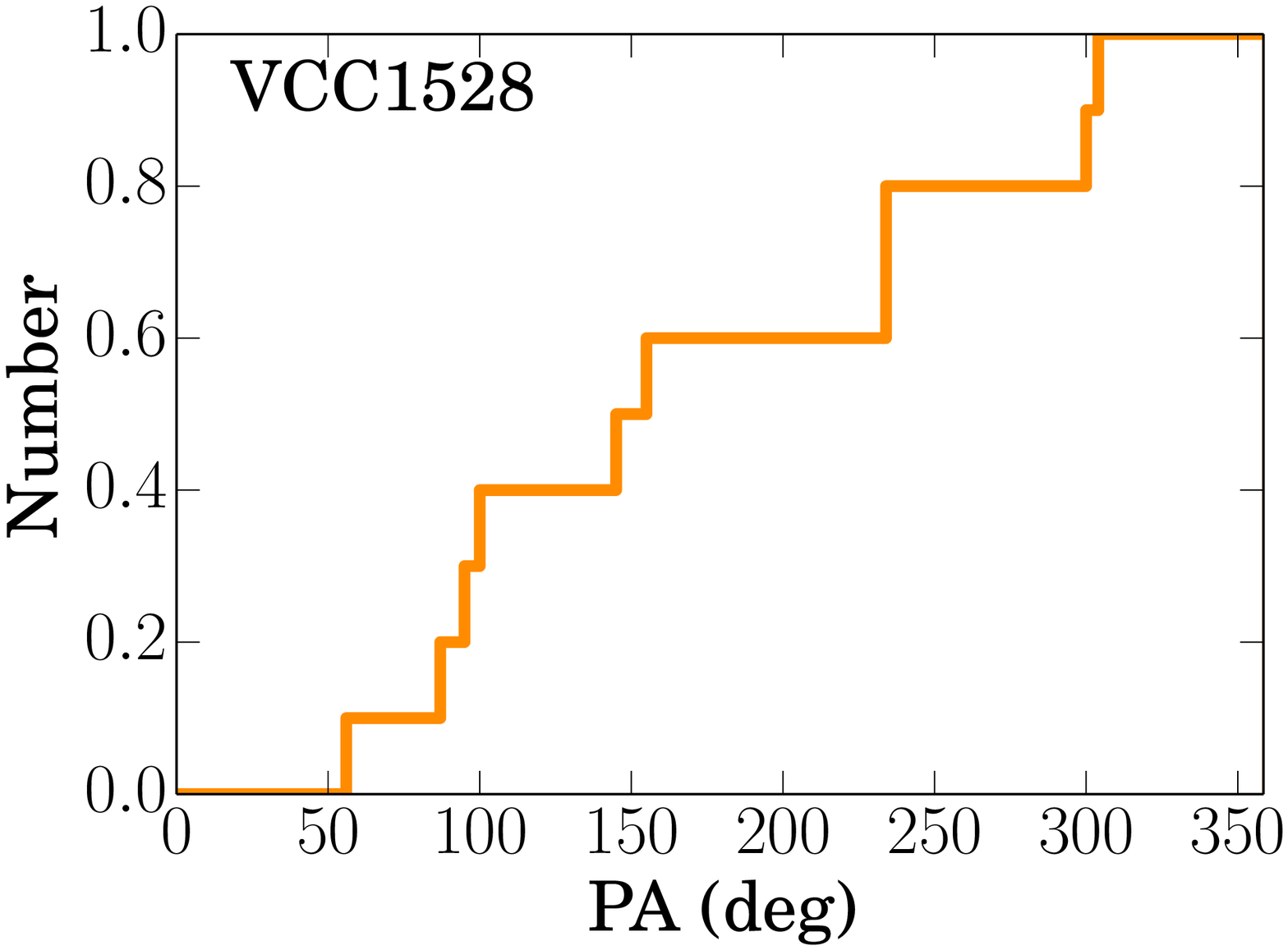}
\caption{Same as Figure \ref{hists} for VCC~1087, VCC~1261, and VCC~1528. The PA distributions are uniform between $0^{\circ}$ and $360^{\circ}$.}\label{Beasley_hists}
\end{figure*}

\begin{figure*}
\centering
\includegraphics[angle=0,width=5.9cm]{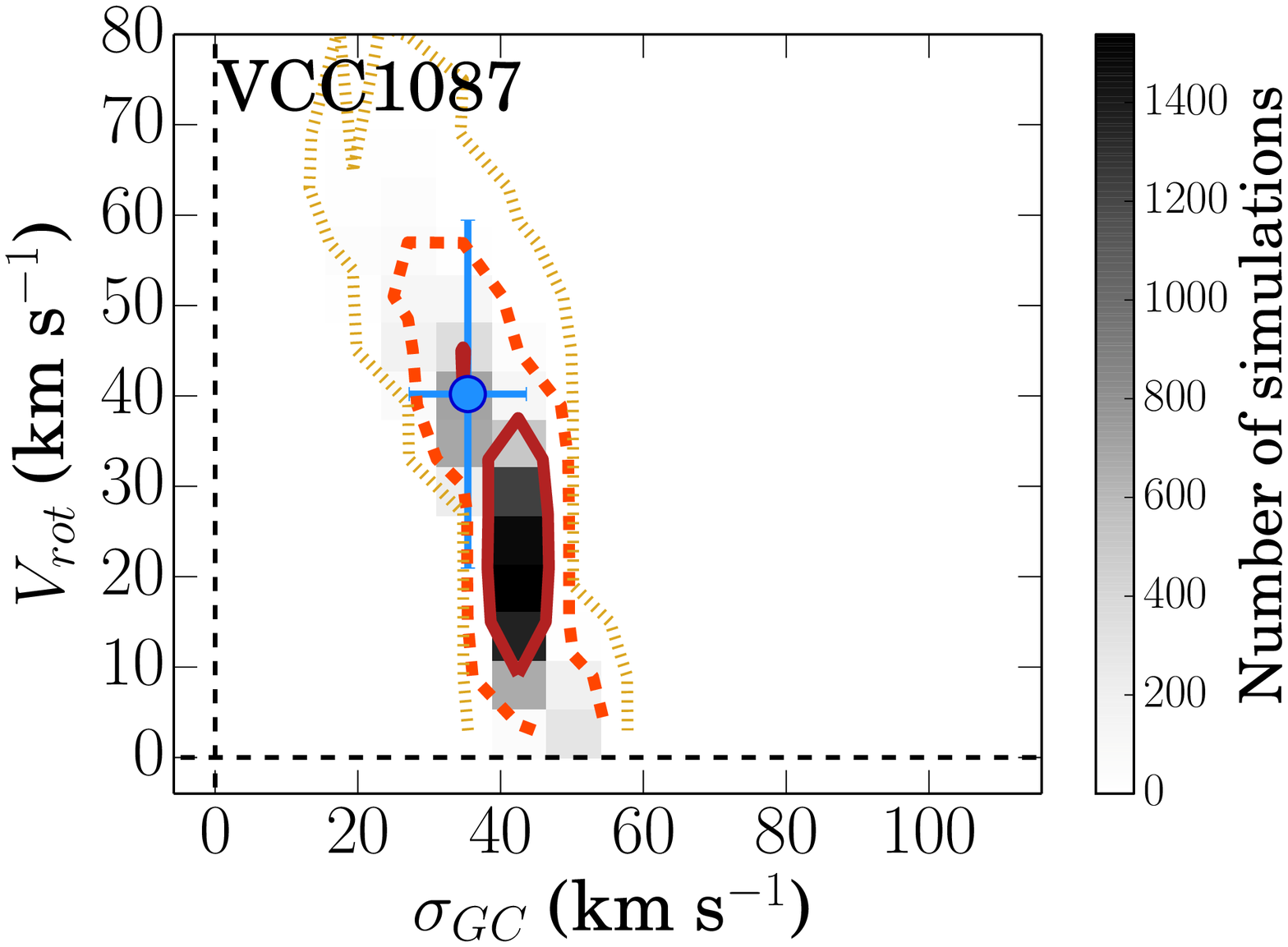}
\includegraphics[angle=0,width=5.9cm]{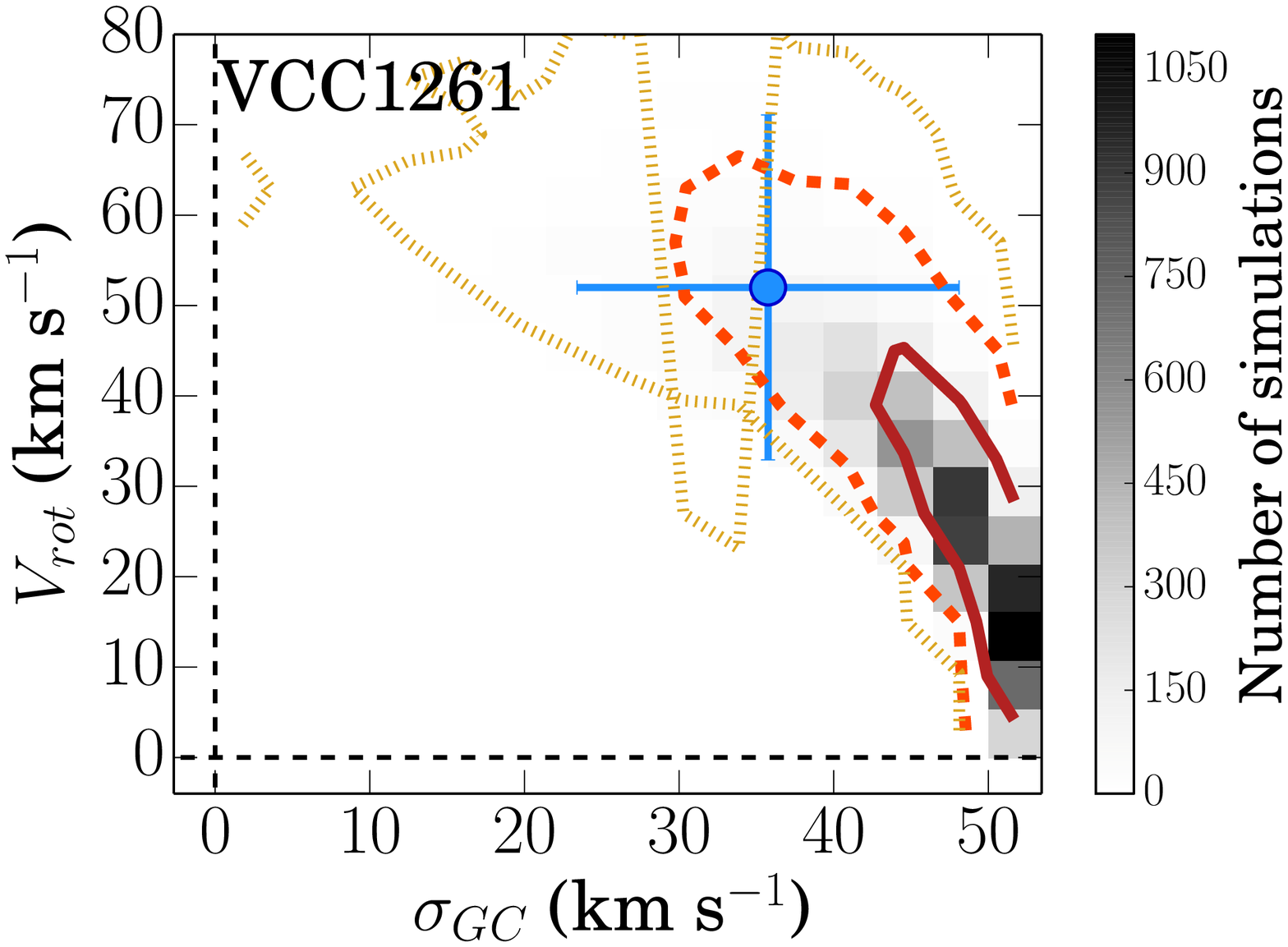}
\includegraphics[angle=0,width=5.9cm]{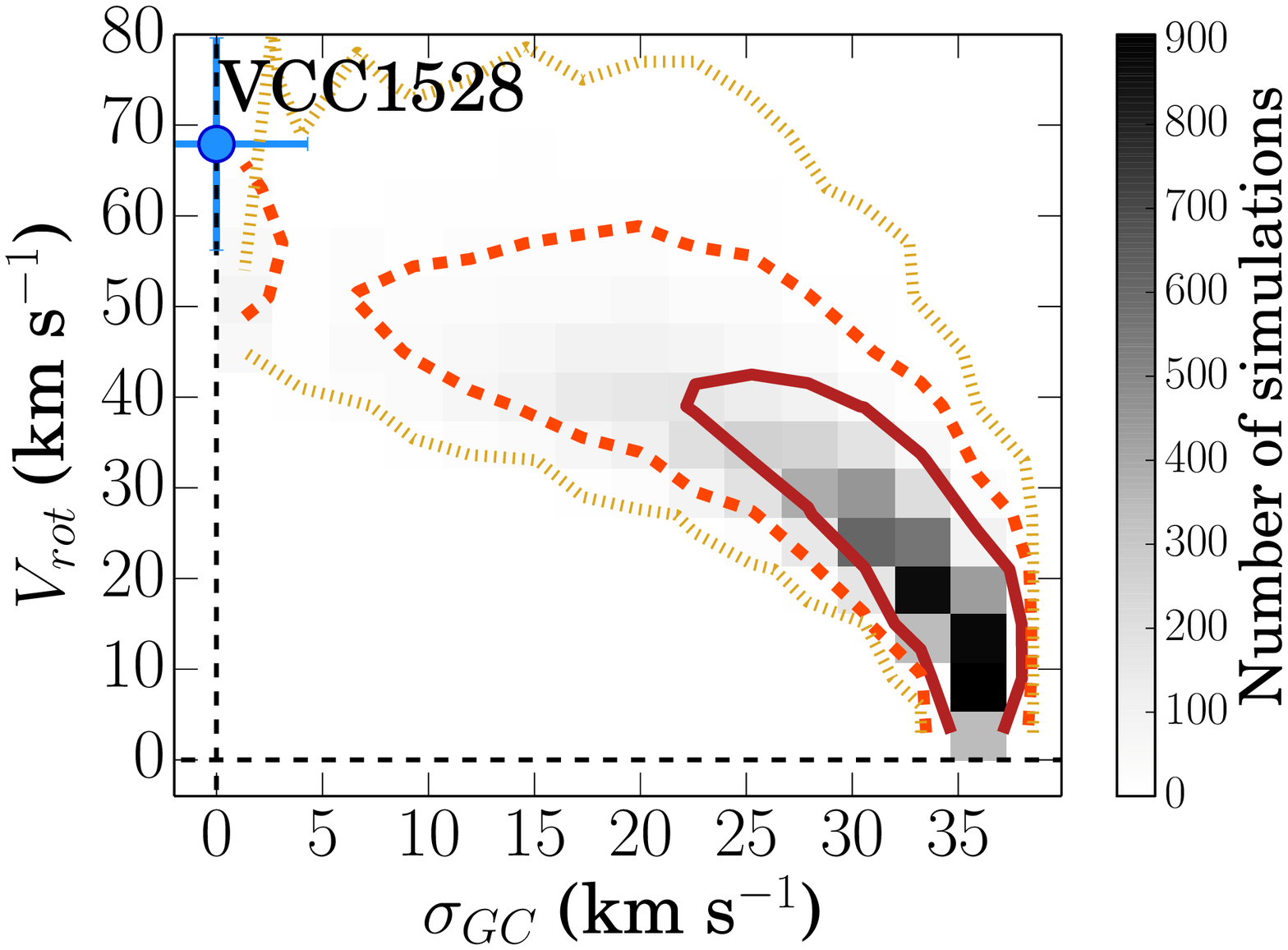}
\includegraphics[angle=0,width=5.9cm]{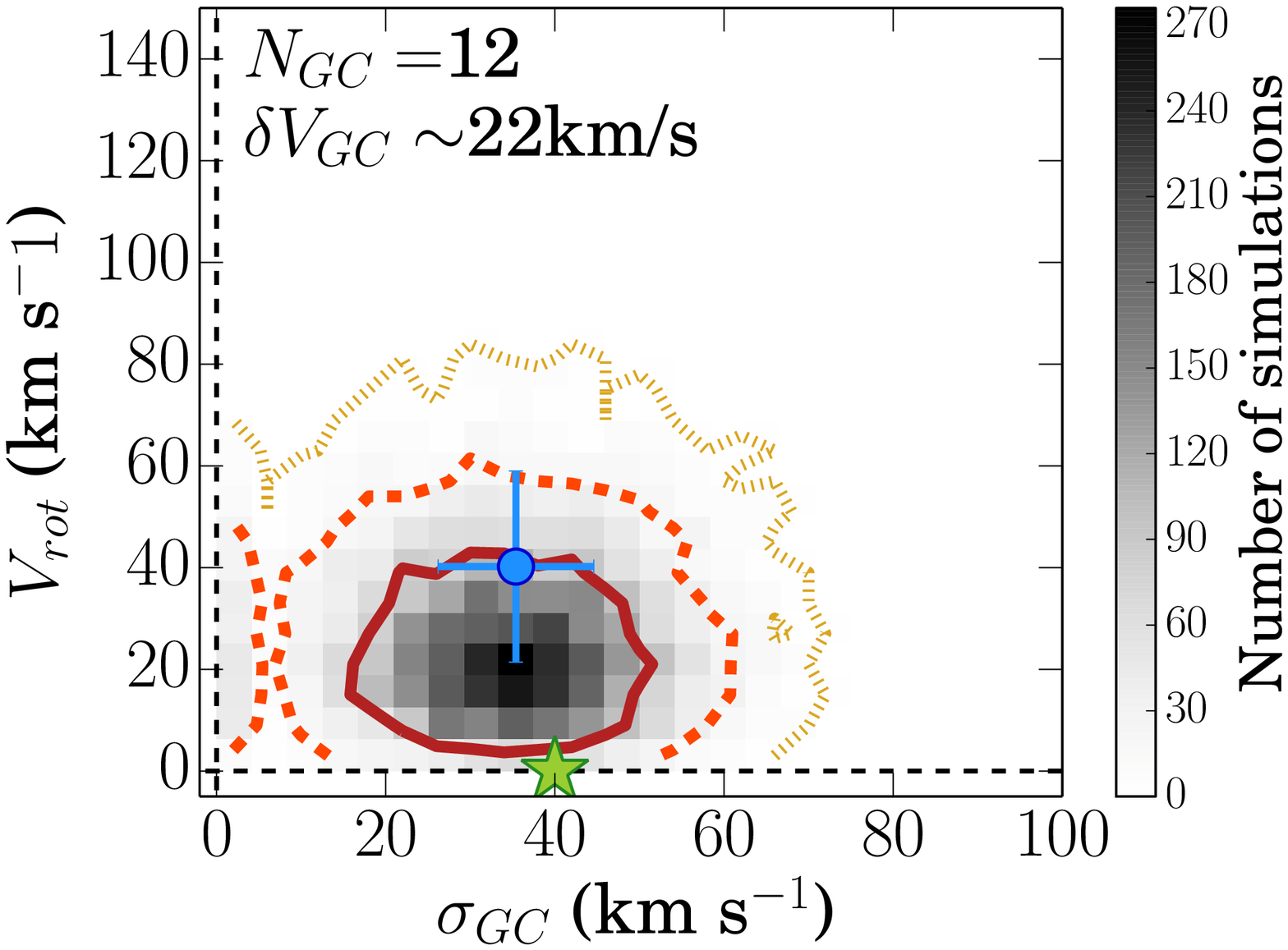}
\includegraphics[angle=0,width=5.9cm]{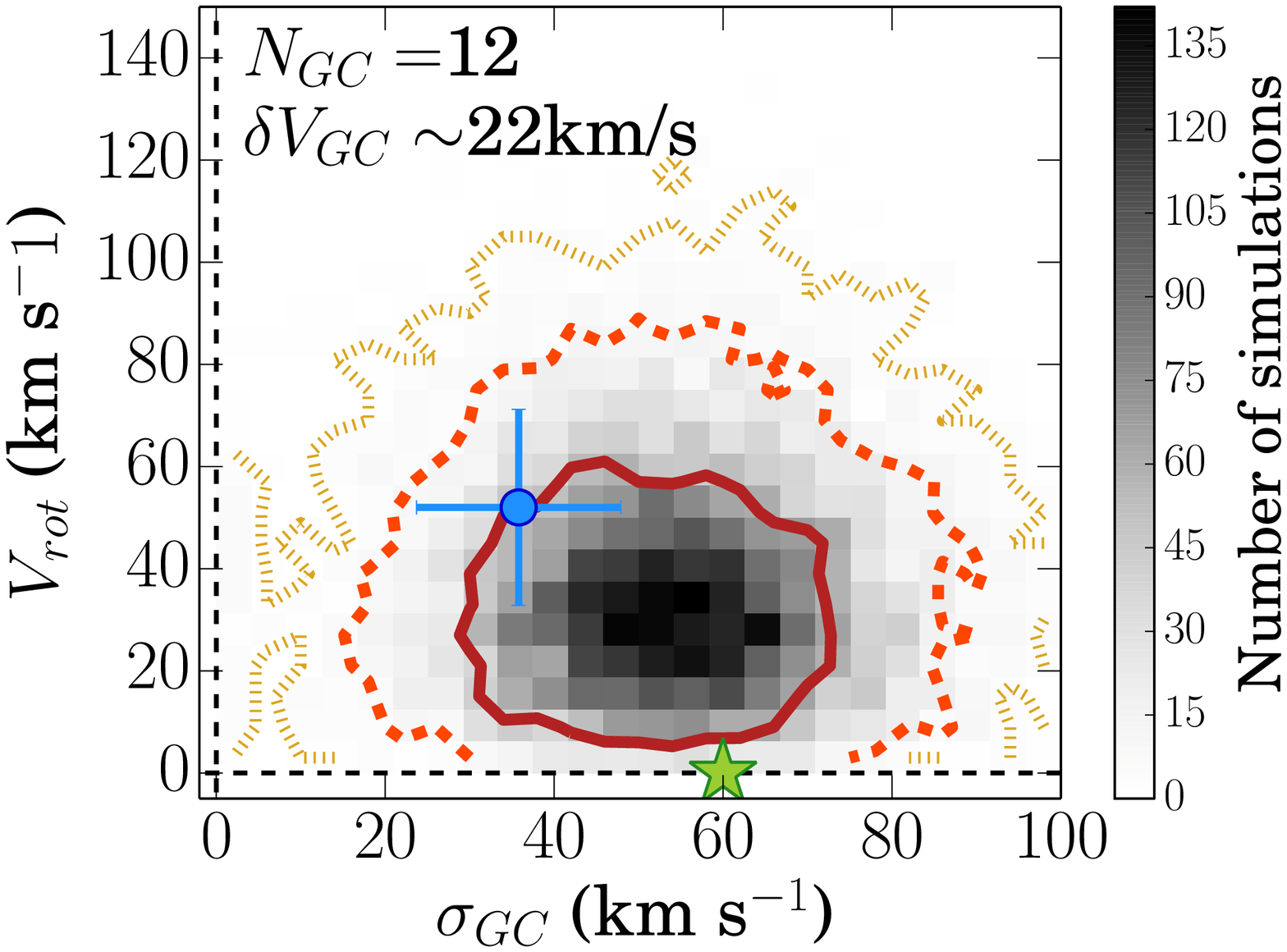}
\includegraphics[angle=0,width=5.9cm]{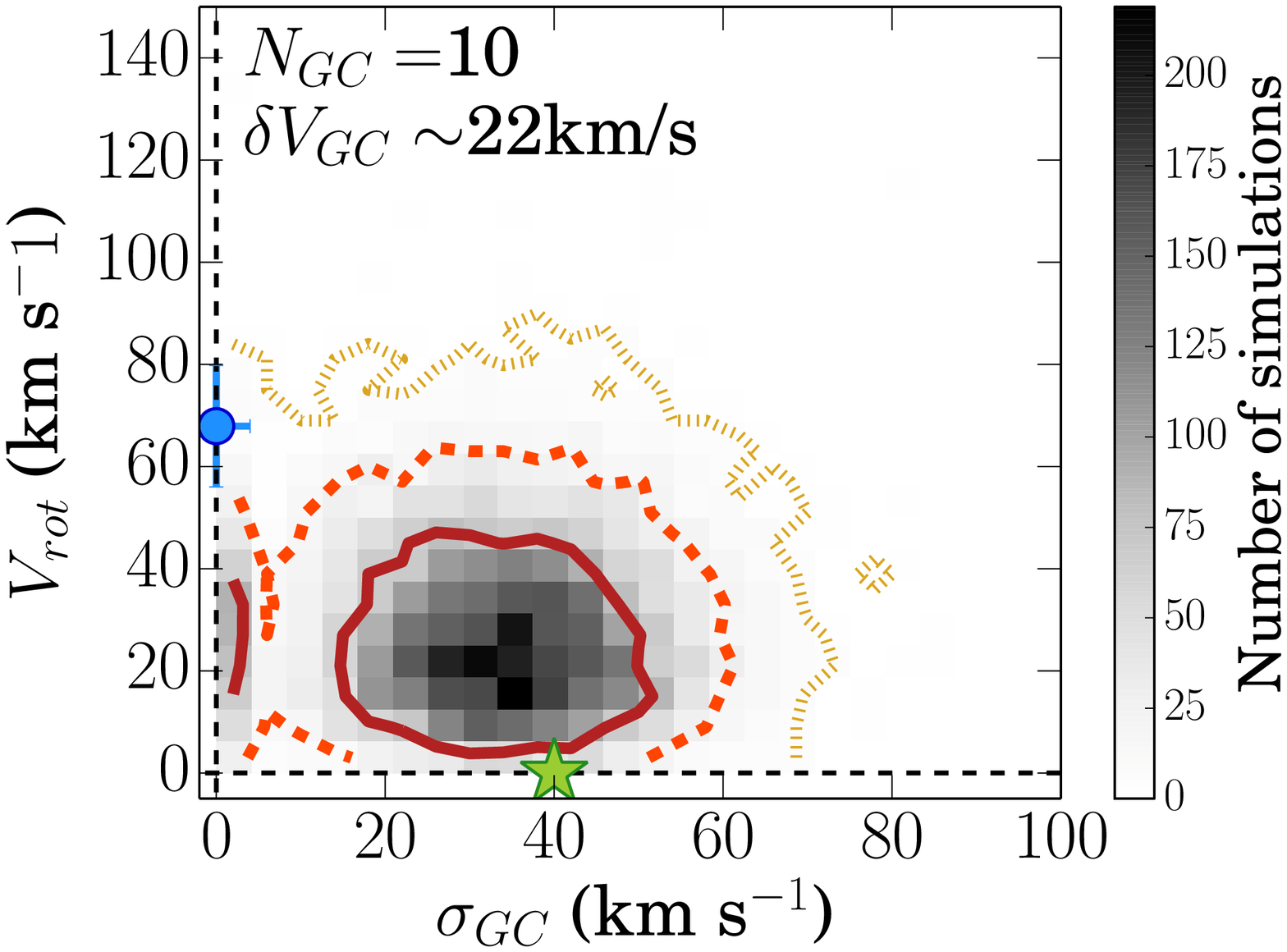}
\caption{Same as Figure \ref{Vsdens} for VCC~1087, VCC~1261, and VCC~1528.}\label{Beasley_Vsdens}
\end{figure*}

We repeat the same exercise as in Section \ref{our_dEs} to test the statistical significance of the measured \Vmax\ by running the VCS and VUS simulations described in Section \ref{stat_sig}. Figure \ref{Beasley_hists} shows that the cumulative distribution of the measured PAs for the GCs in each dE is uniform, thus, the first approach of the VCS simulations is justified. The three different approaches for running the VCS simulations also provide quantitatively the same results. As it happened for VCC~1539, VCC~1545, and VCC~1861, the first and third approaches look nearly identical and the second approach has a more elongated and broader contours. Thus, we chose to show only the first approach in the upper row of Figure \ref{Beasley_Vsdens}.

To run the VUS simulations, we estimate \sigGC\ as in Section \ref{our_dEs}. We also run the VUS simulations in the range \sigGC~$\pm 15$~\kms\ and find that the results are qualitatively the same for all the input velocity dispersions, thus we choose to show in Figure \ref{Beasley_Vsdens} the density map resulting for the estimated \sigGC.

Figure \ref{Beasley_Vsdens} shows that the measured \Vmax\ and \sigGC\ in all three GC systems are between the $1\sigma_G$ and $2\sigma_G$ contours of the VCS and VUS simulations for non-rotating systems. VCC~1528's measurements are right on the $2\sigma_G$ level, so maybe the rotation amplitude is marginally significant for this galaxy although its \sigGC~$=0$~\kms\ is certainly underestimated given that our simulations show that a zero \sigGC\ is the most likely result for systems with large velocity uncertainties and a low number of GCs (see Section \ref{finite_sampling_effects}). The measured \Vmax\ and \sigGC\ for VCC~1087 and VCC~1261 are between the $1\sigma_G$ and $2\sigma_G$ contours of the VCS and VUS simulations for non-rotating systems. Therefore, the null hypothesis of the measured rotation amplitude being consistent with a non-rotating system cannot be rejected. This suggests that the measured \Vmax\ for VCC~1087 and VCC~1261 are not statistically significant, indicating that it is still unknown whether the GC systems of these dEs are rotating or not. In the case of VCC~1528, its GC system {\it may} be rotating and the \Vmax\ estimated should be cautiously considered an upper limit. 

\section{Discussion}\label{discussion}

Kinematics is a powerful tool to study the formation and evolution of galaxies. In the case of Virgo cluster dEs, the amplitude of the rotation curve depends on the position of the galaxy in the cluster, i.e. slow rotators are found in the center of the cluster while the fastest rotators tend to be in the outer regions. This may suggest that the dynamics of these dEs are being transformed by the environment \citep{etj09,etj14c}. 

The kinematics of the outer halos of galaxies can generally be used to constrain their mass assembly history. For example, finding outer halo rotation could be interpreted as evidence for a major merger that transfered angular momentum to the outer regions of the remnant galaxy \citep[e.g.;][]{Cote01,Cote03}. On the other hand, not finding rotation in the outer halo, or finding that the outer halo is dominated by dispersion, can be interpreted as the accretion of multiple GC systems \citep[e.g.;][]{Wang11}. In the case of low mass cluster galaxies, as the Virgo cluster dEs analyzed here, finding or not rotation in the GC system is an indication of the environmental mechanism affecting them \citep[e.g.;][]{Boselli08a,SanchezJanssen12}.

At high masses, several works study the kinematics of red and blue GC satellites (metal-rich and metal poor GCs, respectively) in comparison with the stellar kinematics of the galaxy. Although in many cases it seems that the red GCs are coupled to the stellar kinematics while the blue GCs are decoupled from the stars \citep[e.g.;][]{Romanowsky09,schuberth10,Strader11,Norris12} there are some other cases where all three components are coupled or decoupled \citep[][]{Pota13,Li15}. The number of GC systems analyzed in massive early-type galaxies is still small to make any conclusions based on luminosity, other internal properties, or the environment. This makes it very difficult to guess the expected kinematics of dE GC satellites.

We measure the kinematics of the globular cluster systems bound to six Virgo cluster dEs. We find that the GC system of VCC~1861 is not rotating, in the same way that its stars are not rotating \citep{etj11,etj14b}. The GC system of VCC~1528 {\it may} be rotating. VCC~1528's is a slow rotator based on its stellar dynamics \citep{etj14b,etj14c}. We measure some rotation for the remaining four GC systems which are consistent with previous works, but, after careful simulations we find that these measurements are not statistically significant. Thus, it remains still unknown whether the GC systems of VCC~1087, VCC~1261, VCC~1539, and VCC~1545 are rotating or not.

With secure dynamical information for only one GC system, it is difficult to draw any conclusions on formation scenarios.  However, it is an important first step towards a database of Virgo galaxies, ranging from high to low masses, with stellar and GC kinematic information. Our analysis shows that the current number of observed GC satellites and the velocity uncertainties are not enough to conclude that these dEs have a disky origin as suggested by \citet{Beasley09}. This conclusion should be revisited with an improved data set.

Measuring the rotation speed of GC systems bound to dwarf galaxies is technically very challenging. The RF and RDSF methods, extensively used in the literature, produce some overestimation of \Vmax\ and underestimation of \sigGC, specially when the number of observed GC satellites is small. In the most extreme case of having only 1~GC in the system, the RF and RDSF methods fit a cosine function that goes through this 1~GC finding an infinite rotation amplitude and a zero velocity dispersion.

We study the effects that the number of GC satellites and the velocity uncertainties have on the measured \Vmax\ and \sigGC\ for rotating and non-rotating GC systems. We find that \Vmax\ tends to be overestimated and \sigGC\ tends to be underestimated. These offsets depend on parameters that are unknown, such as \Vmax/\sigGC, and on other parameters that we can control, such as the number of GCs in the sample and the velocity uncertainties. This indicates that the systematic biases cannot be corrected but they can be reduced. For GC systems with \NGC~$\lesssim 10$, the measured \Vmax\ can be considered an upper limit and the measured \sigGC\ a lower limit. For GC systems with \NGC~$\gtrsim 20$, the bias in \Vmax\ and \sigGC\ is close to zero for systems with \Vmax/\sigGC~$\gtrsim 1$, i.e. for systems with \Vmax~$\gtrsim 30$~\kms\ if \sigGC~$\sim 30$~\kms. For systems with \Vmax/\sigGC~$< 1$, the bias is typically half the median velocity uncertainty. These biases can be significantly reduced with small velocity uncertainties, however, they do not fully disappear.

\section{Summary and conclusions}\label{concl}

We have performed a state-of-the-art measurement and compilation of the largest sample of GC velocities in dwarf early-type galaxies. Our catalog consists of 82 new velocity measurements of GC satellites of 21 dEs, and 34 velocity measurements of GC satellites of 3 dEs from the literature. 

We use Keck/DEIMOS spectroscopy to analyze, for the first time, the kinematics of the  GC systems bound to the Virgo cluster dEs VCC~1539, VCC~1545, and VCC~1861. These are the three dEs with the largest number of observed GCs in our catalog \NGC~$\gtrsim 10$. In addition, we re-analyze the kinematics of GC systems bound to three brighter dEs found in the literature, VCC~1087, VCC~1261, and VCC~1528 \citep{Beasley06,Beasley09}. Only the data set for VCC~1861 consists of \NGC~$\sim 20$, for the remaining five dEs \NGC~$\sim 10$.

We apply two independent methods to measure the amplitude of the rotation and the velocity dispersion of the GC system. The first method, RF, is based on first fitting the rotation using a cosine function and then obtaining the velocity dispersion as the departure of the GCs to the best fit rotation curve. The second method, RDSF, is based on a maximum likelihood statistical approach that simultaneously fits the rotation amplitude, using a cosine function, and the velocity dispersion of the GC system. The rotation and velocity dispersion uncertainties obtained in both methods are estimated using a numerical bootstrap procedure. The kinematic properties of the GC systems measured using the RF and RDSF methods are consistent within the error bars.

We evaluate the statistical significance of the measured kinematics of the GC systems by creating artificial non-rotating GC systems based on our data. We model these non-rotating GC systems using two different kinds of simulations: velocity constrained simulations and velocity unconstrained simulations. Both approaches are tuned to reproduce as closely as possible the observed GC systems, i.e. they have the same number of observed GC satellites and the same median velocity uncertainties. In the case of VCC~1861, our largest GC system (\NGC~$=18$), the measured \Vmax\ is consistent with zero and these simulations of non-rotating systems coincide with this measurement. This suggests that the GC system of VCC~1861 is not rotating. The stellar rotation of VCC~1861 is also consistent with zero \citep{etj11,etj14b}. In the case of VCC~1528 (\NGC~$=10$), the measured \Vmax\ is between the $2\sigma_G$ and the $3\sigma_G$ contours of these two kinds of simulated non-rotating systems. This suggests that the GC system of VCC~1528 {\it may} be rotating, and the measured \Vmax~$=67.9 \pm 11.9$ should be considered an upper limit. However, a larger number of GCs is needed to confirm this result. The measured \Vmax\ and \sigGC\ in the remaining four target galaxies are between the $1\sigma_G$ and the $2\sigma_G$ contours of these two kinds of simulated non-rotating systems, which indicates that the measured kinematics are not statistically significant.

We have found that measuring accurately the kinematics of dwarf galaxies based on their GC systems may be more challenging than previously acknowledged. We have carried out careful simulations to explore the parameters that help to make these measurements. We find that there are three key parameters: (1) the intrinsic \Vmax/\sigGC\ of the GC system, (2) the number of observed GC satellites, and (3) the line-of-sight velocity uncertainties. In all cases, \Vmax\ tends to be overestimated and \sigGC\ tends to be underestimated. However, these offsets are larger for GC systems with low \Vmax/\sigGC\ values. Given that \Vmax/\sigGC\ is an unknown parameter, to obtain the smallest rotation and dispersion offsets, the number of observed GC satellites should be \NGC~$\gtrsim 20$. However, many dwarf galaxies do not have such large numbers of GCs, in those cases, the median velocity uncertainties should be as small as possible ($\delta V_{\rm GC} \leq 2$~\kms), although the offsets cannot be reduced to 0. Thus, for data sets with \NGC~$\lesssim 10$, the measured \Vmax\ can be considered an upper limit and the measured \sigGC\ a lower limit. 

We conclude that the current observations, those presented in this paper and in the literature, need to be improved, mainly in terms of number of GCs observed, to be able to know whether these systems are rotating or they are not. The current samples of GC satellites are not complete, i.e. some GCs that appear close in projection to these dEs are not included in the observations here and in the literature due to spatial slit conflicts, i.e. they cannot be observed at the same time in the same slitmask. New observations targeting these same galaxies including these not yet observed GCs would help to address whether these dEs are rotating or not. A major effort is needed to have a database of dynamical halo properties of Virgo cluster dEs to probe their formation scenarios.

\acknowledgments

The authors thank the referee, Igor Chilingarian, for useful suggestions that have helped to improve this manuscript.
E.T. acknowledges the financial support of the Fulbright Program jointly with the Spanish Ministry of Education. PG acknowledges the NSF grants AST-1010039 and AST-1412504. EWP and BL acknowledge support from the National Natural Science Foundation of China under Grant Nos. 11173003 and 11573002, and from the Strategic Priority Research Program, ”The Emergence of Cosmological Structures”, of the Chinese Academy of Sciences, Grant No. XDB09000105. The spectroscopic data presented herein were obtained at the W.M. Keck Observatory, which is operated as a scientific partnership among the California Institute of Technology, the University of California and the National Aeronautics and Space Administration. The Observatory was made possible by the generous financial support of the W.M. Keck Foundation. the photometric data presented herein is based on observations obtained with MegaPrime/MegaCam, a joint project of CFHT and CEA/DAPNIA, at the Canada-France-Hawaii Telescope (CFHT) which is operated by the National Research Council (NRC) of Canada, the Institut National des Sciences de Universe of the Centre National de la Recherche Scientifique (CNRS) of France and the University of Hawaii. This research also made use of the Canadian Astronomy Data Centre, which is operated by the National Research Council of Canada with the support of the Canadian Space Agency. The authors wish to recognize and acknowledge the very significant cultural role and reverence that the summit of Mauna Kea has always had within the indigenous Hawaiian community.  We are most fortunate to have the opportunity to conduct observations from this mountain.

\bibliographystyle{aa}
\bibliography{references}{}

\begin{appendix}

\section{Radial Velocity Measurement and Uncertainty for the Globular Clusters}\label{kin_measurements}

The line-of-sight radial velocities are measured using the penalized pixel-fitting method \citep[pPXF;][]{PPXF}. This software finds the best fit composite stellar template for the target galaxy. This composite stellar template is created as a linear combination of high signal-to-noise ratio (S/N; $100 < {\rm S/N} < 800$~\AA$^{-1}$) stellar templates that best reproduce the target object spectrum by employing non-linear least-squares optimization. Each template is given a different weight.
To overcome the template mismatch that can cause significant velocity uncertainties, we observed 31 stars that are used as templates. These stars were observed with Keck/DEIMOS using the same instrumental setup as described above. We used the LVMslits mask to observe these stars in a long slit. They were also trailed across the slit to uniformly illuminate it. These templates include a variety of stellar types, from B1 to M8, and luminosity classes, from supergiants to dwarfs, and also includes five carbon stars \citep[see][for a comparison of the velocity dispersion measurements obtained using a large set of observed stellar templates or stellar population models]{etj14b}.

Unresolved sources can be off centered across the slit, thus their measured radial velocity differs from the intrinsic value in an amount that depends on the position of the target across the slit. This effect can be corrected using the atmospheric B and A bands at $6850-7020$~\AA\  and $7580-7690$~\AA, respectively, because those will be affected by the same velocity offset. We fit these two bands masking the rest of the spectrum and using the same procedure as described for the target objects.

The radial velocity uncertainties are calculated by running 1000 Monte Carlo simulations. In each simulation, the flux of the spectrum is perturbed within a Gaussian function whose width is the uncertainty in the flux obtained in the reduction process. The radial velocity is measured in each simulation and their uncertainty is defined to be the standard deviation of the Gaussian distribution ($1\sigma_G$). All the radial velocity measurements and the Monte Carlo simulations are visually inspected and only those with a Gaussian distribution shape are included in the analysis.

The final radial velocity uncertainties are the square root of the quadratic addition of four components: (1) the Monte Carlo radial velocity uncertainty of the target object; (2) the Monte Carlo radial velocity uncertainty of the A and B atmospheric bands; (3) the uncertainty in the wavelength solution, which corresponds to $1.4$~\kms; and (4) the uncertainty in the radial velocity of the templates, which is $0.4$~\kms. The stellar templates are not affected by the ghosting effects that affect the science spectra. The wavelength calibration error of the templates is smaller than their radial velocity uncertainty.

We further asses the reliability of the velocity uncertainties by studying the agreement between the repeated measurements of science targets. Fifteen of our objects in the full catalog of GCs and stars are observed in two different slitmasks. For each of these objects we calculate the difference between the duplicated velocities divided by the the square root of the quadratic sum of their uncertainties. These differences follow a Gaussian distribution whose width is unity, which means that both estimations of the velocity are within the error bars, indicating that the velocity uncertainties are reliable.

\section{Radial Velocity and Velocity Dispersion of the dEs}\label{dE_measurements}

Along with the GC candidates, we also placed a short slit in the slitmask targeting the centers of the dEs to measure their line-of-sight radial velocity and velocity dispersion. These measurements and uncertainties are also estimated using the pPXF software and following the same procedure as described in Section \ref{kin_measurements}.

Table \ref{dEs_properties} shows the line-of-sight radial velocity, i.e. systemic velocity \Vsys, and the velocity dispersion measured in the central region, $\sigma_0$, of the dEs using pPXF. This central region corresponds to the extraction window described in Section \ref{obs}. The radius of this window, half its size, is indicated in Table \ref{dEs_properties} as $R_0$.

The heliocentric corrected velocities for VCC~1539, VCC~1545, and VCC~1861 agree within the $1\sigma_G$ uncertainties with the values measured by \citet{GOLDMine} and \citet{etj14b}. 

The central velocity dispersions are only available in the literature for VCC~1545 and VCC~1861. While the values agree within the $1\sigma_G$ uncertainties with those of \citet{etj14b}, they are 2.6 and 2.1 times smaller than those measured by \citet{Chil09} and \citet{Rys13}, respectively. The difference in the central velocity dispersion between the measurement presented here and the one presented in \citet{Chil09} could be due to the better seeing of the data presented here, $0.6''-0.9''$~(FWHM) versus $2.0''$, in combination with the possible kinematically decoupled core that may be present in this galaxy. See \citet{etj14b} for a detailed comparison between our procedure to measure the velocity dispersion and the literature.

\end{appendix}


\end{document}

%% file: table4.tex
\begin{table*} 
\begin{center} 
\caption{Properties of the Globular Cluster Satellites \label{GCsat_table}} 
\resizebox{12cm}{!}{ 
\begin{tabular}{c|c|c|c|c|c|c} 
\hline \hline 
RA         &    DEC    &    $g$   &   $z$   &  Catalog &   $V$         & S/N          \\ 
(hh:mm:ss) & (dd:mm:ss)&   (mag)  &  (mag)  &          & (km~s$^{-1}$) & (\AA$^{-1}$) \\ 
   (1)     &  (2)      &   (3)    &   (4)   &   (5)    &   (6)         &    (7)       \\ 
\hline 
\multicolumn{6}{c}{VCC~1539} \\ 
\hline 
12:34:04.96 & 12:44:32.40 & 22.69 & 21.76 & ACSVCS & 1520.2 $\pm$ 21.4 & 5.2\\ 
12:34:05.84 & 12:44:25.10 & 22.82 & 21.99 & ACSVCS & 1522.6 $\pm$ 23.6 & 5.1\\ 
12:34:06.27 & 12:44:10.50 & 22.54 & 21.59 & ACSVCS & 1485.7 $\pm$ 23.5 & 9.1\\ 
12:34:06.43 & 12:44:23.80 & 23.79 & 22.61 & ACSVCS & 1481.3 $\pm$ 14.5 & 2.8\\ 
12:34:06.67 & 12:44:34.00 & 23.79 & 23.25 & NGVS   & 1566.1 $\pm$ 19.2 & 3.5\\ 
12:34:07.20 & 12:44:42.36 & 21.25 & 20.37 & ACSVCS & 1548.9 $\pm$ 19.9 & 12.0\\ 
12:34:07.27 & 12:44:40.10 & 23.62 & 22.47 & ACSVCS & 1528.4 $\pm$ 22.5 & 2.8\\ 
12:34:07.39 & 12:44:25.60 & 23.13 & 22.22 & ACSVCS & 1512.2 $\pm$ 4.2 & 3.6\\ 
12:34:07.61 & 12:44:17.35 & 23.65 & 22.60 & ACSVCS    & 1515.8 $\pm$ 23.3 & 2.0\\ 
\hline 
\multicolumn{6}{c}{VCC~1545} \\ 
\hline 
12:34:08.69 & 12:01:59.80 & 23.55 & 22.69 & NGVS   & 2064.9 $\pm$ 7.4 & 3.7\\ 
12:34:09.77 & 12:03:28.60 & 23.50 & 22.66 & ACSVCS & 1977.4 $\pm$ 9.3 & 3.3\\ 
12:34:10.42 & 12:03:56.20 & 23.73 & 22.86 & ACSVCS & 2036.8 $\pm$ 15.6 & 2.9\\ 
12:34:10.56 & 12:02:59.30 & 24.52 & 23.37 & ACSVCS & 2056.2 $\pm$ 24.4 & 2.0\\ 
12:34:10.88 & 12:03:26.90 & 23.56 & 22.81 & ACSVCS & 1970.1 $\pm$ 12.0 & 2.9\\ 
12:34:11.34 & 12:02:58.63 & 24.02 & 22.68 & ACSVCS & 2060.0 $\pm$ 13.5 & 6.0\\ 
12:34:11.37 & 12:01:51.40 & 23.85 & 22.97 & ACSVCS & 2167.2 $\pm$ 19.6 & 2.9\\ 
12:34:11.55 & 12:02:21.40 & 22.22 & 21.31 & ACSVCS & 2053.0 $\pm$ 16.5 & 10.8\\ 
12:34:11.56 & 12:03:22.50 & 23.92 & 22.97 & ACSVCS & 2012.9 $\pm$ 13.4 & 2.7\\ 
12:34:11.83 & 12:02:49.60 & 22.50 & 21.51 & ACSVCS & 2060.0 $\pm$ 18.6 & 9.5\\ 
12:34:11.93 & 12:03:12.00 & 24.15 & 23.12 & ACSVCS & 2032.5 $\pm$ 21.3 & 1.7\\ 
12:34:14.03 & 12:02:40.10 & 22.61 & 21.64 & ACSVCS & 2079.2 $\pm$ 11.3 & 8.5\\ 
12:34:15.37 & 12:02:55.30 & 23.70 & 22.82 & ACSVCS & 2112.4 $\pm$ 7.1 & 3.0\\ 
\hline 
\multicolumn{6}{c}{VCC~1861} \\ 
\hline 
12:40:51.72 & 11:10:44.20 & 23.67 & 22.79 & NGVS   & 558.2 $\pm$ 14.4 & 3.5\\ 
12:40:55.95 & 11:11:00.00 & 23.07 & 22.21 & ACSVCS & 670.8 $\pm$ 17.2 & 4.5\\ 
12:40:57.08 & 11:11:11.00 & 23.57 & 22.44 & ACSVCS & 715.4 $\pm$ 24.9 & 3.6\\ 
12:40:57.10 & 11:10:31.40 & 23.65 & 22.47 & ACSVCS & 622.8 $\pm$ 23.5 & 4.3\\ 
12:40:57.14 & 11:11:37.50 & 22.86 & 22.04 & ACSVCS & 627.4 $\pm$ 7.6 & 5.7\\ 
12:40:57.37 & 11:11:01.80 & 23.65 & 22.66 & ACSVCS & 628.1 $\pm$ 17.0 & 2.1\\ 
12:40:57.62 & 11:10:05.60 & 23.29 & 22.33 & ACSVCS & 650.6 $\pm$ 13.8 & 5.1\\ 
12:40:57.73 & 11:10:37.70 & 23.47 & 22.26 & ACSVCS & 644.1 $\pm$ 6.0 & 5.0\\ 
12:40:58.51 & 11:11:09.80 & 23.63 & 22.24 & ACSVCS & 629.6 $\pm$ 21.9 & 3.7\\ 
12:40:58.71 & 11:11:06.40 & 23.69 & 22.35 & ACSVCS & 634.6 $\pm$ 24.6 & 4.6\\ 
12:40:59.73 & 11:11:17.20 & 22.76 & 21.83 & ACSVCS & 615.0 $\pm$ 13.5 & 6.1\\ 
12:40:59.98 & 11:11:17.00 & 22.94 & 21.65 & ACSVCS & 665.2 $\pm$ 13.0 & 8.3\\ 
12:41:00.01 & 11:11:37.30 & 23.57 & 22.53 & ACSVCS & 645.4 $\pm$ 20.4 & 2.8\\ 
12:41:00.08 & 11:10:36.90 & 21.51 & 20.65 & ACSVCS & 571.4 $\pm$ 11.9 & 18.0\\ 
12:41:00.17 & 11:11:02.60 & 23.61 & 22.38 & ACSVCS & 595.1 $\pm$ 32.3 & 4.0\\ 
12:41:00.27 & 11:10:48.90 & 23.10 & 22.28 & ACSVCS & 639.8 $\pm$ 23.4 & 5.2\\ 
12:41:00.47 & 11:10:24.50 & 21.75 & 20.91 & ACSVCS & 594.1 $\pm$ 42.2 & 15.1\\ 
12:41:03.00 & 11:10:44.90 & 21.97 & 21.17 & ACSVCS & 662.7 $\pm$ 16.1 & 14.5\\ 
\hline 
\end{tabular} 
} 
\end{center} 
\tablecomments{Columns 1 and 2: coordinates in J2000. Columns 3 and 4: $g$ and $z$ band magnitudes in the AB system. Column 5: photometric catalog from which the magnitudes are taken. In the case of ACSVCS, the magnitudes come from HST/ACS observations \citep{Peng06}, otherwise they come from the NGVS (Peng et al., in prep.). The magnitudes from the ACSVCS catalog are corrected to the NGVS astrometric zeropoint. Column 6: heliocentric line-of-sight radial velocity of the GC. Column 7: signal-to-noise ratio measured in the CaT region of the spectrum. See \citet{Beasley06,Beasley09} for the properties of the GC satellites of VCC~1087, VCC~1261, and VCC~1528.} 
\end{table*}